\documentclass[fleqn,usenatbib]{mnras}

\usepackage{newtxtext, newtxmath}

\usepackage[T1]{fontenc}

\DeclareRobustCommand{\VAN}[3]{#2}
\let\VANthebibliography\thebibliography
\def\thebibliography{\DeclareRobustCommand{\VAN}[3]{##3}\VANthebibliography}

\usepackage{graphicx}	
\usepackage{amsmath}	
\usepackage{multicol}
\usepackage{physics}
\usepackage{bm}
\usepackage{ulem}
\usepackage{comment} 
\usepackage{physics}
\usepackage{xcolor}
\usepackage{mathrsfs}
\usepackage{tabularx}
\usepackage{threeparttable}
\usepackage[title]{appendix}
\allowdisplaybreaks
\usepackage{CJKutf8}

\newcommand{\sign}[1]{\mathrm{sgn}\qty(#1)}
\graphicspath{{./}{figures/}}

\title{The Cosmic Ray Staircase: the Outcome of the Cosmic Ray Acoustic Instability}

\author[Tsung et al.]{
Tsun Hin Navin Tsung,$^{1}$\thanks{E-mail: ttsung@ucsb.edu}
S. Peng Oh,$^{1}$
Yan-Fei Jiang(姜燕飞)$^{2}$
\\
$^{1}$Dept. of Physics, University of California, Santa Barbara, CA 93106, USA\\
$^{2}$Center for Computational Astrophysics, Flatiron Institute, New York, NY 10010, USA\\
}

\date{Accepted XXX. Received YYY; in original form ZZZ}

\pubyear{2020}

\begin{document}
\begin{CJK*}{UTF8}{gbsn}
\label{firstpage}
\pagerange{\pageref{firstpage}--\pageref{lastpage}}
\maketitle

\end{CJK*}

\begin{abstract}
Recently, cosmic rays (CRs) have emerged as a leading candidate for driving galactic winds. Small-scale processes can dramatically affect global wind properties. We run two-moment simulations of CR streaming to study how sound waves are driven unstable by phase-shifted CR forces and CR heating. We verify linear theory growth rates. As the sound waves grow non-linear, they steepen into a quasi-periodic series of propagating shocks; the density jumps at shocks create CR bottlenecks. The depth of a propagating bottleneck depends on both the density jump and its velocity; $\Delta P_c$ is smaller for rapidly moving bottlenecks. A series of bottlenecks creates a CR staircase structure, which can be understood from a convex hull construction. The system reaches a steady state between growth of new perturbations, and stair mergers. CRs are decoupled at plateaus, but exert intense forces and heating at stair jumps. The absence of CR heating at plateaus leads to cooling, strong gas pressure gradients and further shocks. If bottlenecks are stationary, they can drastically modify global flows; if their propagation times are comparable to dynamical times, their effects on global momentum and energy transfer are modest. The CR acoustic instability is likely relevant in thermal interfaces between cold and hot gas, as well as galactic winds. Similar to increased opacity in radiative flows, the build-up of CR pressure due to bottlenecks can significantly increase mass outflow rates, by up to an order of magnitude. It seeds unusual forms of thermal instability, and the shocks could have distinct observational signatures. 
\end{abstract}




\section{Introduction} \label{sec:introduction}

It is generally believed that cosmic rays (CR) should play crucial dynamical roles in the interstellar and circumgalactic medium (ISM, CGM) because the energy density of these high-energy particles is comparable to the thermal energy of the gas or the magnetic field \citep{blandford87}. The coupling between CRs and the thermal plasma is believed to be mediated through the streaming instability \citep{kulsrud69} in which CRs pitch-angle scattered by hydromagnetic waves causes the waves to grow and thus lead to more scattering. This wave-particle interaction causes energy and momentum to be transferred between the gas and CRs. On global scales, the interaction of waves with CRs are key to the transport and confinement of CRs in a galaxy. Cosmic Rays can provide a significant amount of non-thermal support \citep{ji20,crocker21a} and is a strong candidate for driving galactic winds \citep{ipavich75,breitschwerdt91,uhlig12,ruszkowski17,crocker21b,hopkins21a}. On smaller scales, CRs accelerated by shocks can modify shock structures \citep{blandford87,drury81,volk84,haggerty20,tsung20} and impact the entrainment, survival and destruction of cold clouds \citep{bruggen20,bustard20}. Thus CRs can significantly affect the multiphase structure of the ISM and CGM.

Even though details of the wave-particle interaction are inherently kinetic, in the limit of strong scattering a fluid description is possible and more practical for galaxy (or cosmological) scale simulations. CRs, treated as a bulk fluid, have the following general transport modes: 1. Wave-particle interactions lock the bulk of CRs with the local Alfven wave, causing them to advect at the Alfven speed along magnetic fields (streaming). 2. Slippage from perfect wave locking causes CRs to diffuse relative to the local Alfven wave frame, down the CR pressure gradient (diffusion). More detailed transport models in the presence of various wave damping mechanisms have been studied (e.g. ion-neutral damping \citep{farber18,bustard20}, turbulent damping \citep{holguin19}, dust damping \citep{squire21} or some combination thereof \citep{hopkins21b}). There is, however, no consensus within the community as to the correct form of CR transport in the ISM and CGM. One important observational constraint lies in reconciliation with gamma ray observations. Gamma-ray emission from pion production by CRs is over-produced in simulations unless CRs can be rapidly transported out of dense star forming regions \citep{chan19}. \citet{thomas20} modeled harp-like structures in radio synchrotron maps of the Galactic center. Their analysis suggested streaming dominated transport rather than diffusion. 

In the fluid description, CRs have been found to modify well-known fluid instabilities such as the Parker instability \citep{ryu03,rodrigues16,heintz18,heintz20}, magneto-rotational instability \citep{kuwabara15}, thermal instability \citep{shadmehri09,kempski20,butsky20}, Kelvin-Helmholtz instability \citep{suzuki14}, etc., while driving some entirely new instabilities, such as the CR acoustic instability \citep{drury86,begelman94}. The CR acoustic instability arises when CRs amplify sound waves, via CR pressure forces and/or CR heating of the gas. This causes acoustic waves to increase in amplitude and steepen into shocks. In this paper, we generalize and test previous linear theory predictions for the CR acoustic instability, and study its non-linear saturation. We find a characteristic staircase structure in the CR pressure profile-- a new feature in CR transport -- and explain its physical origin. 

In the diffusion dominated regime, \citet{drury86} found that the acoustic instability occurs when the CR pressure scale height $L_{\rm c} \equiv P_c/\nabla P_c$ is shorter than the diffusion length $l_{\rm diff} \sim \kappa/c_s$ (where $\kappa$ is the diffusion coefficient and $c_s$ is the gas sound speed), a condition not easily met except at shock precursors (see \citet{quataert21} for application to galactic winds, where they find the instability to be unimportant). \citet{kang92} performed simulations of its non-linear growth at shocks and found that acoustic waves can steepen into many small scale shocks, resulting in enhanced particle acceleration. \citet{ryu93} found, in a 2D shock setup, that the steepened acoustic waves can create density inversions, trigger a secondary Rayleigh-Taylor instability and generate turbulence in the downstream. All in all, the CR diffusion driven acoustic instability is mostly relevant at shocks. 

On the other hand, \citet{begelman94} found that in the streaming dominated regime, CR heating can cause acoustic modes to become unstable even without a sharp CR pressure gradient. They speculated that the acoustic modes would, in the non-linear regime, generate constant CR pressure regions (CR plateaus) separated by sudden drops, although they were unable to test this. We shall see in this paper, fulfilment of their prescient predictions.

Numerical simulation of this streaming driven acoustic instability have not yet been conducted to date. In the past, such simulations were infeasible due to a numerical instability which arises at CR pressure gradient zeros. Regularization of this instability \citep{sharma09} requires very high resolution and short time-steps, making the calculation infeasibly expensive. In recent years, a new two-moment method \citep{jiang18,thomas19} now makes this calculation possible. The two moment method has already been deployed in FIRE simulations of galaxy formation \citep{chan19,hopkins21a}. 

We will, in this paper, utilize this relatively new tool to study the linear and non-linear growth of the streaming driven acoustic instability. We begin, in \S\ref{sec:analytic}, with an analytic discussion of the CR acoustic instability and relevant physics. In \S\ref{sec:simulation} we describe our simulation setup and results in the linear and non-linear regime. We proceed in \S\ref{sec:discussion} a discussion of its astrophysical significance and conclusions. In Appendix \ref{app:linear_growth_rates}, we derive the linear growth rate of the CR acoustic instability. A resolution study is conducted in Appendix \ref{app:resol_rspl}.

\section{Analytic Considerations} \label{sec:analytic}

The two-moment equations governing the dynamics of a CR-MHD coupled fluid is given by \citealt{jiang18}
\begin{gather}
    \pdv{\rho}{t} + \nabla\cdot\qty(\rho\vb{v}) = 0, \label{eqn:continuity} \\
    \pdv{\qty(\rho\vb{v})}{t} + \nabla\cdot\qty(\rho\vb{v}\vb{v} - \vb{B}\vb{B} + P^*\vb{I}) = \bm{\sigma}_c\cdot\qty[\vb{F}_c - \qty(E_c + P_c)\vb{v}] + \rho\vb{g}, \label{eqn:momentum} \\
    \pdv{E}{t} + \nabla\cdot\qty[\qty(E + P^*)\vb{v} - \vb{B}\qty(\vb{B}\cdot\vb{v})] = \qty(\vb{v} + \vb{v}_s)\cdot\bm{\sigma}_c\cdot \nonumber \\
    \qquad\qty[\vb{F}_c - \qty(E_c + P_c)\vb{v}] + \rho\vb{g}\cdot\vb{v} + \mathcal{L}, \label{eqn:energy} \\
    \pdv{\vb{B}}{t} = \nabla\cross\qty(\vb{v}\cross\vb{B}), \label{eqn:induction} \\
    \pdv{E_c}{t} + \nabla\cdot\vb{F}_c = -\qty(\vb{v} + \vb{v}_s)\cdot\bm{\sigma}_c\cdot\qty[\vb{F}_c - \qty(E_c + P_c)\vb{v}], \label{eqn:cr_energy} \\
    \frac{1}{c^2}\pdv{\vb{F}_c}{t} + \nabla P_c = -\bm{\sigma}_c\cdot\qty[\vb{F}_c - \qty(E_c + P_c)\vb{v}], \label{eqn:cr_flux}
\end{gather}
where $c$ is the speed of light, $\mathcal{L} = \mathcal{H} - \mathcal{C}$ is gas heating minus cooling, $\vb{v}_s = -\vb{v}_A\sign{\vb{B}\cdot\nabla P_c}$ is the streaming velocity, $P^* = P_g + B^2/2$, $E = \rho v^2/2 + P_g/\qty(\gamma_g - 1) + B^2/2$ and $\bm{\sigma}_c$ is the interaction coefficient defined by
\begin{gather}
    \bm{\sigma}_c^{-1} = \bm{\sigma}_d^{-1} + \frac{\vb{B}}{\abs{\vb{B}\cdot\nabla P_c}}\vb{v}_A\qty(E_c + P_c), \label{eqn:interaction_coef} \nonumber \\
    \bm{\sigma}_d^{-1} = \frac{\bm{\kappa}}{\gamma_c - 1}. \label{eqn:diffusion_coef}
\end{gather}
where $\kappa$ is the CR diffusion coefficient. For simplicity we assume $\kappa$ to be constant and time-steady, ignoring the dynamics of magnetic waves (see \citealt{thomas19} for a full inclusion). This assumption can be relaxed by using the equilibrium $\kappa$ calculated from linear theory (see the appendix of \citealt{jiang18}, and \citealt{bustard20} for an implementation of ion-neutral damping). 
CRs exchange momentum according to the source term $\bm{\sigma}_c\cdot\qty[\vb{F}_c - \qty(E_c + P_c)\vb{v}]$ and energy according to $\qty(\vb{v} + \vb{v}_s)\cdot\bm{\sigma}_c\cdot\qty[\vb{F}_c - \qty(E_c + P_c)\vb{v}]$. We shall call these the generalized CR forcing and heating terms respectively. Microscopically, some degree of anisotropy in the CR distribution is required to trigger the streaming instability; macroscopically, this translates to requiring a finite $P_c$ gradient. As $\nabla P_c \rightarrow 0$, the interaction coefficient $\sigma_{c} \rightarrow 0$ (equation \ref{eqn:interaction_coef}), and CRs can free stream at the speed of light, as encapsulated by the time-dependent term in equation \ref{eqn:cr_flux}. The condition for the time-dependent term in equation \ref{eqn:cr_flux} to be negligible is:
\begin{equation}
    L_c = \frac{P_c}{\nabla P_c} \ll \frac{c^2}{v_A^2} v_A\Delta t \label{eqn:coupling_cond}.
\end{equation}
where $\Delta t$ is a dynamical time. This sets a condition on the strength of the $P_c$ gradient. If it is fulfilled, the equations reduce to the standard one-moment equations \citep{skilling75a,breitschwerdt91}, and the CR flux, from equation \ref{eqn:cr_flux}, reduces to
\begin{equation}
    \vb{F}_c = \qty(\vb{v} + \vb{v}_s)\qty(E_c + P_c) - \frac{1}{\gamma_c - 1}\nabla\cdot\bm{\kappa}\cdot\nabla P_c, \label{eqn:steady_state_flux}
\end{equation}
which shows that in the well-coupled limit, CR transport is given as a sum of advection, streaming and diffusion processes. The CR energy equation (equation \ref{eqn:cr_energy}) reduces to:
\begin{equation}
\pdv{E_c}{t} + \nabla\cdot\vb{F}_c = -\qty(\vb{v} + \vb{v}_s)\cdot \nabla P_c
\label{eqn:well-coupled-CRs} 
\end{equation}
where $\vb{F}_c$ is given by equation \ref{eqn:steady_state_flux}. The RHS, written in this form shall be called the coupled CR heating term, while the coupled CR forcing term is $\nabla P_c$. In \S\ref{subsec:linear_theory}, we will use this canonical form of the CR equations in the well-coupled limit. 

In this study we ignore any CR collisional losses due to Coulomb collisions and hadronic interactions. These losses are important in dense gas, but are unlikely to be important in the diffuse halo gas.

We now discuss two key pieces of physics: linear growth rates for the CR acoustic instability (\S\ref{subsec:linear_theory}), and the CR bottleneck effect (\S\ref{subsubsec:bottleneck}). 

\subsection{CR Acoustic Instability: Linear Theory} \label{subsec:linear_theory}

\begin{figure}
    \centering
    \includegraphics{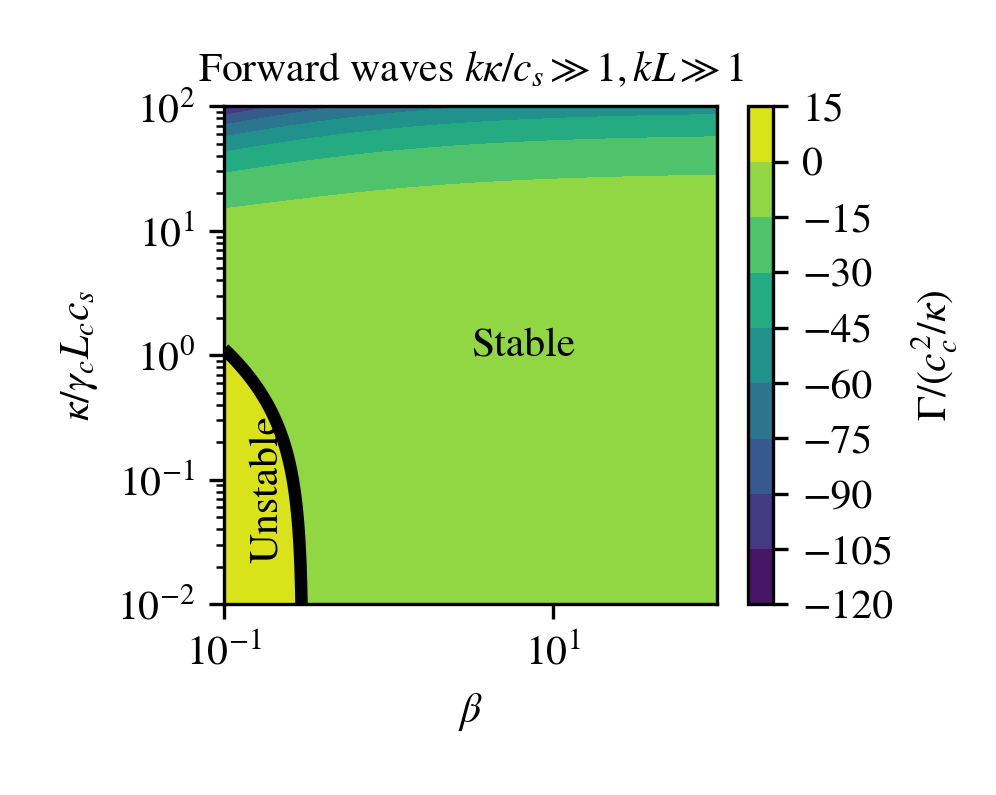} \\
    \includegraphics{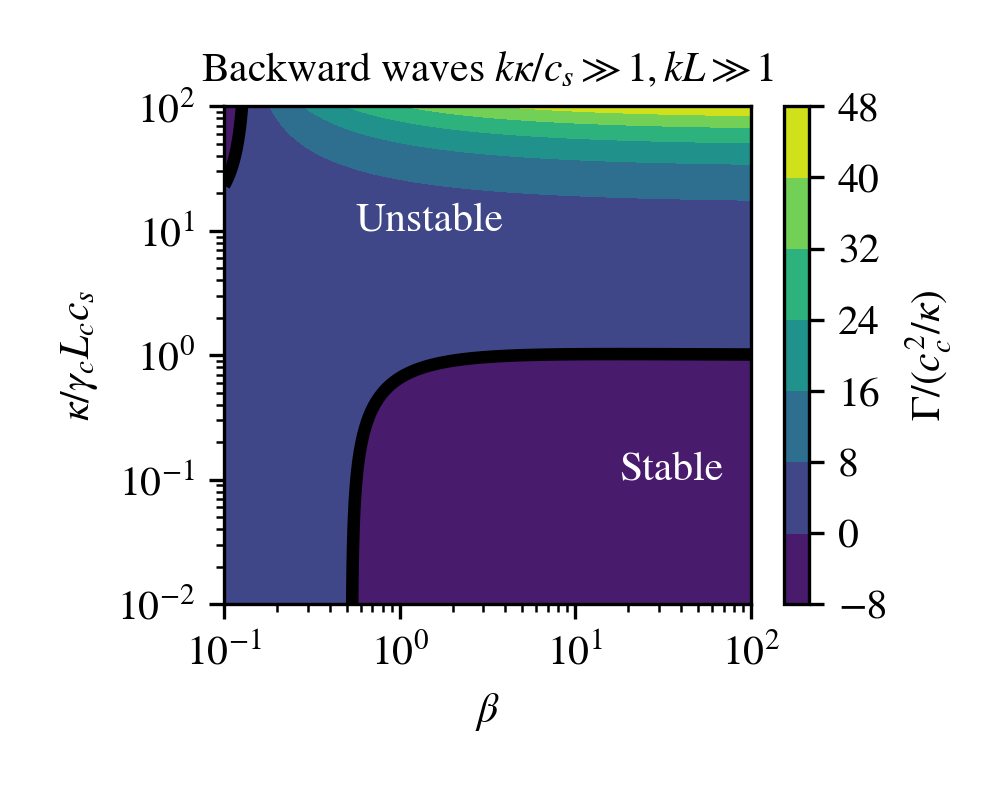}
    \caption{Growth rate (in units of $c_c^2/\kappa$) as function of $\eta = \kappa/\gamma_c L_c c_s$ and $\beta$ of the forward (top) and backward (bottom) acoustic waves in the short wavelength limit ($k\kappa/c_s\gg 1$, $k L\gg 1$). The stable and unstable regimes are demarcated by a thick black line.}
    \label{fig:growth}
\end{figure}

In this section, we make order of magnitude arguments for the threshold and growth rate of instabilities driven primarily by CR diffusion and streaming respectively, in the most physically relevant asymptotic limits for the CGM. The detailed dispersion relations are derived in Appendix \ref{app:linear_growth_rates}, and solutions to these dispersion relations give the growth rates shown in Fig. \ref{fig:growth}. Broadly speaking, in this section we seek to understand the features seen in Fig. \ref{fig:growth}. The reader can get a feel for the physics of the instability here, which are relevant to understanding the simulation results in \S\ref{sec:simulation}; only those interested in the technical details need to consult Appendix \ref{app:linear_growth_rates}. 

If CRs were completely locked to the gas, the system would simply behave as a single fluid with adiabatic index intermediate between $\gamma_c = 4/3$ and $\gamma_g=5/3$, depending on $\alpha=P_c/P_g$. However, CRs can both stream and diffuse relative to the gas, which leads to a phase offset between gas CR pressure and gas density perturbations. In addition, CRs affect gas pressure perturbations by heating the gas as they stream. Acoustic waves thus experience external forcing. If this forcing is in phase with wave oscillations, they grow; otherwise, they damp.  

There are several characteristic lengthscales in the problem: 
\begin{itemize} 
\item{The mode wavelength, $\lambda \sim k^{-1}$.} 
\item{The CR diffusion length $l_\mathrm{diff} \sim \kappa/c_s$. This is the lengthscale over which the sound crossing time  $t_{\rm sc} \sim L/c_{\rm s}$ and the diffusion time $t_{\rm diff} \sim L^{2}/\kappa$ are comparable. On scales below $l_\mathrm{diff}$, diffusion is faster than advection.}
\item{The CR scale height $L_c = \abs{P_c/\nabla P_c}$, as well as the gas pressure scale height $L_{\rm g}$ and the density scale height $L_{\rho}$, defined similarly.}
\end{itemize}
Additionally, there are two important dimensionless parameters: $\beta=P_{g}/P_{B}$, and $\alpha =P_c/P_g$. Finally, the direction of the sound wave, and in particular whether the sound wave propagates down (`forward' wave) or up (`backward' wave) the CR pressure gradient also affects instability and growth rates.  

We work in the WKB approximation $k L_c \gg 1$. Furthermore, we ignore background gas pressure and density gradients, i.e. we assume a uniform background $L_{g}, L_{\rho} \rightarrow \infty$. In Appendix A, we show that our results are unchanged even if we allow for non-zero gas pressure and density gradients. Essentially, this is because in the WKB approximation, $k L_g, k L_{\rho} \gg 1$, the background looks locally uniform. We still retain the CR scale height $L_c$ because there is an additional lengthscale in CR dynamics, the diffusion length $l_{\mathrm{diff}}  \sim \kappa/c_s$. The ratio $\eta\equiv l_\mathrm{diff}/L_c \sim \kappa/L_c c_s$ impacts CR dynamics and instability growth. If we work in the approximation where besides $k L_c \gg 1$, $k l_\mathrm{diff} = k \kappa/c_s \gg 1$ (i.e. the diffusion time is much shorter than the wave period), then the ratio $\eta = \kappa/L_c c_s$ is the only dimensionless parameter involving lengthscales which is important. For the purposes of this subsection, we will work in the limit where $L_c$ is small enough that CRs are well-coupled to the thermal gas, and equations \ref{eqn:steady_state_flux} and \ref{eqn:well-coupled-CRs} apply. 

For simplicity, we discuss regimes where either CR diffusion and streaming dominate. The diffusion coefficient $\kappa$ is assumed constant in space and time. Since diffusion rates are independent of B-field strength, while streaming velocities and heating rates are both proportional to $v_{\rm A} \propto B$, we expect that diffusion and streaming dominated regimes correspond to high and low $\beta$ respectively, a notion we shall quantify.  

\subsubsection{Diffusion dominated} 

{\it Damping.} `Drag' against CRs provides a frictional force which damps sound waves, a phenomenon known as Ptuskin damping \citep{ptuskin81}. The physics is very similar to radiative damping of sound waves, which famously leads to Silk damping of acoustic waves in the early universe. We can estimate the damping rate as follows. Sound waves are just a simple harmonic oscillator (SHO), where the restoring force is proportional to displacement $F \propto - x$. If CR diffusion produces a perturbed CR force which is instead proportional to velocity, $F \propto -v$, then just as for the SHO, this force will damp oscillations, since it is $\pi/2$ radians out of phase with the restoring force\footnote{Mathematically, this must be true since the diffusion operator brings down an additional factor of $i$ compared to the gradient operator.}. Since we work in the limit $k \kappa/c_s \gg 1$, where diffusion is much more rapid than advection on scales of the wave period, the Lagrangian time derivative in the CR energy equation (equation \ref{eqn:well-coupled-CRs}) can be ignored, and CR compression is balanced by diffusion: 
$i \gamma_{c} P_{\rm c,0} k u_1 \sim - \kappa k^{2} P_{c,1}$, which gives rise to an acceleration:
\begin{equation}
    \dot{u}_1 \sim -\frac{1}{\rho} \nabla P_{c,1} \sim - \frac{P_{\rm c,0} u_1}{\rho \kappa} \sim - \frac{c_{c}^{2}}{\kappa} u_1 
\label{eq:ptuskin_force} 
\end{equation}
which is indeed proportional to velocity ($\dot{u}_{1} \propto - u_{1}$), and damps the wave, with damping rate: 
\begin{equation}
    \Gamma_{\rm damp}  \sim \frac{\dot{u_1}}{u_1} \sim - \frac{c_{c}^{2}}{\kappa}.
    \label{eq:diffusion_damping} 
\end{equation}
Note that the frictional force, and hence the damping rate, is independent of wavelength in this limit. Using $|P_{c,1}/P_{c,0}| \sim u_1/(\kappa k)$, $P_{\rm g,1}/P_{\rm g,0} \sim u_1/c_s$, we find that rapid diffusion causes the CR pressure perturbation to be suppressed:
\begin{equation}
    \left| \frac{P_{\rm c,1}}{P_{\rm g,1}} \right| \sim \frac{c_{\rm s}}{k \kappa} \left( \frac{P_{\rm c,0}}{P_{\rm g    ,0}} \right) \ll 1.
\end{equation}
Since CR pressure perturbations do not provide a restoring force but a damping force, the acoustic mode is driven by gas pressure perturbations, and propagates at the gas sound speed $c_s$. Furthermore, since the cosmic ray pressure perturbations are so small, the damping time is much longer than the wave period, $1/t_{\rm damp} c_s k \sim c_{s}/(k \kappa) (P_{\rm c,0}/P_{\rm g,0}) \ll 1$, even if equipartition holds $P_{\rm c,0} \sim P_{\rm g,0}$. Note also from equation \ref{eq:diffusion_damping} that damping is stronger for a weaker diffusion coefficient: the CR pressure perturbations are still $\pi/2$ out of phase, but now have larger amplitude. However, they can now only suppress smaller scale perturbations. 

{\it Growth with a Background gradient.} If sound waves propagate in a medium with a background CR gradient, they can be driven unstable \citep{drury86}. This can be understood as follows. Consider the limit of rapid diffusion. In this case, the CR gradient is time-steady and simply given by the background gradient, which is much larger than the perturbed CR gradients due to sound waves\footnote{This is no longer true in the non-linear phase of the instability; we address this in numerical simulations.}, which are strongly suppressed by diffusion. Since the CR gradient $-\nabla P_c$ is independent of density, any fluctuations in density will result in a differential acceleration, since underdense regions receive a larger force per unit mass: 
\begin{equation}
    \dot{u}_{1} \sim \frac{\rho_1}{\rho^{2}} \nabla P_{c,0} \sim \mp \frac{u_1}{c_s} \frac{P_{c,0}}{\rho L_c} \sim \mp \frac{u_1}{c_s} \frac{c_c^2}{L_c}
\end{equation}
where we have used $\rho_1/\rho \sim u_1/c_s$, and the $\mp$ sign refers to forward and backward waves respectively. Thus, underdense (overdense) regions having relative acceleration down (up) the gradient. The above force is proportional to velocity, and can either drive or damp sound waves. Consider density maxima, where the velocity perturbation $u_1$ has the same direction as wave propagation. In a forward wave, the wave and hence $u_1$ point down the CR gradient, but dense regions are accelerated up the gradient. We have $\dot{u_1} \propto - u_1$, and the wave is damped. Conversely, for a backward wave, $\dot{u_1} \propto u_1$, and the wave grows in amplitude. The growth rate is: 
\begin{equation}
    \Gamma_{\rm growth,diffuse} \sim \frac{\dot{u_1}}{u_1} \sim  \frac{c_c^{2}}{c_s L_c}. 
    \label{eq:diffusion_growth} 
\end{equation}
For growth driven by a background CR gradient to overcome Ptuskin damping, we see from equation \ref{eq:diffusion_damping} and \ref{eq:diffusion_growth} that we require: 
\begin{equation} 
    \frac{\kappa}{c_s L_c} > 1 \ \ {\rm (growth)} 
\label{eq:growth_criterion_diffusion}
\end{equation} 
For the sound wave to see a steady CR gradient $\nabla P_c$ independent of density, the diffusion time must be shorter than the sound crossing time across a scale height $L_c$, which is equivalent to equation \ref{eq:growth_criterion_diffusion}. 

\subsubsection{Streaming dominated}

We now consider the streaming dominated regime. For simplicity, and similar to \citet{begelman94}, we consider a weak background gradient ($L_c$ large) which is sufficient to couple CRs to the gas and give the streaming velocity a definite sign\footnote{CRs are assumed to always stream down the background gradient, which is presumed to be larger than any gradients induced by the sound wave. If this is no longer true, very interesting consequences arise, which we explore in \S\ref{sec:simulation}.}, but otherwise does not affect CR dynamics. In particular, the force and heating from the background gradient is assumed to be negligible. We will relax this assumption shortly. The magnitude of the background gradient has important implications for CR coupling and non-linear saturation, but here it just complicates matters. We do include diffusion in our WKB analysis, which is essential because otherwise there is no $\pi/2$ phase offset between CRs and density perturbations; streaming with flux $F_{\rm c} \propto P_c$ (rather than $F_{\rm c} \propto \nabla P_c$) cannot introduce a $\pi/2$ phase shift\footnote{Importantly, stratification can introduce phase shifts, so that sound waves can be destabilized for the pure streaming case in a stratified background \citep{quataert21}. The instability discovered by \citet{quataert21} is driven purely by phase shifts and does not rely on heating; hence it can operate even in isothermal gas.}. For any finite scattering rate, CRs are imperfectly locked to the Alfven wave frame, and will always diffuse relative to the wave frame. 

CR streaming has two effects. First, it introduces an additional advective component to CR transport which can be either aligned or anti-aligned with gas motions. Thus, it modulates the amplitude and even the sign of CR perturbations. Since the phase shift between CRs and gas depends on the competition between advective and diffusive transport, we might expect that as before, growth/damping depends on whether the wave is forward or backward. Second, CR streaming heats the gas, at a rate $v_{\rm A} \cdot \nabla P_c$, which perturbs the gas pressure. Both of these processes are only important if the streaming velocity $v_{\rm A}$ is large compared to the gas sound speed $c_s$, or at low $\beta \sim (c_s/v_A)^2$. 

Heating is a new consideration, particular to CR streaming. Does it drive growth or damping? CR compression followed by gas heating as CRs stream out of an overdensity is a situation where the  adiabatic index of the system is increasing, as energy is transferred from CRs (more compressible) to gas (less compressible). This stiffening of the equation of state is equivalent to a secular increase in the spring constant of a simple harmonic oscillator, which drives overstable oscillations. The peak pressure (arising from CR heating as CRs stream out of an overdensity) lags the peak density, and so work is done on the fluid during the expansion phase. CRs give up more energy streaming out of an overdensity than they receive during compression, and there is net energy transfer from CRs to gas\footnote{This is in contrast to the diffusion case, where CRs expand `for free', without transferring energy to the gas. In this case, there is net energy transfer from the gas to the CRs, and the wave damps.}. Unlike the perturbed CR force, these effects are independent of the direction of wave propagation, so we expect heating to be destabilizing for both forward and backward waves. 

We can make order of magnitude estimates for these remarks. Let us write the perturbed acceleration $\dot{u}_1 \approx \dot{u}_{1,a} + \dot{u}_{1,b}$, where $\dot{u}_{1,a}$ arises due to the phase-shifted CR force and $\dot{u}_{1,b}$ arises from gas pressure gradients due to CR heating. The calculation of the phase-shifted, perturbed CR force is the same as for Ptuskin damping, where compression and diffusion balance, except that now: 
\begin{equation}
    u_{1} \rightarrow u_{1} + v_{\rm A,1} = u_1 - \frac{1}{2} \frac{\rho_{1}}{\rho_0} v_A = u_1 \left(1 \mp \frac{v_{\rm A}}{c_s} \right)  
\end{equation}
where we have used $\rho_1/\rho \sim \pm u_1/c_s$, and $\mp$ sign is for forward and backward waves respectively ($v_{\rm A,1}$ always points down the CR gradient, whereas $u_1$ depends on whether the wave is forward or backward). From substituting this replacement for $u_1$ into equation \ref{eq:ptuskin_force}, we obtain a perturbed acceleration from the phase-shifted CR force:
\begin{equation}
    \dot{u}_{1,a} \sim - \frac{1}{\rho} \nabla P_{c,1} \sim - \frac{c_{c}^{2}}{\kappa} \left(1 \mp \frac{v_{\rm A}}{2c_{s}} \right) u_1 
    \label{eq:u1a} 
\end{equation}
The perturbed gas pressure from heating is $\dot{E}_g \sim \omega P_{g,1}/(\gamma_g-1) \sim v_A \cdot \nabla P_c \sim \pm i v_A k P_c$. Solving for $P_{g,1}$, and using $\omega \sim k c_s$, we obtain a perturbed acceleration from CR heating: 
\begin{equation} 
\dot{u}_{1,b} \sim - \frac{1}{\rho} \nabla P_{g,1} \sim \pm (\gamma_g -1) \frac{v_A}{c_s} \dot{u}_{1,a}. 
\end{equation}
We thus obtain a net acceleration: $\dot{u}_1= \dot{u}_{1,a} + \dot{u}_{1,b} = (1 \pm (\gamma_g -1) v_A/c_s) \dot{u}_{1,a}$. Using equation \ref{eq:u1a} and $\Gamma=u_1/\dot{u}_1$, we obtain: 
\begin{equation} 
\Gamma_{\rm stream} = -\frac{c_c^{2}}{2 \kappa} \left( 1 \mp \frac{1}{2 \beta^{1/2}} \right) \left( 1 \pm \frac{(\gamma_g -1)}{\beta^{1/2}} \right) 
\label{eq:growth-rate-stream}
\end{equation}
as derived by \citet{begelman94}. Note that instability arises for both forward waves (if $\beta < 0.25$) and backward waves (if $\beta < (\gamma_g-1)^2 = 0.5$). The thresholds differ because $u_1$ and $v_{\rm A,1}$ can be either aligned or anti-aligned, depending on the direction of wave propagation. The perturbed CR force only destabilizes forward waves, while at sufficiently low $\beta$, CR heating destabilizes waves independent of wave direction (as can be seen if the second terms in the two brackets in equation \ref{eq:growth-rate-stream} dominate).   

The growth rate is proportional to the Ptuskin damping rate due to diffusion, $\Gamma_{\rm stream} \sim - \beta^{-1} \Gamma_{\rm damp}$. The diffusive flux $F_{\rm d} \propto \nabla P_c$ is important since it causes a $\pi/2$ phase shift, so that perturbed forces are proportional to velocity rather than displacement. The diffusion time of CRs thus still sets a characteristic timescale. However, by changing the amplitude and sign of total pressure perturbations, CR streaming converts Ptuskin damping ($\dot{u}_1 \propto -u_1$) to a destabilizing force ($\dot{u}_1 \propto u_1$), with a growth rate which depends on the rapidity of streaming and hence heating. 

Broadly speaking, in the WKB regime $k L_c \gg 1$ and $k \kappa/c_s \gg 1$, there are two instability regimes, the streaming dominated regime $\beta < 0.5$, which is unstable regardless of $\kappa/c_s L_c$, and the diffusion dominated regime, $\kappa/c_s L_c > 1$, which is unstable regardless of $\beta$. Growth rates, however, can depend on the secondary parameter. This is essentially what we see in Fig \ref{fig:growth}. In both cases, the instability threshold does not depend on $P_c/P_g$, although growth rates do. The growth rates are simply $c_c^2/{\min}(2 c_s L_c, 6 \beta \kappa)$. Where both instabilities are allowed, we anticipate that diffusion dominates when $c_s L_c < 3 \beta \kappa$, and vice-versa. 

For completeness, we derive in Appendix \ref{app:linear_growth_rates} an equation governing the growth of an acoustic perturbation as it propagates in an arbitrary background profile in the limit $k\kappa/c_s\gg 1$. Its growth rate can be expressed as
\begin{align}
    \Gamma_\mathrm{grow} &= -\frac{c_c^2}{2\kappa}\Bigg\{\qty[1\pm\qty(\gamma_g - 1)\frac{v_A}{c_s}]\qty(1\mp\frac{v_A}{2 c_s}) \nonumber \\
    &\pm\frac{\kappa}{\gamma_c L_c c_s}\qty(1\pm\qty(\gamma_g - 1)\frac{v_A}{2 c_s})\Bigg\}. \label{eqn:growth_rate}
\end{align}
This quantity has to be greater than zero for growth. In the absence of streaming, we recover the instability condition $\kappa/\gamma_c L_c c_s > 1$ for backward waves as estimated in equation \ref{eq:growth_criterion_diffusion}. In the streaming dominated regime, where $\kappa/\gamma_c L_c c_s \ll 1$, we recover the growth condition in equation \ref{eq:growth-rate-stream}.

\subsection{CR Bottleneck Effect}
\label{subsubsec:bottleneck}

\begin{figure}
    \centering
    \includegraphics{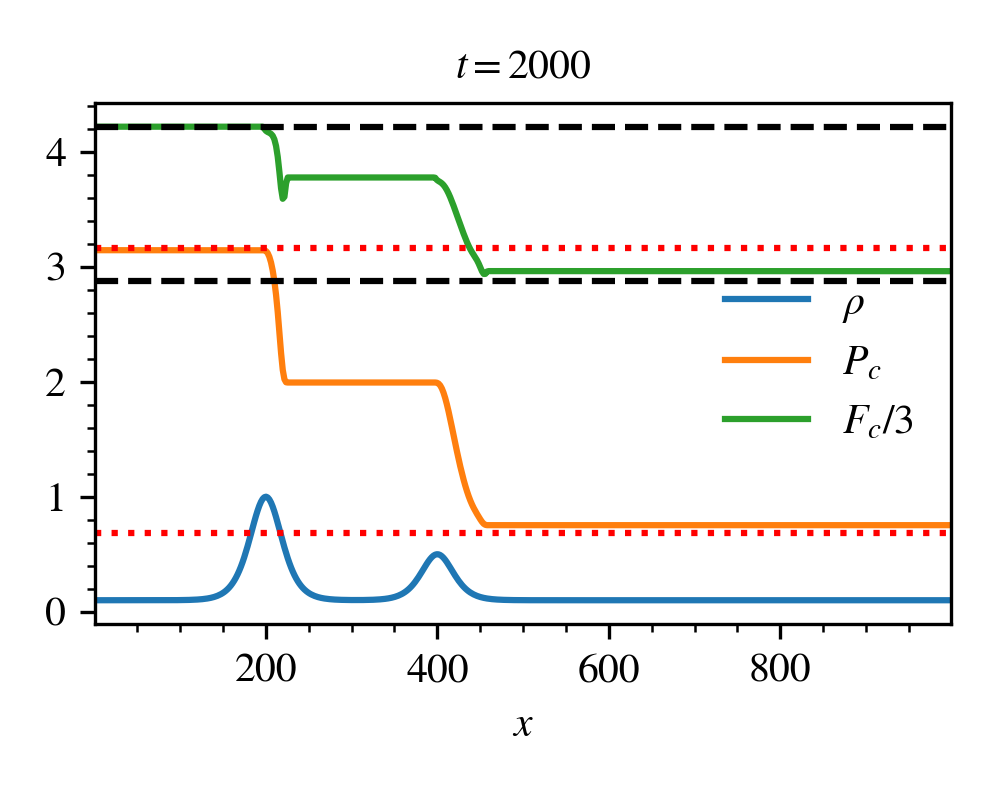} \\
    \includegraphics{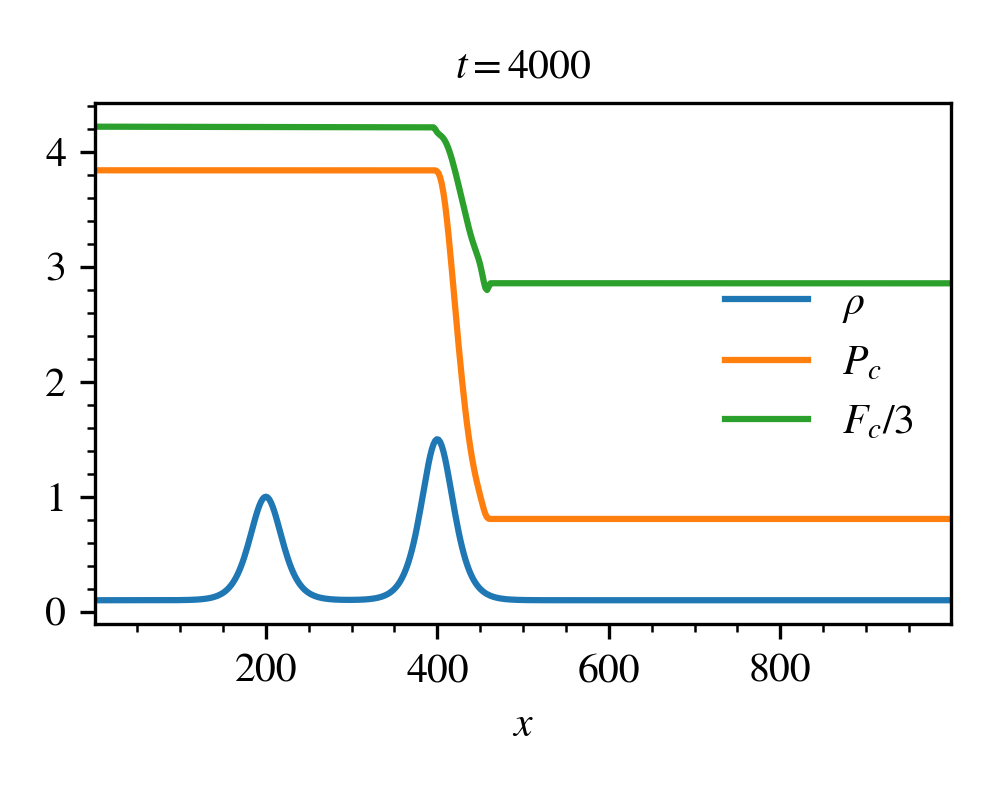} \\
    \caption{\emph{The bottleneck effect}. Only CR variables are evolved while the gas variables are held fixed. A double peak is initialized in the static density field. CR pressure responds with a double plateau. At $t=2000$ the peak at $x=400$ is manually pushed up to $1.5$. The two $P_c$ plateaus merge. The two panels show the equilibriated $P_c$, $\rho$ and $F_c$ profiles before and after the push. Note that $F_c$ has been rescaled for comparison. The region enclosed by the black dashed lines represents $\Delta F_c$ predicted using the density profile and equation \ref{eq:Delta_Fc}. Similarly, the region enclosed by the red dotted lines represents $\Delta P_c$ predicted using the density profile and equation \ref{eq:Delta_Pc}. Both are in good agreement with simulation. If instead we start out with the bump structure in lower panel and manipulate the bumps to end up with that in the upper panel, the CR pressure and flux profiles adjust accordingly to give the results in the upper panel.}
    \label{fig:bottle_merge}
\end{figure}


\begin{figure}
    \centering
    \includegraphics{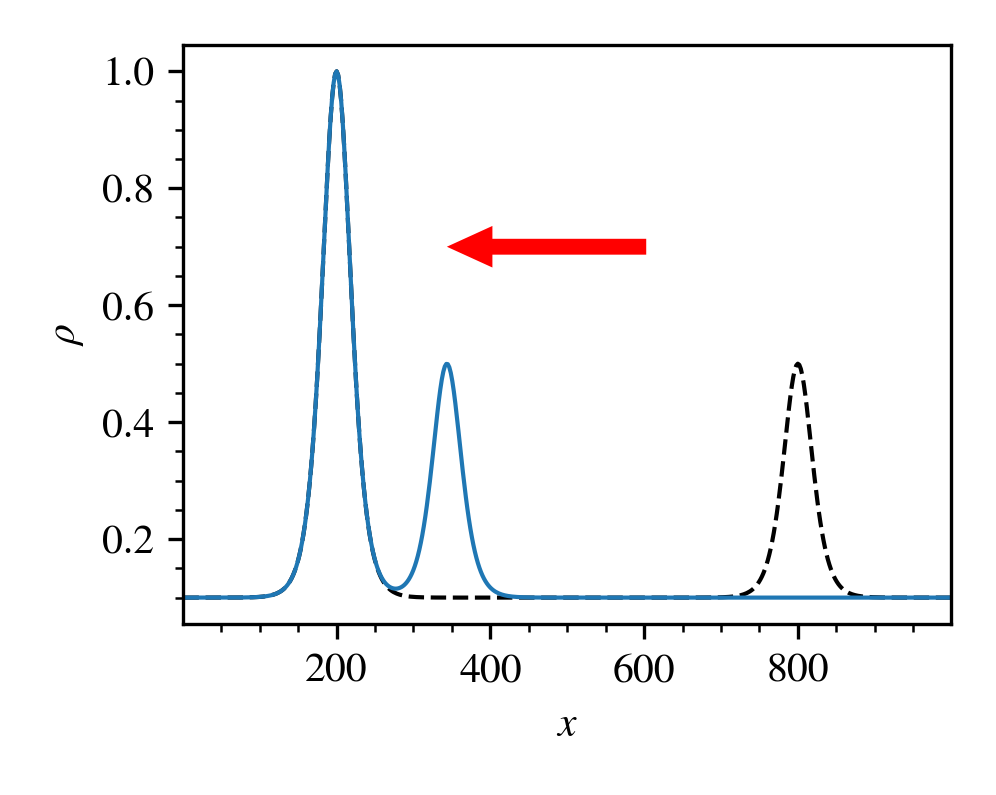} \\
    \includegraphics{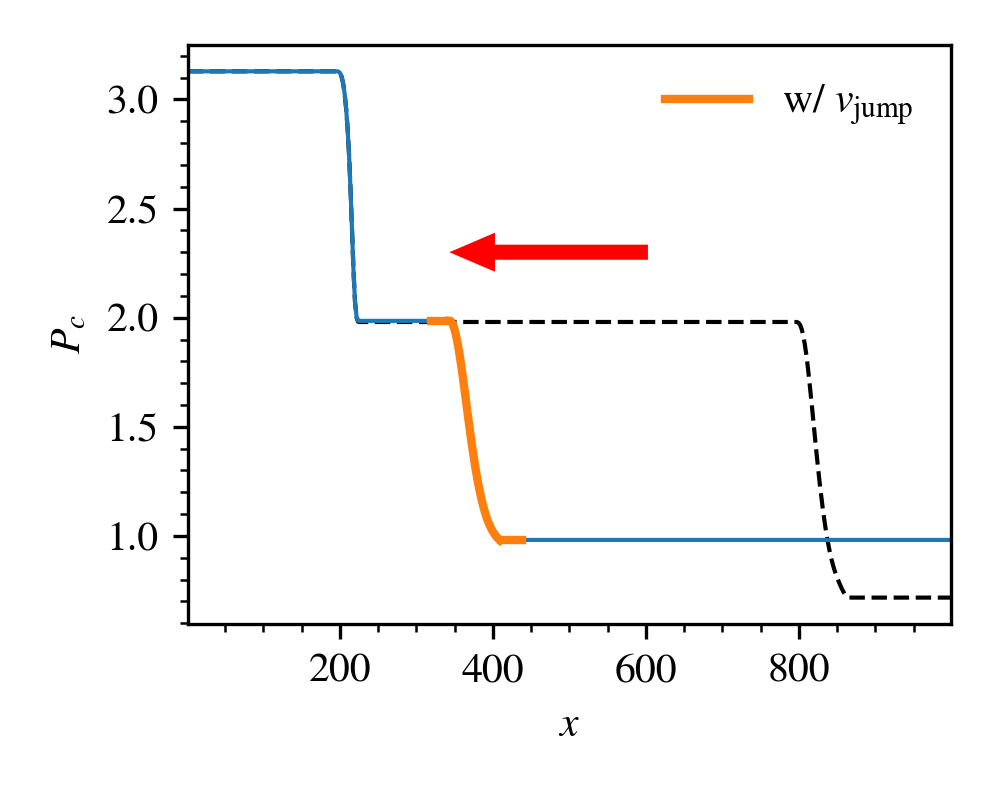}
    \caption{These two panels denote the possible effect of a moving bump on the $P_c$ jumps. Similar to the setup in fig.\ref{fig:bottle_merge}, only the CR variables are evolved while the gas variables remain decoupled. Two density peaks are placed apart and kept stationary. The density and equilibrated $P_c$ profiles are expressed by dashed lines. At $t=2500$ the peak at $x=800$ is moved manually at constant speed towards the right while the peak at $x=200$ remains fixed. The red arrow indicates the direction of motion. The subsequent density and $P_c$ profiles are indicated by the solid blue line. The orange line denotes the $P_c$ profile across the second bump evaluated using equation \ref{eqn:new_bottleneck_with_flow}, including the effect of $v_\mathrm{jump}$.}
    \label{fig:bottle_move}
\end{figure}

\begin{figure}
    \centering
    \includegraphics{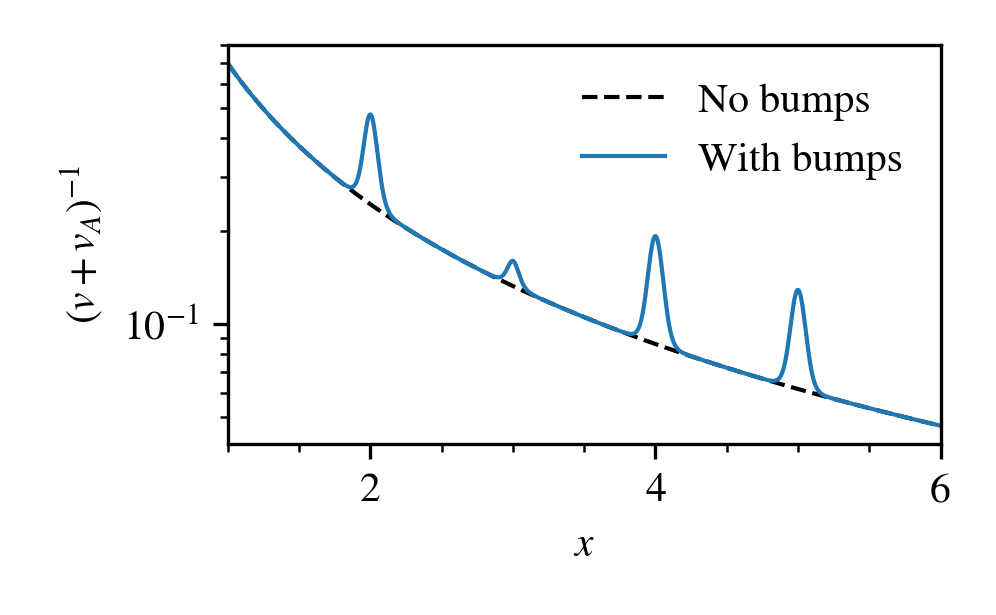} \\
    \includegraphics{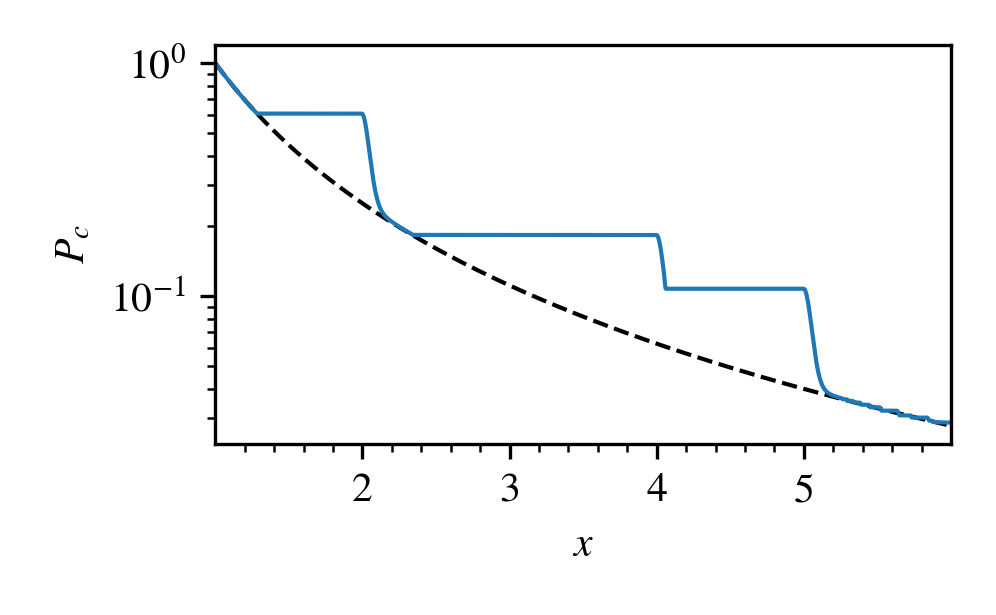} \\
    \includegraphics{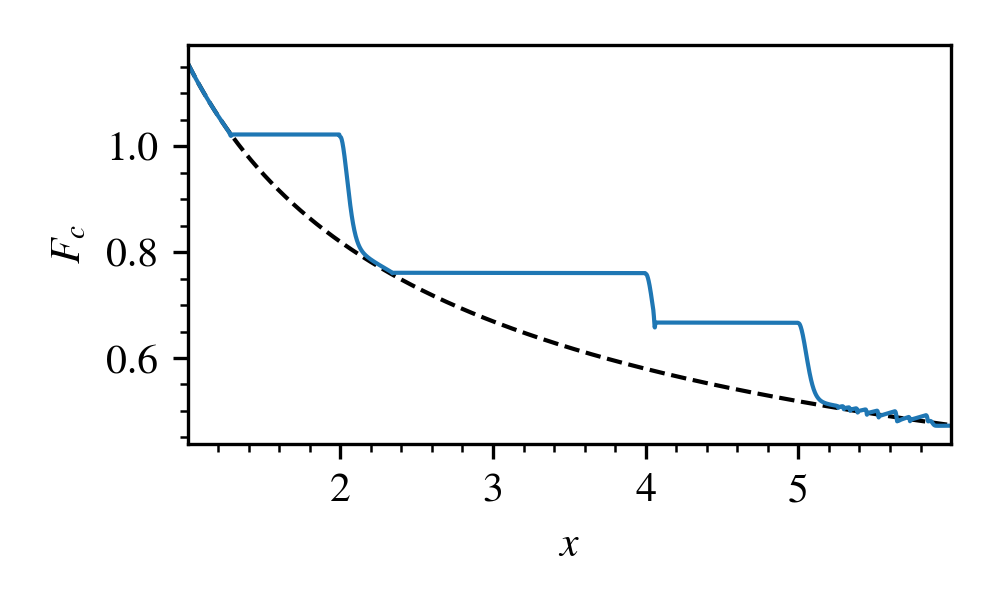}
    \caption{Steady state profiles of CR sub-system which in one case, denoted by black dashed lines, the $\qty(v + v_A)^{-1}$ profile is initiated without bumps and the other case, denoted by solid blue lines, it is initiated with several bumps. None of the bumps rise above the global maximum of the background profile. The overall $\Delta P_c$ and $\Delta F_c$ with and without bumps are the same.}
    \label{fig:bottle_bumps}
\end{figure}

\begin{figure}
    \centering
    \includegraphics{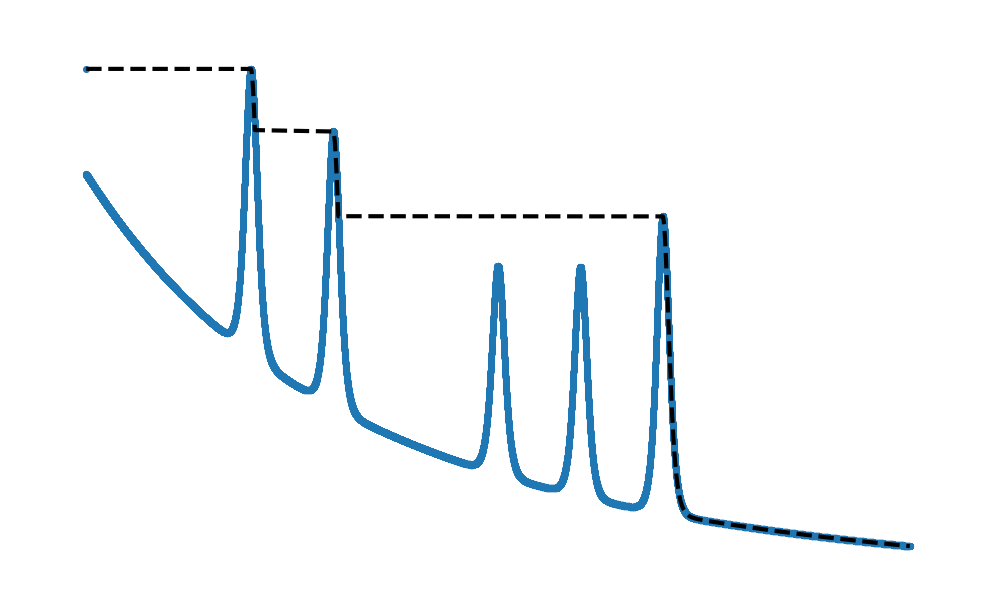}
    \caption{Constructing a convex hull over wiggly curve, surface, etc. is similar to covering it with a rubber band that connects all the highest peaks. Fluctuations lying in the valley between the ridges are swept under the rubber band. See \citet{vergassola94}. Note that the convex hull described here is slightly different from the canonical mathematical definition.}
    \label{fig:convex_hull}
\end{figure}

A streaming dominated fluid fully coupled with CRs should in steady state obey the the following\footnote{This conserved quantity is geometry dependent. In spherically symmetric geometry, for example, the conserved quantity is $r^2 P_c\qty(v + v_A)^{\gamma_c}$.} \citep{breitschwerdt91}
\begin{equation}
    P_c\qty(v + v_A)^{\gamma_c} = \mathrm{const} \label{eqn:bottleneck_with_flow}
\end{equation}
This relation can be derived by setting the time-dependent and CR diffusion terms to zero in equation \ref{eqn:cr_energy} and \ref{eqn:cr_flux} and integrating. For a static fluid and for constant B-field (true in our 1D simulations), this reduces to: 
\begin{equation}
    P_c\rho^{-\gamma_c/2} = \textrm{const}. \label{eqn:bottleneck}
\end{equation}
The CR pressure rises with density.

This property, together with the requirement that CRs can only stream down their gradient, leads to an unusual feature of CR transport known as the `bottleneck effect', predicted analytically by \citet{skilling71} and first simulated by \citet{wiener17_cloud}. For simplicity, consider a 1D setup with constant B-field, where the gas variables are held fixed, and CRs stream from left to right. Now consider an overdense bump. Equation \ref{eqn:bottleneck} demands that $P_c$ increases at the rising edge with the density. This contradicts the requirement that CRs only streams down the $P_c$ gradient. The resolution (seen in Fig \ref{fig:bottle_merge}) is for $P_c$ to flatten on the incoming side and for CRs to decouple from that gas in that region; they free-stream at the speed of light. CRs recouple to the gas and obey equation \ref{eqn:bottleneck_with_flow} on the far side of the density bump, where gas density and hence $P_c$ falls, with CRs streaming down the gradient.  Physically, the decrease in Alfven speed as the density rises causes a CR traffic jam at the bump, causing CR pressure to build up and flatten out. Simulations of this bottleneck effect in the presence of a single bump have been conducted in 1D by \citet{wiener17_cloud}, \citet{jiang18}, and in 2D by \citet{bustard20}.  

Here, we follow a similar setup as in \citet{jiang18} in considering a CR sub-system (i.e. keeping the gas variables fixed and allow only the CR variables to vary). However, here we consider the impact of multiple density peaks. Two density bumps are placed apart from each other, one higher than the other. The initial CR pressure is set to near zero and the CR flux to zero. CRs are injected by fixing the CR flux at the inner boundary, while keeping inner CR pressure boundary free. At some time well after the $P_c$ profile has equilibriated, the second density bump is pushed manually down to lower than the first and the $P_c$ profile allowed to adjust and equilibrate. The result is shown in Fig \ref{fig:bottle_merge}, and can be described as follows: CRs always bottleneck behind the highest density peak they see from the incoming direction. Specifically, incoming CRs would bottleneck and form a plateau all the way up to the highest density peak they see, and upon climbing down in a fully coupled manner (for which equation \ref{eqn:bottleneck} holds), bottleneck up the next highest peak and so on and so forth, forming a staircase. Should the order of peak heights be changed, manually in fig.\ref{fig:bottle_merge}, or (in our simulations of the CR acoustic instability) due to rise of some newly seeded unstable modes, for example, then the $P_c$ profile will adjust accordingly such that the above holds true in steady state. Thus, if instead we start out with the bump structure in lower panel of fig.\ref{fig:bottle_merge} and manipulate the bumps to end up with that in the upper panel, the CR profiles adjust to give the results in the upper panel. If the fluid has a background flow or variable B-fields, equation \ref{eqn:bottleneck_with_flow} holds, with CR bottlenecks at the deepest minima of $(v+v_A)$.  

How is the net momentum and energy transfer from CRs affected by the presence of a staircase? The spatial distribution is obviously affected, since there is no momentum and energy transfer at the plateaus; instead, these only happen at the staircase jumps, where the CRs are coupled \footnote{In our subsequent simulations of the acoustic instability, the jumps propagate and eventually all gas fluid elements experience a force and CR heating.}. However, we shall now show that in a static setup, the total momentum and energy transfer from CRs to the gas only depends on the net change in Alfven speed across the profile. If the bump structure does not change this, then even if a CR staircase develops, the total momentum and energy transfer is unaffected.

Consider the previous setup in the coupled limit. The net momentum transfer by CR forces, integrated over the profile, is: 
\begin{equation}
\int dx \, \nabla P_{c} = -\Delta P_c
\end{equation}
Similarly, in our static setup, the net amount of CR heating in steady-state is: 
\begin{equation} 
\int dx \, v_A \cdot \nabla P_c =  \int dx \, \nabla \cdot F_c = -\Delta F_c.  
\end{equation}
Since we deal exclusively with decreasing $P_c$ and $F_c$ profiles and will make use of $\Delta P_c$ and $\Delta F_c$ frequently in the following, we defined $\Delta P_c = P_\mathrm{c,left} - P_\mathrm{c,right}$ and $\Delta F_c = F_\mathrm{c,left} - F_\mathrm{c,right}$ to ensure positive definiteness, hence the minus sign. 

Fig.\ref{fig:bottle_bumps} shows a smooth density profile and the associated  background $P_c$ profile (without bumps) and the steady-state $P_c$ profile in the presence of bumps. Again, we decouple the hydrodynamics so that the gas distribution does not evolve. While the spatial distribution of $P_c$ (and hence the spatial distribution of CR momentum and energy transfer) is strongly affected by the presence of bumps, the net momentum/energy transfer ($\Delta P_c$ and $\Delta F_c$ respectively) is almost unchanged. See also \citet{wiener17_cloud} for similar conclusions (their sections 3.6, 3.7). The CR pressure only changes where CRs are coupled; there, $P_c \propto v_{\rm A}^{-\gamma_c}$. Thus, $\Delta P_c \propto \Delta [v_A^{-\gamma_c}]$. Since the net density drop is the same, so is the net change in $v_A$ and hence $P_c$. Similarly, the net change in the flux is given by $\Delta F_c \approx  \Delta (P_c v_A) \propto \Delta [v_A^{1-\gamma_c}]$, so the net heating is also determined by the initial and final Alfven speeds (in our 1D sims with constant B-field, this is equivalent to the net density jump). Since these are almost unchanged by the presence of bumps, the net heating rate is similar. 

The net momentum transfer in Fig \ref{fig:bottle_merge}, $\Delta P_c \propto \Delta [v_A^{-\gamma_c}]$, is similarly given by the net change in the Alfven speed: 
\begin{equation} 
\Delta P_{\rm c} = P_{\rm c,left} \left[1 - \qty(\frac{v_{A,\mathrm{min}}}{v_{A,\mathrm{right}}})^{\gamma_c} \right]
\label{eq:Delta_Pc} 
\end{equation}
where $P_{c,\mathrm{left}} = \qty(\gamma_c - 1) F_{c,\mathrm{inj}}/\gamma_c v_{A,\mathrm{min}}$. The net energy transfer is likewise $\Delta F_c \approx  \Delta (P_c v_A) \propto \Delta [v_A^{1-\gamma_c}]$, or 
\begin{equation}
    \Delta F_c = F_{c,\mathrm{inj}}\qty[1 - \qty(\frac{v_{A,\mathrm{min}}}{v_{A,\mathrm{right}}})^{\gamma_c - 1}]. \label{eq:Delta_Fc} 
\end{equation}
We show $\Delta P_{\rm c}, \Delta F_c$ calculated using these equations as dashed black lines in Fig. \ref{fig:bottle_merge}; they agree well with the simulations. When the second peak is pushed up in the lower panel of fig.\ref{fig:bottle_merge} there is an increase in $\Delta P_c$ and $\Delta F_c$, as expected.

In many realistic applications (and certainly in the CR acoustic instability) the density profile is not static but dynamic, and the density peaks are seldom stationary. As we will see in \S\ref{sec:simulation} the non-linear evolution of the CR acoustic instability often involves density bumps propagating up the CR pressure gradient. The $P_c$ profile develops into a propagating staircase in which equation \ref{eqn:bottleneck_with_flow} holds only in the respective rest frames of the jumps. The motion of the jumps will have non-negligible effect on the $P_c$ jumps and hence the overall energy and momentum transfer. A simple illustration is given in Fig.\ref{fig:bottle_move}, again evolving only the CR sub-system, in which a density peak manually moved at constant speed to the left, incurs a reduced $P_c$ jump at the moving peak. 

How can we understand this? The key is to realize that equation \ref{eqn:bottleneck_with_flow} only holds in the rest frame of the jumps, which is the frame where the density (and hence $P_c$) is time-steady. In the lab frame, the conserved quantity is therefore: 
\begin{equation}
    P_c\qty(v + v_A - v_\mathrm{bump})^{\gamma_c} = \text{const} \label{eqn:new_bottleneck_with_flow}
\end{equation}
instead, where $v$ is the lab frame velocity profile and $v_\mathrm{bump}$ is the propagation velocity of the bump. In Fig \ref{fig:bottle_move}, we show the result of applying equation \ref{eqn:new_bottleneck_with_flow}, which matches the simulation results well. 

The conservation law in equation \ref{eqn:new_bottleneck_with_flow} has 3 asymptotic limits, when each of the 3 terms $v,v_A,v_{\rm bump}$ dominates. When the CR flux $F_c \sim 4 P_c v$ is dominated by gas flows, and the CRs simply advect with the gas, then $P_c \propto v^{-\gamma_c} \propto \rho^{\gamma_c}$, i.e. the CRs are adiabatic with an adiabatic index $\gamma_c =4/3$ for a relativistic fluid. When the CR flux is dominated by streaming $F_c \sim 4 P_c v_A$, then $P_c \propto v_A^{-\gamma_c} \propto \rho^{\gamma_c/2}$ (for constant $B$), which is a limit most studied in the literature for the bottleneck effect \citep{wiener17_cloud,bustard20}. When $v_{\rm bump} \gg v,v_A$, then the CR flux in the frame of the bump is $F_c \sim 4 P_c v_{\rm bump}$, which is constant. As $\nabla \cdot F \rightarrow 0$, from equation \ref{eqn:well-coupled-CRs}, $\nabla P_c \rightarrow 0$, i.e. $P_c \rightarrow$const at the moving bump, as is also given by equation \ref{eqn:new_bottleneck_with_flow}. The motion of the bump reduces CR heating of the gas, and when $v_{\rm bump} \gg v,v_A$, there is almost no apparent energy exchange between the two fluids! In this limit, the heating time $\sim l_{\rm bump}/v_A$ is much longer than the bump propagation time $\sim l_{\rm bump}/v_{\rm bump}$ (where $l_{\rm bump}$ is the bump size), so before the CRs have a chance to transfer much energy, the bump has already moved on. 

Another perspective is to see that the motion of the density bump weakens the minimum in $(v+v_A-v_{\rm bump})$, and reduces the strength of the bottleneck. The moving bump makes a net time-averaged contribution to the density profile which is much smoother than the density profile of the stationary bump, and approaches the background profile for a rapidly moving bump. If the background profile is already flat, as in this example, then coupling between the CRs and gas becomes weak and there is hardly any CR staircase. In this way, the motion of density bumps alters the CR staircase (and energy and momentum transfer) compared to the stationary case, where staircase heights are maximized. We will return to this when interpreting our simulation results. Note that if bumps are propagating at different velocities, then one must apply a different frame transformation for each bump. Although one can still infer the CR staircase structure given velocity information, the lack of a global reference frame means that it is no longer possible to write $\Delta P_c, \Delta F_c$ in terms of endpoint quantities, as in equation \ref{eq:Delta_Pc} and \ref{eq:Delta_Fc}. 

These properties are the basis for the `staircase' features seen in the non-linear outcome of the CR acoustic instability, and discussed further in \S\ref{subsec:non-linear}. Interestingly, such staircase features are also seen in Lagrangian maps (i.e., correspondence between initial (Lagrangian) and final (Eulerian) particle positions) in adhesion models of cosmological structure formation \citep{vergassola94}. They are also seen in doubly diffusive fluids, such as sea water where both salt and heat diffuse \citep{radko07}. However, we caution that while some mathematical machinery can be used in common, the origin and physics of these staircases is quite different. In particular, the CR staircase arises from features peculiar to CR transport -- namely, the bottleneck effect in a two-fluid system. 

Mathematically, the $P_c$ staircase is similar to constructing a convex hull (see fig.\ref{fig:convex_hull}) of $\rho$ (or $\qty(v + v_A)^{-1}$ for non-zero flow) and then determining $P_c$ from equation \ref{eqn:bottleneck} (or \ref{eqn:bottleneck_with_flow}). A convex hull is the smallest convex set that encloses a particular shape. For our purposes, given a plot of $(v+v_A)^{-\gamma_c}$ as a function of position, the convex hull of this structure is the non-increasing set of lines of minimal length which encloses all points, including the peaks. As shown in Fig \ref{fig:convex_hull}, it is equivalent to connecting the peaks with rubber bands, via horizontal ridge lines. 

The reasoning above did not take into account multi-dimensional effects, e.g. that due to magnetic field draping around density enhancements \citep{sparre20}. \citet{bustard20} show in 2D that magnetic field strength can affect the path CRs choose around density peaks. Particularly, it was shown that a higher magnetic field facilitates penetration of CRs into density peaks, since magnetic tension causes the field lines to resist draping. The bottleneck effect can be important in this case.

\section{Simulation} \label{sec:simulation}

The following simulations were performed with Athena++ \citep{stone20}, an Eulerian grid based MHD code using a directionally unsplit, high order Godunov scheme with the constrained transport (CT) technique. CR streaming was implemented with the two moment method introduced by \citet{jiang18}. This code solves equations \ref{eqn:continuity}-- \ref{eqn:diffusion_coef}. Cartesian geometry is used throughout.

\subsection{Setup} \label{subsec:setup}

Our 1D setup consists of a set of initial profiles, source terms and appropriate boundary conditions. Magnetic field is constant both in space and time in 1D Cartesian geometry (as required to maintain $\nabla \cdot B =0$). Both CR transport modes (streaming and diffusive) are present. We assume that CRs stream at the local Aflven speed $v_A$. Slippage from perfect wave locking gives rise to CR diffusion, which in the absence of a model for damping, is assumed constant in space and time. In this study we focus mostly on streaming dominated transport; the CR diffusion coefficient is usually taken to be small (in a sense we shall quantify). 

The CR acoustic instability is a small scale instability that only depends on local conditions. In the following we will frequently reference our setup to actual galactic halo conditions, obtained mostly from galaxy scale simulations. The purpose of doing so is to provide a context for which this instability could act. Our 1D Cartesian setup can be crudely thought of as mimicking the vertical profile of disk galaxies, though obviously it is highly idealized. However, it allows for high resolution and a detailed scrutiny of the physics in this first study. 

\subsubsection{Initial Profiles} \label{subsubsec:initial}

\begin{figure}
    \centering
    \includegraphics{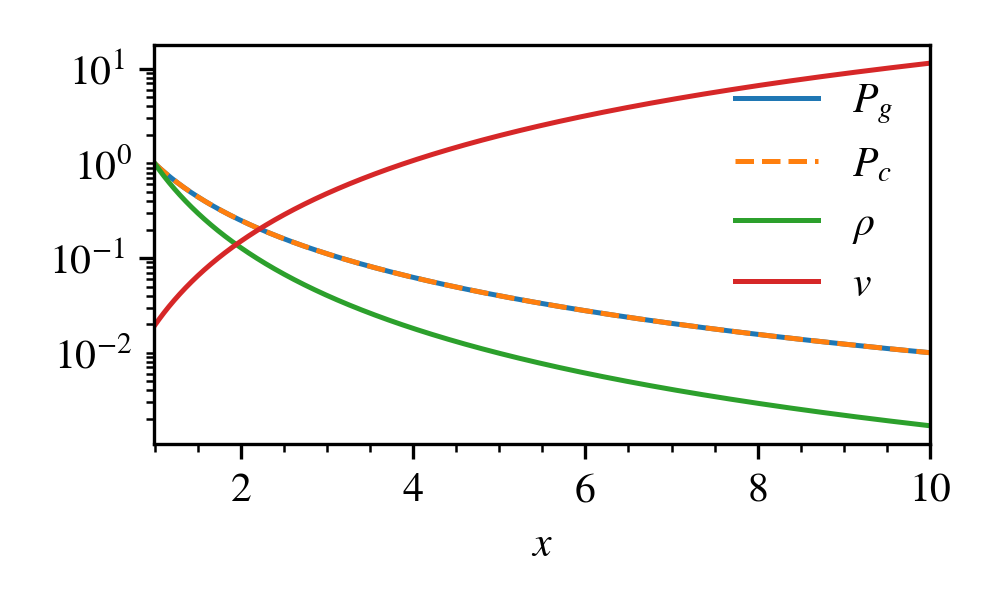} \\
    \includegraphics{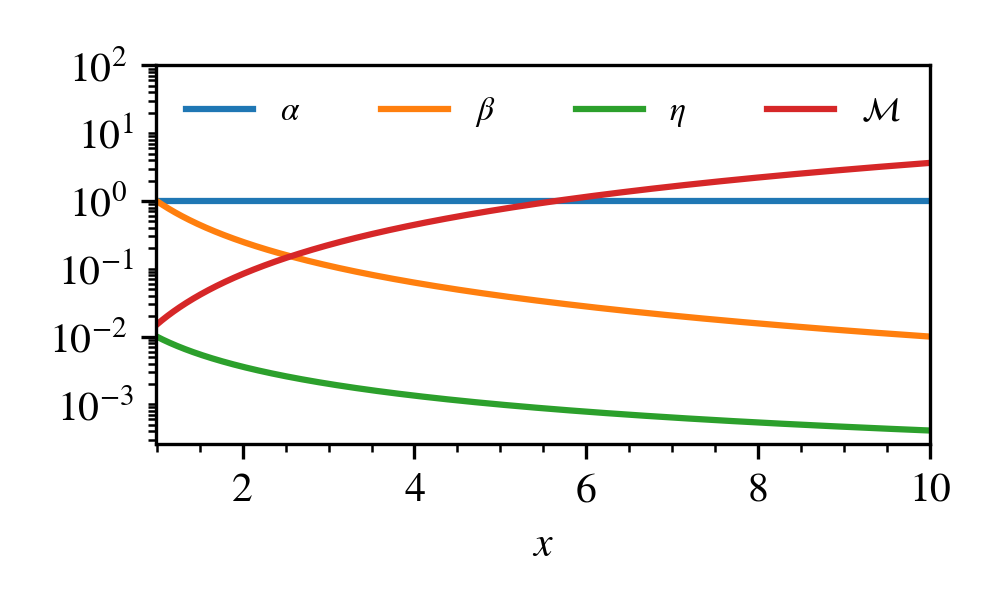}
    \caption{Top: Typical initial $\rho, v, P_g, P_c$ profiles found by integrating eqn.\ref{eqn:integrate_eqn} from $x=1$ to $10$. The profiles shown are obtained with $\alpha_0 = 1, \beta_0 = 1, \eta_0 = 0.01, \mathcal{M}_0 = 0.015, \phi = 2$. Bottom: Typical variation of $\alpha, \beta, \eta, \mathcal{M}$ with $x$.}
    \label{fig:initial_profile}
\end{figure}

The initial profiles are calculated by solving a set of ODE's assuming time steadiness of the fluid equations. In the absence of any instability, the initial profiles will remain steady in simulations. To simplify our calculations, we assume a power law profile in the gas and CR pressure and calculate the required density, velocity profiles and gravity, cooling/heating source terms required for these profiles to remain time-steady. The pressure profiles take the form:
\begin{align}
    P_g &= P_{g0}\qty(\frac{x}{x_0})^{-\phi}, \label{eqn:powerlaw_pg_profile} \\
    P_c &= \alpha_0 P_{g0}\qty(\frac{x}{x_0})^{-\phi}, \label{eqn:powerlaw_pc_profile}
\end{align}
for some specified $\phi, x_0, P_{g0}$ and $\alpha_0$. For pressure to decrease with radius, $\phi > 0$. A power law pressure profile is motivated by galaxy scale simulations (e.g. \citealt{voort12}) and its simplicity in describing a generic halo profile. Since magnetic fields are constant in our model, this implies that $\beta \propto x^{-\phi}$, i.e. the gas becomes magnetically dominated at large $x$. Physically, magnetically dominated halos can arise in simulations \citep{pakmor20,voort20} and in analytic solutions \citep{ipavich75}; we discuss this further in \S\ref{sec:discussion}.    

Specifying the density $\rho_0$, velocity $v_0$ and Alfven speed $v_{A0}=B/\sqrt{4\pi\rho_0}$ at $x_0$, the velocity $v$ and density $\rho$ profiles are found by integrating from $x_0$ the equations
\begin{gather}
    \dv{x}\qty(v + v_A) = \frac{\kappa P''_c - \qty(v + v_A)P'_c}{\gamma_c P_c}, \label{eqn:integrate_eqn} \\
    \rho v = \text{constant} \label{eqn:const_mass_flux},
\end{gather}
where the first equation is the steady state version of equation \ref{eqn:well-coupled-CRs}, and the second from mass conservation. Each prime means an additional derivative with respect to $x$. An example of the initial profiles is shown in fig.\ref{fig:initial_profile}. Using the steady state profiles calculated, the gravity source term $g$ is defined as
\begin{equation}
    g = \qty(\rho v\dv{v}{x} + \dv{P_g}{x} + \dv{P_c}{x})/\rho
\end{equation}
to ensure momentum balance. To have a sense of what functional form $\rho$ and $g$ have, consider the sub-sonic and sub-Alfvenic limit where we can ignore terms involving the velocity $v$ (for a galactic halo/wind profile this would hold near the base of the profile). For streaming dominated transport the diffusive term in equation \ref{eqn:integrate_eqn} can be ignored, which then reduces to  equation \ref{eqn:bottleneck}. We obtain, for the density, a power law profile:
\begin{equation}
    \rho = \rho_0\qty(\frac{x}{x_0})^{-3\phi/2}, \quad v\ll c_s, v_A. \label{eqn:sub_density_profile}
\end{equation}
The gravity term has a power law form too:
\begin{equation}
    g = \frac{\phi P_{g0} \qty(1 + \alpha_0)}{\rho_0 x_0}\qty(\frac{x}{x_0})^{\phi/2 - 1}, \quad v\ll c_s, v_A, \label{eqn:sub_gravity}
\end{equation}
where we have used $\gamma_c = 4/3$. In our fiducial setup ($\phi=2$), gravity is constant until the critical point (see discussion below equation \ref{eq:c_eff}). 

For cooling, adiabatic processes and CR heating is balanced by a time-independent heating/cooling term $\mathcal{H}\qty(x)$, defined using the steady state profiles,
\begin{align}
    \mathcal{H} &= \Bigg[v\dv{P_g}{x} + \gamma_g P_g\dv{v}{x} \nonumber \\
    &+ \qty(\gamma_g - 1) v_A\dv{P_c}{x}\Bigg]/\qty(\gamma_g - 1). \label{eqn:cooling_caseI}
\end{align}
In the subsonic and sub-Alfvenic limit this approximates to
\begin{equation}
    \mathcal{H} = -\frac{\alpha_0\phi P_{g0} v_{A0}}{x_0}\qty(\frac{x}{x_0})^{-\phi/4 - 1}, \quad v\ll c_s, v_A. \label{eqn:sub_cooling_caseI}
\end{equation}

Although not fully realistic, it is a simple and attractive setup in global force and energy balance. Note that it {\it does} have cooling, which in the background profile offsets CR Alfven heating. However, this cooling is simply a function of spatial position, rather than thermodynamic variables. This simplification allows us to initialize arbitrary profiles which are still in energy balance. 

Thus, in each scenario the initial profile is determined by the parameters: 
\begin{itemize}
    \item $\rho_0, \mathcal{M}_0$, $P_{g0}, \alpha_0, \beta_0, \eta_0, \phi$. 
\end{itemize}
The subscripts 0 all indicate they are defined at $x_0$. The dimensionless parameters are defined as
\begin{gather}
    \alpha_0 = P_{c0}/P_{g0}, \quad \beta_0 = 8\pi P_{g0}/B^2, \quad \eta_0 = \kappa/\gamma_c L_{c0} c_\textrm{s0}, \nonumber \\ \mathcal{M}_0 = v_0/c_{s0}, \label{eqn:dimensionless}
\end{gather}
where $c_\textrm{s0}=\sqrt{\gamma_g P_{g0}/\rho_0}$ is the adiabatic sound speed and $L_{c0} = \abs{P_c/P'_c}_0$ is the CR scale height. Note that $L_\mathrm{c,0} = x_0/\phi$, so $x_0$ can also be interpreted as a CR pressure scale-height. In general, $\alpha, \beta, \eta, \mathcal{M}$ (defined similarly as \ref{eqn:dimensionless} but without the subscript $0$) vary along the profile. Their typical variation is given by the bottom plot of fig.\ref{fig:initial_profile}. $\beta$ and $\eta$ usually decrease as $x$ increases while $\mathcal{M}$ increases. $\alpha$, by construction of the power law pressure profile equations \ref{eqn:powerlaw_pg_profile} and \ref{eqn:powerlaw_pc_profile}, is a fixed quantity throughout. Unless otherwise specified, we set $\rho_0 = 1, P_{g0} = 1$ and $x_0 = 1$. 

One issue in 1D Cartesian geometry is the transition to supersonic flow. If we combine the Euler equation with equation \ref{eqn:bottleneck_with_flow} (in the streaming dominated regime), we obtain, after some manipulations, the wind equation
\begin{equation}
    \dv{v}{x} = \frac{g\qty(x)}{v\qty(v^2 - c^2_\mathrm{eff} - c_s^2)}, \quad \text{1D Cartesian}
\end{equation}
where
\begin{equation}
    c_\mathrm{eff}^2 = \frac{\gamma_c P_c}{\rho}\frac{v + v_A/2}{v + v_A}, \quad c_s^2 = \dv{P_g}{\rho}.
    \label{eq:c_eff} 
\end{equation}
As usual with wind equations, there is some critical point where the wind becomes transonic (i.e. $v^2 = c_\mathrm{eff}^2 + c_s^2$ in this case). To avoid any singularity, $g\qty(x)$ has to go through zero at the critical point, and indeed it must change sign if the wind is to keep acceleration ($\dv*{v}{x} > 0$). This is obviously unphysical. We cannot focus on subsonic flow alone in our simulations; the flow must be supersonic at large $x$ to avoid boundary problems (see \S\ref{subsubsec:boundaries}). In reality, at large radii disk winds transitions to a more spherical geometry, where this problem no longer occurs. But for simplicity, we simply solve for the gravitational field $g(x)$ which maintains a steady wind solution through the sonic point in Cartesian geometry. Our conclusions are unchanged if we focus solely on the subsonic portion of the flow, where the gravitational field is fully realistic (e.g., constant or power law up to the sonic point).  

We shall try to answer the following questions with this 1D setup: 1. Verify the linear growth of the CR acoustic instability and study the non-linear growth and saturation. Since we find that the non-linear CR profile exhibits a staircase structure, we follow up with the questions below: 2. How can we understand the staircase structure and characteristic scales? 3. How does the staircase affect the time-averaged momentum and energy transfer between the gas and CR? 

Our simulations focus on situations where streaming dominates CR transport, i.e. $\kappa/ L_c c_s \sim \eta \ll 1$. The CR diffusion dominated limit (with $\eta\gtrsim 1$) has already been studied \citep{drury86,kang92,ryu93,drury12,quataert21}. In actual simulations using the two-moment formalism, the diffusion coefficient $\kappa$ is never set to zero (as that would give $\texttt{nan}$ in the calculation of $\sigma_c$, equation \ref{eqn:diffusion_coef}). Also, with our boundary conditions, the very fast growth rates for small $\kappa$ cause the simulations to crash. We find that for stability and numerical convergence, the diffusion length $l_{\rm diff} \sim \kappa/c_s$ has to be resolved with $\gtrsim 4$ grid cells. Thus, the minimum value of $\kappa$ in our simulations is dictated by resolution. Since the diffusion length is resolved, the fastest growing, small-scale modes in our simulation are in the limit $k \kappa/c_s > 1$. As discussed in Appendix \ref{subsec:adiabatic_small_kappa}, on scales below the diffusion length, growth rates are independent of wave number. In addition, the acoustic mode dominates, $\omega \approx \pm  k c_s$, i.e. the wave propagation speed is simply the gas sound speed.

\subsubsection{Static and Outflow Setup and Boundary Conditions} \label{subsubsec:boundaries}

{\it Linear Growth.} To evaluate linear growth rates, we will (mostly) adopt a static background. The initial profiles are first evaluated up to the boundary ghost zones and input into the simulation box. Then an acoustic wave is generated from a boundary  and its amplitude tracked as it propagates. We perturb the velocity, gas density and pressure as follows:
\begin{equation}
    \delta v = A\zeta\qty(t)\sin(\mp k c_s t), \quad \delta\rho = \pm\rho\frac{\delta v}{c_s}, \quad \delta P_g = \pm\gamma_g P_g \frac{\delta v}{c_s},
\end{equation}
where $A$ is some injection amplitude and $\rho, P_g, c_s$ are evaluated at the boundary with the top sign for forward propagating waves, and bottom sign for backward waves. The perturbations are multiplied by a buffer function $\zeta\qty(t)$, given by
\begin{equation}
    \zeta\qty(t) = 1 - e^{-t/\tau} \label{eqn:buffer}
\end{equation}
where $\tau$ is small (around one wave-crossing time), to ensure the wave profile and its derivatives are continuous when the acoustic perturbation is injected.

{\it Non-Linear Growth.} When studying the non-linear growth and saturation, we include a background flow. As we shall explain, this is important to avoid boundary effects; it also mimics a disk wind. We impose the initial density, gas pressure and CR flux onto the inner ghost zones while keeping the CR pressure free by linearly extrapolating from the active zones. The inner velocity is determined by maintaining constant mass flux. For the outer boundary, we copy the density, gas pressure and CR flux from the last active zone and linearly extrapolate the CR pressure. The velocity is again determined from constant mass flux. This set of boundary conditions mimics a stratified disk atmosphere with the inner boundary fixed by galactic disk properties and the outer boundary kept free. To limit boundary effects, a buffer zone with viscosity is added near the boundaries to damp out inbound or outbound unstable acoustic waves\footnote{Specifically, we add the term $\nu\nabla^2 v$ to the momentum equation, where $\nu$ is chosen to be small enough not to affect the overall profile, but large enough to damp out high frequency sound waves.}. Still, it is important, when the outer boundary is kept free, to initiate a background velocity such that the flow near the outer boundary is supersonic, as otherwise inbound unstable sound wave can cause unphysical effects\footnote{In keeping the boundary free, the values at the ghost zones should depend on the last active zones. Instead, inbound sound waves carry information from outside in. This usually isn't a problem when the inbound sound waves are stable, but here they are problematic.}.  (e.g. spurious shocks). Despite requiring the flow near the outer boundary to be supersonic, it is possible to initiate the flow at the inner boundary to be highly subsonic (see the bottom figure of \ref{fig:initial_profile}). To further ensure our discussion will not be affected by outer boundary conditions, we focus on the inner (subsonic) half of the simulation domain. Unlike the linear setup, where we explicitly perturb the profile, here all growth is seeded by numerical noise. 

\subsection{Acoustic Instability: Comparison with Linear Theory} \label{subsec:linear}

\begin{table*}
    \centering
    \begin{threeparttable}
        \begin{tabular}{c|c|c|c|c|c|c|c|c}
            \hline 
            \hline
            Identifier & Dir. of prop. & $\alpha_0$ & $\beta_0$ & $\eta_0$ & $\phi$ & $\lambda$ ($l_\mathrm{diff,0}$) & Inj. amp. & resolution ($\lambda/\Delta x$) \\
            \hline
            alpha1beta1eta.01phi2 & Up & 1 & 1 & 0.01 & 2 & 1 & $1.84\times10^{-5}$ & 109 \\
            alpha1beta1eta.1phi2 & Up & 1 & 1 & 0.1 & 2 & 0.1 & $1.99\times10^{-4}$ & 109 \\
            alpha10beta1eta.1phi2 & Up & 5 & 1 & 0.1 & 2 & 0.1 & $1.99\times10^{-5}$ & 109 \\
            alpha1beta.1eta1phi2 & Up & 1 & 0.1 & 1 & 2 & 0.01 & $2.35\times10^{-4}$ & 109 \\
            alpha1beta.01eta10phi2 & Up & 1 & 0.01 & 10 & 2 & 0.003 & $3.47\times10^{-4}$ & 328 \\
            alpha1beta.1eta.1phi1 & Up & 1 & 0.1 & 0.1 & 1 & 0.1 & $1.57\times10^{-4}$ & 219 \\
            alpha1beta.1eta.1phi.5 & Down & 1 & 0.1 & 0.1 & 0.5 & 0.1 & $1.44\times10^{-4}$ & 437 \\
            alpha1beta.1eta.1phi2 & Up & 1 & 0.1 & 0.1 & 2 & 0.1 & $1.29\times10^{-5}$ & 109 \\
            alpha1beta.5eta.1phi2ms.03\tnote{a} & Up & 1 & 0.5 & 0.1 & 2 & 0.1 & $1.87\times10^{-5}$ & 109 \\
            \hline
        \end{tabular}
        \begin{tablenotes}
            \item[a] A background flow with $\mathcal{M}_0 = 0.03$ (see eqn.\ref{eqn:dimensionless}) is initiated for this case.
        \end{tablenotes}
    \end{threeparttable}
    \caption{Parameters for simulation of linear growth of acoustic waves. Column 1: Case identifier. Column 2: Direction of propagation up or down the CR pressure gradient. Column 3-5: Parameters defined in equation \ref{eqn:dimensionless}. Column 6: Power-law index of the background $P_c$ profile defined in eqn.\ref{eqn:powerlaw_pc_profile}. Column 7: Wavelength of the acoustic wave in units of $l_\mathrm{diff,0}\equiv\kappa/c_{s,0}$. Column 8: Injection amplitude. Column 9: Resolution, the number of grids each wavelength is resolved with.}
    \label{tab:linear_params}
\end{table*}

\begin{figure*}
    \centering
    \includegraphics[width=0.33\textwidth]{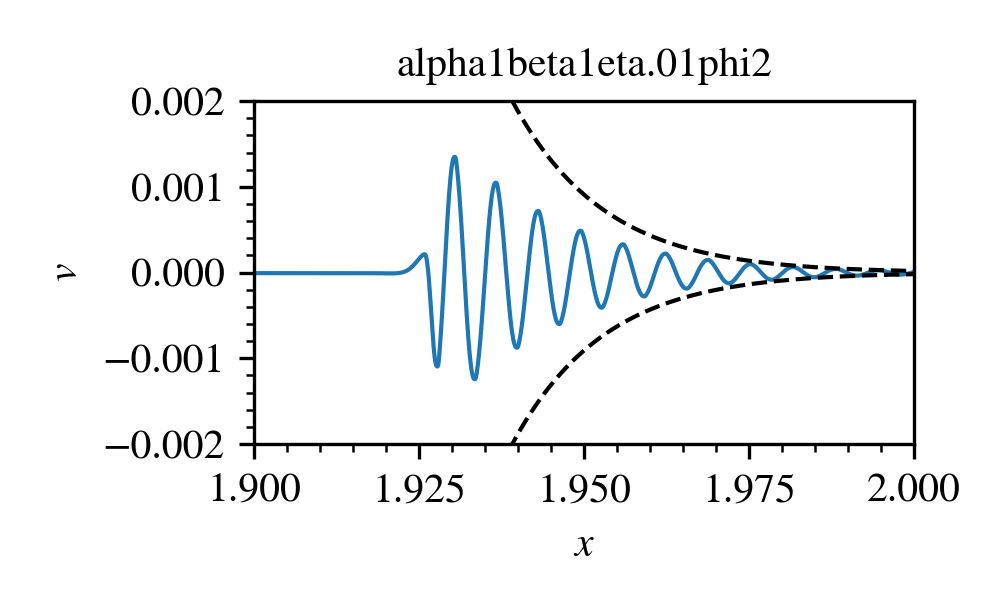}
    \includegraphics[width=0.33\textwidth]{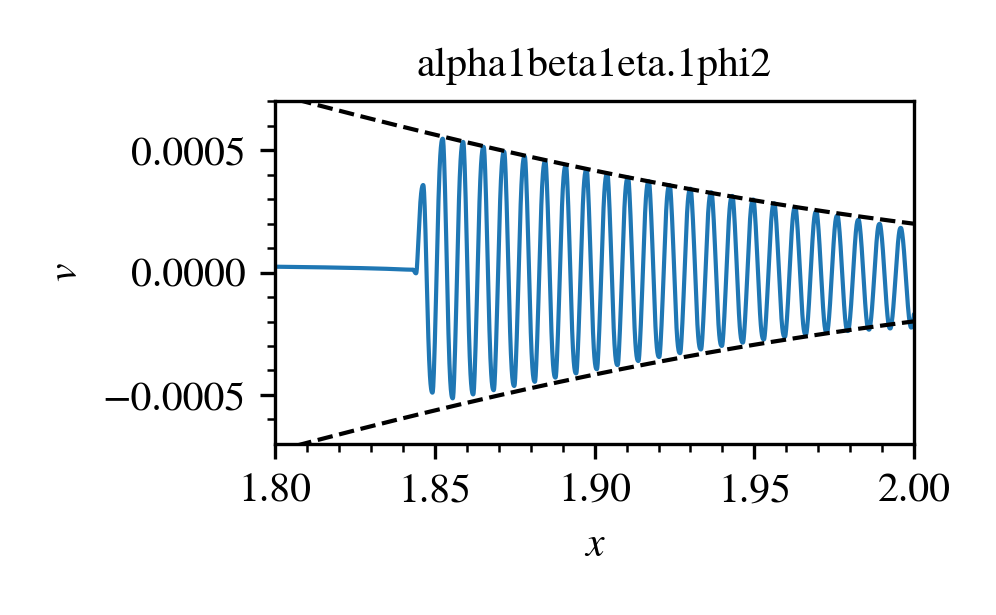}
    \includegraphics[width=0.33\textwidth]{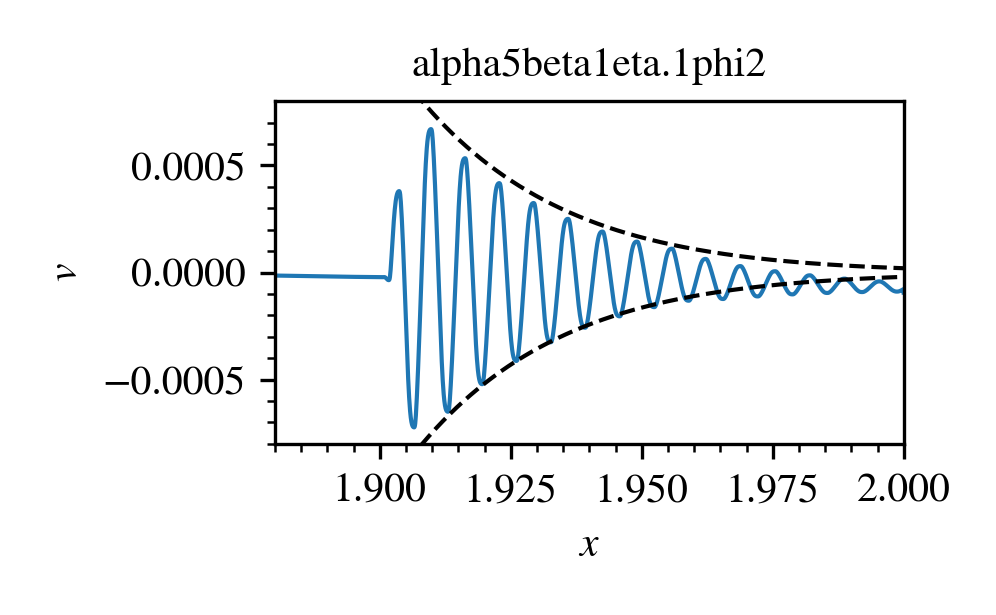} \\
    \includegraphics[width=0.33\textwidth]{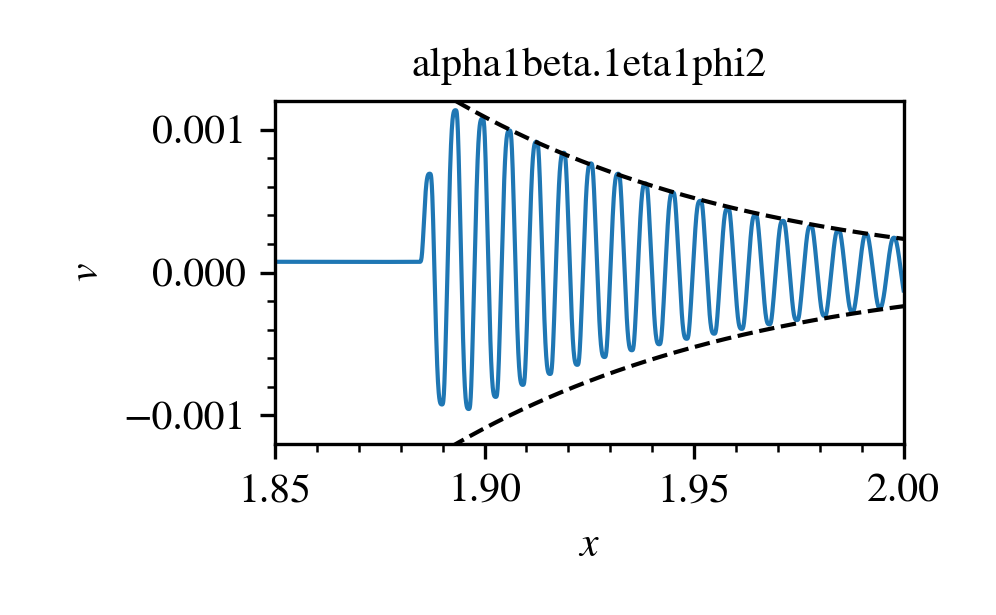}
    \includegraphics[width=0.33\textwidth]{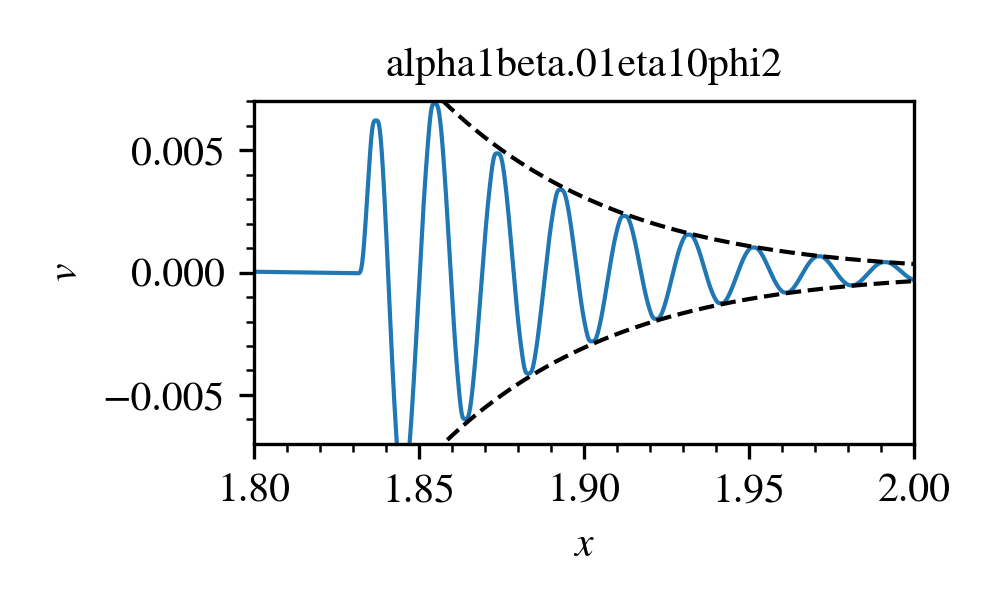}
    \includegraphics[width=0.33\textwidth]{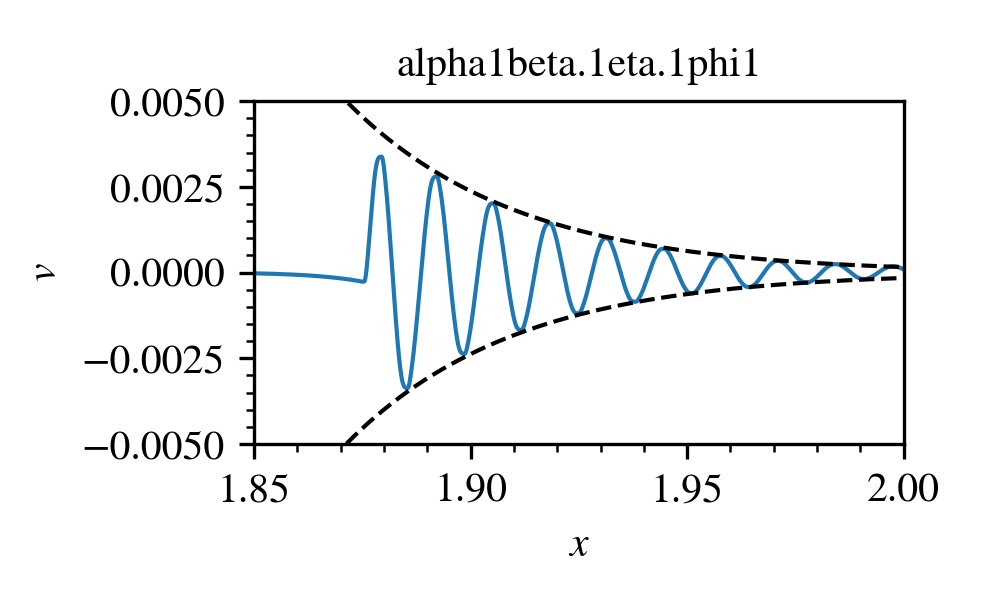} \\
    \includegraphics[width=0.33\textwidth]{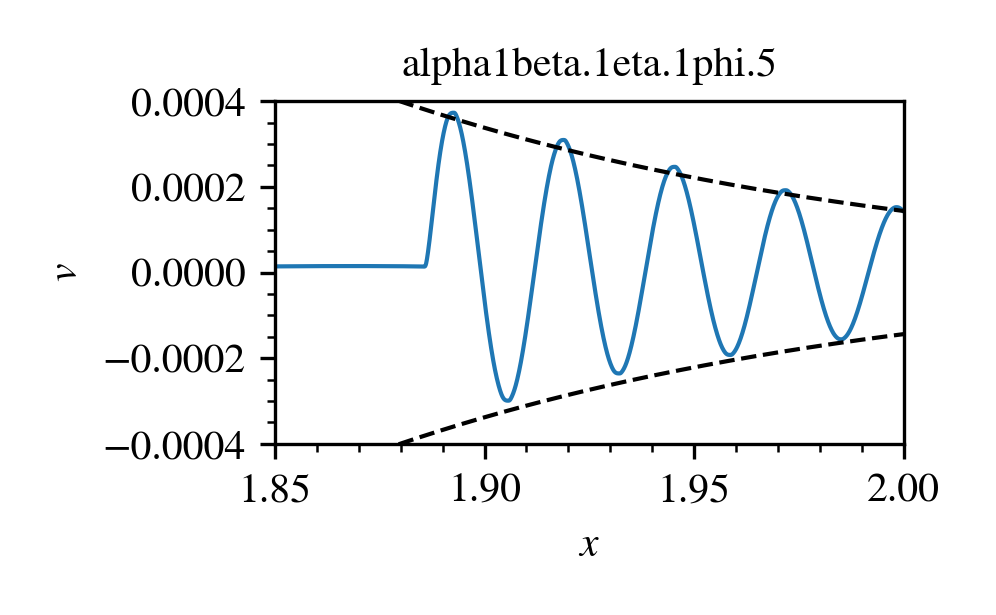} 
    \includegraphics[width=0.33\textwidth]{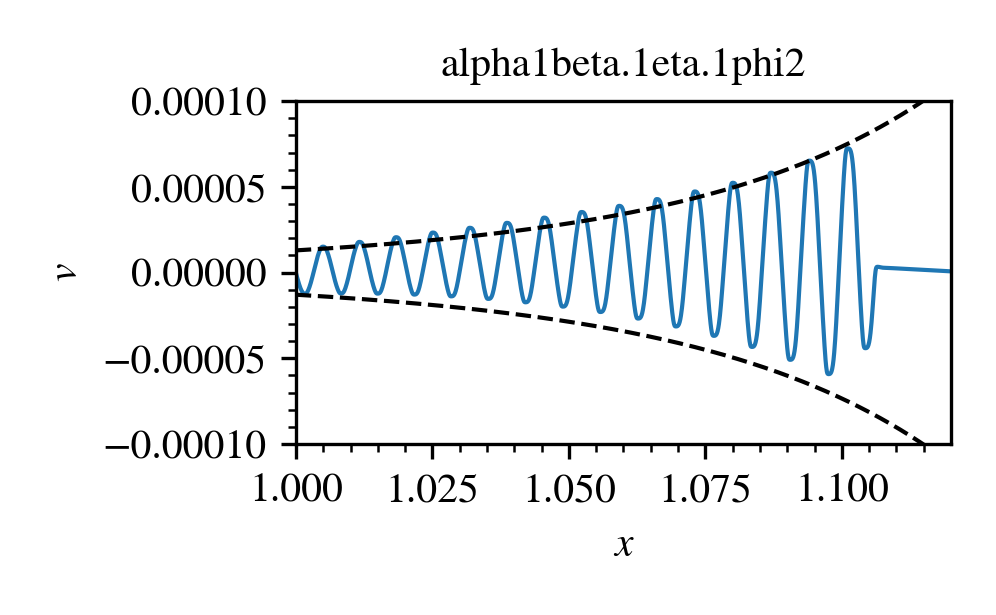}
    \includegraphics[width=0.33\textwidth]{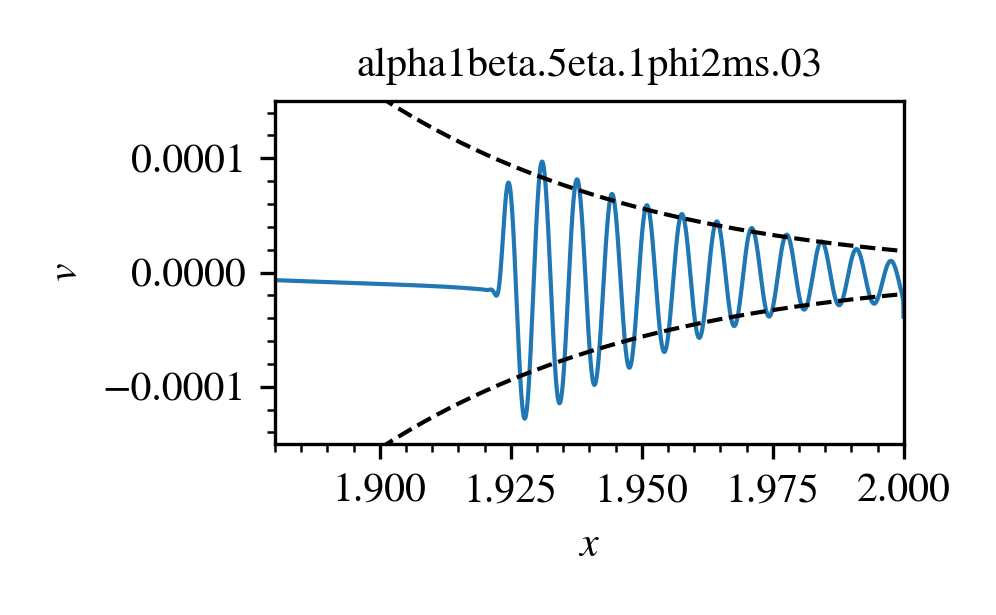} 
    \caption{Simulation of linear growth of acoustic waves. An acoustic wave is injected according to the description described in \S\ref{subsubsec:boundaries} with the parameters listed in table \ref{tab:linear_params}. In each panel, the identifier is given at the top. The blue curve shows the simulated velocity profile of the growing acoustic wave. The analytically predicted amplitude (using equation \ref{eqn:amplitude_track}) is displayed in black dashed line for comparison.}
    \label{fig:amp_compare}
\end{figure*}

\begin{figure}
    \centering
    \includegraphics{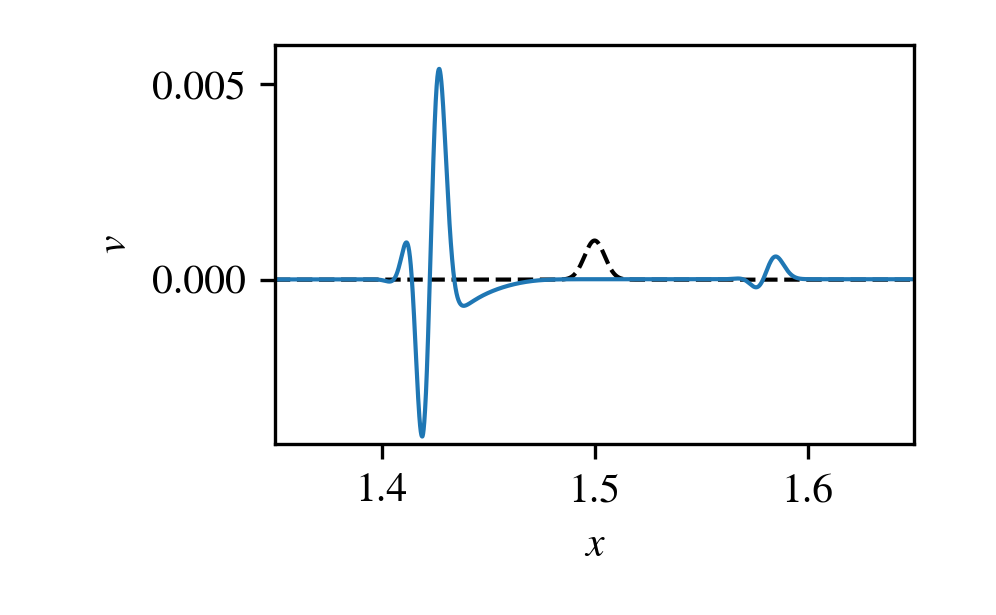}
    \caption{Growth comparison of forward and backward propagating waves. The black dashed line shows the initial velocity profile with a slight Gaussian perturbation. The blue solid line shows the linear growth of the forward (right) and backward (left) propagating modes. As expected, the latter grow more rapidly. The background $\alpha_0 = 1, \beta_0 = 0.5, \eta_0 = 0.01, \phi = 2$. The Gaussian bump has amplitude $10^{-3}$ and characteristic width of $\kappa/c_{s0}$.}
    \label{fig:alpha1beta_5eta_01psi0_gauss}
\end{figure}

Table \ref{tab:linear_params} lists the parameters used for simulating the linear growth of acoustic waves. In each case, an acoustic wave with a specified amplitude and wavelength (expressed in units of diffusion length) is injected by a boundary perturbation as described in \S\ref{subsubsec:boundaries}. The background profile spans $1<x<2$. The resolution is given in number of grids used to resolve each wavelength, the whole domain is typically resolved with 16384 grids. The reduced speed of light is $c=1000$. The results are displayed in fig.\ref{fig:amp_compare}. In each panel, the velocity profile is given by the blue solid curve. In the linear growth phase, the velocity amplitude of the acoustic perturbation can be analytically expressed, to first order approximation, as
\begin{equation}
\hat{v}\qty(x) = \hat{v}\qty(x_\mathrm{inj})\exp{\frac{1}{2}\ln{\frac{\rho_\mathrm{inj}}{\rho}} + \frac{1}{2} \mathcal{I}\qty(x, x_\mathrm{inj})}, \label{eqn:amplitude_track_in_text}    
\end{equation}
where $\mathcal{I}\qty(x,x_\mathrm{inj})$, given in equation \ref{eqn:integral_tracking}, is an integral involving the growth rate from the location of injection $x_\mathrm{inj}$ to some point $x$ along the path of propagation. Overall there is good agreement between the simulated amplitude growth and analytics, except in the case where $\lambda = l_\mathrm{diff,0}$ (case \texttt{alpha1beta1eta.01phi2}, panel in the upper left corner), for which $k\kappa/c_s\sim 1$ and the growth rate formula (equation \ref{eqn:growth_rate}) is no longer valid. In particular, for $k\kappa/c_s\lesssim 1$ the acoustic mode bifurcates into additional hybrid modes (appendix \ref{app:linear_growth_rates}). These modes have lower growth rates than the asymptotic small wavelength $k\kappa/c_s\gg 1$ limit.

In Fig.\ref{fig:alpha1beta_5eta_01psi0_gauss}, rather than injecting a sound wave from the right boundary, we set up a Gaussian perturbation of amplitude $10^{-3}$ and characteristic width $\kappa/c_{s0}$ in the middle of the simulation domain. Both the forward and backward acoustic modes are unstable at the Gaussian bump. The background mode clearly grows faster than the forward mode, as expected.

All in all, we have shown that acoustic perturbations can be amplified by CRs in various settings and the growth rate is consistent with that expected from linear theory. In particular, in the fluid rest frame, waves propagating up the CR gradient are more unstable.  

\subsection{Acoustic Instability: Non-Linear Outcome} \label{subsec:non-linear}

\begin{table*}
    \centering
    \begin{threeparttable}
        \begin{tabular}{c|c|c|c|c|c|c|c|c|c|c}
            \hline 
            \hline
            Identifier & $\alpha_0$ & $\beta_0$ & $\eta_0$ & $\phi$ & $c$ & Resolution $\qty(\Delta x)$ & $\langle\dot{M}\rangle/\dot{M}_0$ & $\langle\Delta P_c\rangle/\Delta P_{c0}$ & $\langle\Delta F_c\rangle/\Delta F_{c0}$ & $\gamma_\mathrm{eff}$ \\
            \hline
            NL4096alpha.5beta1eta.01phi2c200 & 0.5 & 1 & 0.01 & 2 & 200 & $2.20\times 10^{-3}$ & 0.969 & 1.120 & 0.947 & 1.28 \\
            NL4096alpha.6beta1eta.01phi2c200 & 0.6 & 1 & 0.01 & 2 & 200 & $2.20\times 10^{-3}$ & 0.977 & 1.184 & 0.932 & 1.20 \\
            NL4096alpha.7beta1eta.01phi2c200 & 0.7 & 1 & 0.01 & 2 & 200 & $2.20\times 10^{-3}$ & 1.063 & 1.207 & 0.911 & 1.17 \\
            NL4096alpha.8beta1eta.01phi2c200 & 0.8 & 1 & 0.01 & 2 & 200 & $2.20\times 10^{-3}$ & 1.123 & 1.230 & 0.915 & 1.20 \\
            NL4096alpha.9beta1eta.01phi2c200 & 0.9 & 1 & 0.01 & 2 & 200 & $2.20\times 10^{-3}$ & 1.175 & 1.234 & 0.899 & 1.19 \\
            NL4096alpha1beta1eta.01phi2c200 & 1 & 1 & 0.01 & 2 & 200 & $2.20\times 10^{-3}$ & 1.388 & 1.304 & 0.896 & 1.24 \\
            NL4096alpha2beta1eta.01phi2c200 & 2 & 1 & 0.01 & 2 & 200 & $2.20\times 10^{-3}$ & 1.713 & 1.269 & 0.852 & 1.16 \\
            NL4096alpha3beta1eta.01phi2c200 & 3 & 1 & 0.01 & 2 & 200 & $2.20\times 10^{-3}$ & 1.825 & 1.210 & 0.844 & 1.12 \\
            NL4096alpha4beta1eta.01phi2c200 & 4 & 1 & 0.01 & 2 & 200 & $2.20\times 10^{-3}$ & 1.861 & 1.186 & 0.844 & 1.10 \\
            NL4096alpha5beta1eta.01phi2c200 & 5 & 1 & 0.01 & 2 & 200 & $2.20\times 10^{-3}$ & 1.890 & 1.187 & 0.848 & 1.09 \\
            NL4096alpha6beta1eta.01phi2c200 & 6 & 1 & 0.01 & 2 & 200 & $2.20\times 10^{-3}$ & 1.901 & 1.175 & 0.846 & 1.09 \\
            NL4096alpha7beta1eta.01phi2c200 & 7 & 1 & 0.01 & 2 & 200 & $2.20\times 10^{-3}$ & 1.925 & 1.158 & 0.848 & 1.09 \\
            NL4096alpha8beta1eta.01phi2c200 & 8 & 1 & 0.01 & 2 & 200 & $2.20\times 10^{-3}$ & 1.944 & 1.141 & 0.843 & 1.09 \\
            NL4096alpha9beta1eta.01phi2c200 & 9 & 1 & 0.01 & 2 & 200 & $2.20\times 10^{-3}$ & 1.366 & 1.120 & 0.813 & 1.09 \\
            NL4096alpha10beta1eta.01phi2c200 & 10 & 1 & 0.01 & 2 & 200 & $2.20\times 10^{-3}$ & 1.579 & 1.107 & 0.825 & 1.09 \\
            NL1024alpha1beta.02eta.01phi2c4000 & 1 & 0.02 & 0.01 & 2 & 4000 & $8.79\times 10^{-3}$ & 5.635 & 1.408 & 0.671 & 1.22 \\
            NL1024alpha1beta.04eta.01phi2c3000 & 1 & 0.04 & 0.01 & 2 & 3000 & $8.79\times 10^{-3}$ & 4.318 & 1.393 & 0.739 & 1.25 \\
            NL4096alpha1beta.05eta.01phi2c2000 & 1 & 0.05 & 0.01 & 2 & 200 & $8.79\times 10^{-3}$ & 4.232 & 1.423 & 0.752 & 1.25 \\
            NL1024alpha1beta.06eta.01phi2c3000 & 1 & 0.06 & 0.01 & 2 & 3000 & $8.79\times 10^{-3}$ & 3.943 & 1.376 & 0.727 & 1.25 \\
            NL1024alpha1beta.08eta.01phi2c2000 & 1 & 0.08 & 0.01 & 2 & 2000 & $8.79\times 10^{-3}$ & 3.354 & 1.364 & 0.783 & 1.27 \\
            NL2048alpha1beta.1eta.01phi2c1000 & 1 & 0.1 & 0.01 & 2 & 1000 & $4.39\times 10^{-3}$ & 3.078 & 1.666 & 0.858 & 1.31 \\
            NL2048alpha1beta.3eta.01phi2c550 & 1 & 0.3 & 0.01 & 2 & 550 & $4.39\times 10^{-3}$ & 2.140 & 1.500 & 0.888 & 1.26 \\
            NL2048alpha1beta.5eta.01phi2c400 & 1 & 0.5 & 0.01 & 2 & 400 & $4.39\times 10^{-3}$ & 1.680 & 1.463 & 0.919 & 1.26 \\
            NL4096alpha1beta.6eta.01phi2c200 & 1 & 0.6 & 0.01 & 2 & 200 & $2.20\times 10^{-3}$ & 1.685 & 1.433 & 0.889 & 1.25 \\
            NL16384alpha1beta.6eta.01phi2c200 & 1 & 0.6 & 0.01 & 2 & 200 & $5.49\times 10^{-4}$ & 1.685 & 1.505 & 0.926 & - \\
            NL4096alpha1beta.8eta.01phi2c200 & 1 & 0.8 & 0.01 & 2 & 200 & $2.20\times 10^{-3}$ & 1.466 & 1.352 & 0.908 & 1.26 \\
            NL4096alpha1beta2eta.01phi2c200 & 1 & 2 & 0.01 & 2 & 200 & $2.20\times 10^{-3}$ & 1.091 & 1.117 & 0.864 & 1.17 \\
            NL4096alpha1beta3eta.01phi2c200 & 1 & 3 & 0.01 & 2 & 200 & $2.20\times 10^{-3}$ & 0.937 & 1.053 & 0.914 & 1.17 \\
            NL4096alpha1beta4eta.01phi2c200 & 1 & 4 & 0.01 & 2 & 200 & $2.20\times 10^{-3}$ & 0.896 & 1.036 & 0.953 & 1.16 \\
            NL4096alpha1beta1eta.02phi2c200 & 1 & 1 & 0.02 & 2 & 200 & $2.20\times 10^{-3}$ & 1.378 & 1.299 & 0.879 & 1.23 \\
            NL4096alpha1beta1eta.04phi2c200 & 1 & 1 & 0.04 & 2 & 200 & $2.20\times 10^{-3}$ & 1.312 & 1.271 & 0.880 & 1.21 \\
            NL4096alpha1beta1eta.06phi2c200 & 1 & 1 & 0.06 & 2 & 200 & $2.20\times 10^{-3}$ & 1.209 & 1.271 & 0.899 & 1.21 \\
            NL4096alpha1beta1eta.08phi2c200 & 1 & 1 & 0.08 & 2 & 200 & $2.20\times 10^{-3}$ & 1.290 & 1.255 & 0.871 & 1.18 \\
            NL4096alpha1beta1eta.1phi2c200 & 1 & 1 & 0.1 & 2 & 200 & $2.20\times 10^{-3}$ & 1.211 & 1.260 & 0.884 & 1.18 \\
            \hline
        \end{tabular}
    \end{threeparttable}
    \caption{Simulation parameters for non-linear evolution of the acoustic instability. We have listed out only the test cases explicitly mentioned or used for figures in this paper. Column 1: Identifier of the test cases. Column 2-7: $\alpha_0,\beta_0,\eta_0,\phi$ defined in \ref{eqn:powerlaw_pc_profile} and \ref{eqn:dimensionless}. Column 8: Resolution given in grid size. Column 9-11: Ratio of the time averaged mass flux, $\langle\Delta P_c\rangle$ and $\langle\Delta F_c\rangle$ to the initial values. Column 12: Effective CR adiabatic index (defined by eqn.\ref{eqn:eff_adiabatic_index}).}
    \label{tab:test_cases}
\end{table*}

We list, in Table \ref{tab:test_cases} the simulations we have used to probe the non-linear regime, the parameters used and some relevant results. These include the change in mass flux, as well as $\Delta P_c$ and $\Delta F_c$ of the time averaged profiles. As discussed in \S\ref{subsubsec:bottleneck}, $\Delta P_c$ and $\Delta F_c$ probe the net momentum and energy transfer. We show the ratios $\Delta P_c/\Delta P_{c0}, \Delta F_c/\Delta F_{c0}$  between the non-linear staircase and the background profile. 

\subsubsection{General observation of the nonlinear behavior} \label{subsubsec:general}

\begin{figure*}
    \centering
    \includegraphics[width=\textwidth]{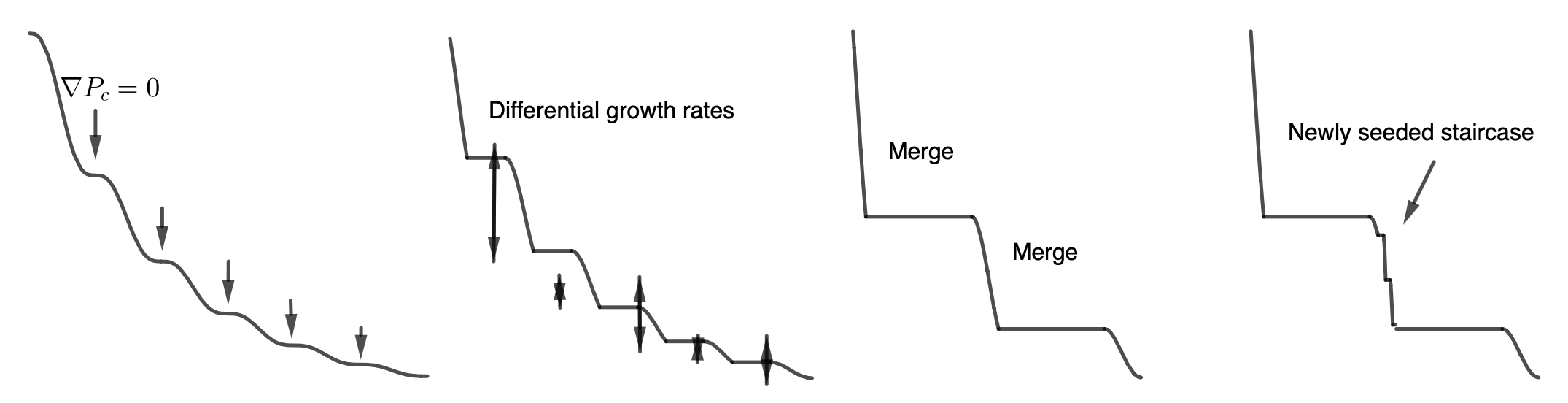}
    \caption{Non-linear growth and generation of the staircase. Time proceeds from the left panel to the right. Initial growth of acoustic waves generates a series of $\nabla P_c$ zeros, which then expand to form a series of staircases. Differential non-linear growth rates of the jumps causes stronger jumps to expand at the expense of weaker jumps, merging into bigger jumps. Subsequently, as merging slows down and new modes grow, the stair jumps fragment into smaller sub-steps.}
    \label{fig:non-linear_staircase}
\end{figure*}

The following proceeds after the linear growth phase. Growth of acoustic waves is slowed when the amplitude becomes large enough such that the CR pressure gradient becomes zero at the wave extrema (left most panel of fig.\ref{fig:non-linear_staircase}). At these locations, CRs decouple from the gas, truncating CR heating, which is the source of energy driving the instability. Elsewhere gas and CRs are still coupled, so growth continues, though growth rates become strongly inhomogeneous. The local patches of CR gradient zeros expand, forming a series of CR plateaus separated by jumps in CR pressure, i.e. a staircase structure that travel up the $P_c$ gradient (second left of fig.\ref{fig:non-linear_staircase}). Gas and CR remain coupled at the jumps, so the instability continues to act, stretching the jump heights. Each CR jump can be seen to associate with a density spike. Local conditions drive a differential in non-linear growth for each jump, causing the CR plateaus to rise or drop at varying rates. When one plateau levels with another, the jump between them vanishes, they merge and move thereafter as one (second right of fig.\ref{fig:non-linear_staircase}). Occasionally, newly seeded modes with wavelengths at or smaller than the jump width would arise at a stair jump, breaking it up into a series of sub-staircases (right most of fig.\ref{fig:non-linear_staircase}). When a stair propagates into a region for which $\beta\gtrsim 0.5$, where acoustic waves are damped, the jump will shrink. As the instability saturates, we see continual staircase propagation, breaking and merging of the staircase jumps in an overall time-steady manner. 

\begin{figure*}
    \centering
    \includegraphics{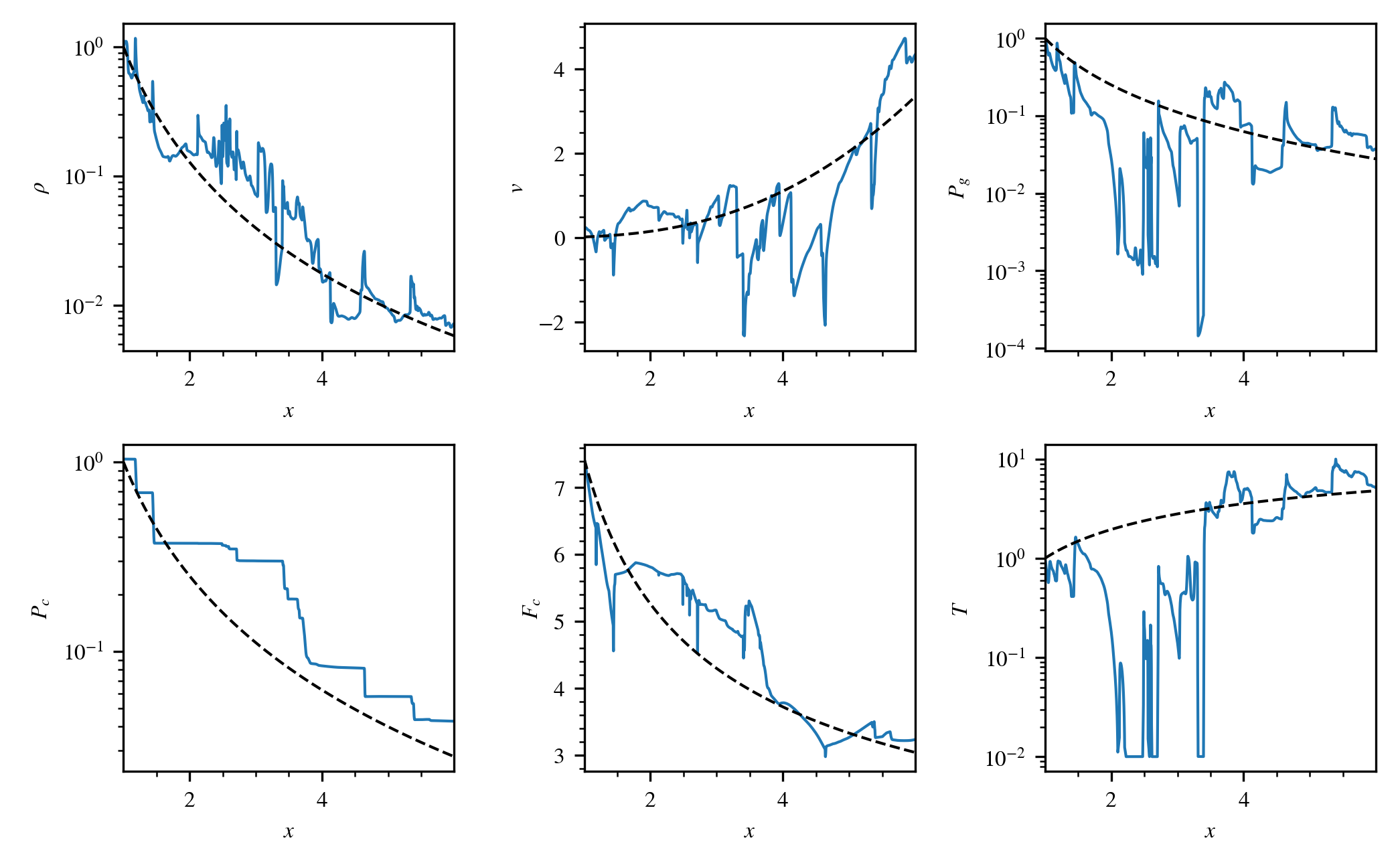}
    \caption{Density (top left), velocity (top middle), gas pressure (top right) and CR pressure (bottom left), CR flux (bottom middle), temperature (bottom right, defined in code units by $T=P_g/\rho$) plots of the non-linear evolution of the acoustic instability at $t=2.84$ (blue solid lines). The initial profiles are shown by black dashed lines for comparison ($t=0$). A staircase structure can be seen in the CR pressure. Plasma $\beta$ decreases from 0.6 to 0.017 from $x=1$ to $x=6$, going below the stability threshold $\beta = 0.53$ at $x\approx 1.1$. The case shown is a time slice taken from \texttt{NL4096alpha1beta.6eta.01ms.015psi0c200}.}
    \label{fig:time_snapshot}
\end{figure*}

Fig.\ref{fig:time_snapshot} depicts a snapshot which clearly shows the aforementioned staircase structure in the $P_c$ profile. The morphology of the $P_c$ profile is distinct from the other profiles, particularly the gaseous profiles, in several ways. First, $P_c$ decreases monotonically whereas the density exhibits small scale shocks. Second, whereas the $P_c$ jumps, as well as gas density and velocity fluctuations are of order $\Delta P_c/ P_c \sim \Delta \rho/\rho \sim \Delta v/ v \sim 1$, the gas pressure and temperature exhibits much greater fluctuations, $\Delta P_g/ P_g \sim \Delta T/ T \gg 1$. 

\begin{figure}
    \centering
    \includegraphics[width=0.4\textwidth]{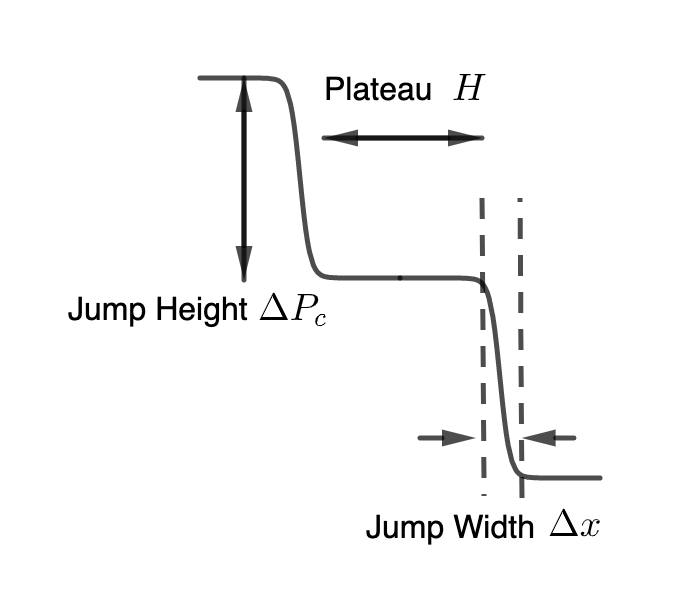}
    \caption{Clarification of jump width, height and plateau.}
    \label{fig:jump_define}
\end{figure}

In Fig. \ref{fig:jump_define}, we illustrate the meaning of the terms \emph{jump width} $\Delta x$, \emph{jump height} $\Delta P_c$ and \emph{plateau} $H$, which we use throughout the rest of this paper. We often express the jump width as $w\equiv\Delta x/l_\mathrm{diff}$, normalized with respect to the local diffusion length, while the jump height is often expressed as $h\equiv\Delta P_c/P_c$, i.e. the logarithmic change in $P_c$.

\subsubsection{Zoom-in of staircase jumps} \label{subsubsec:zoom-in}

\begin{figure}
    \centering
    \includegraphics{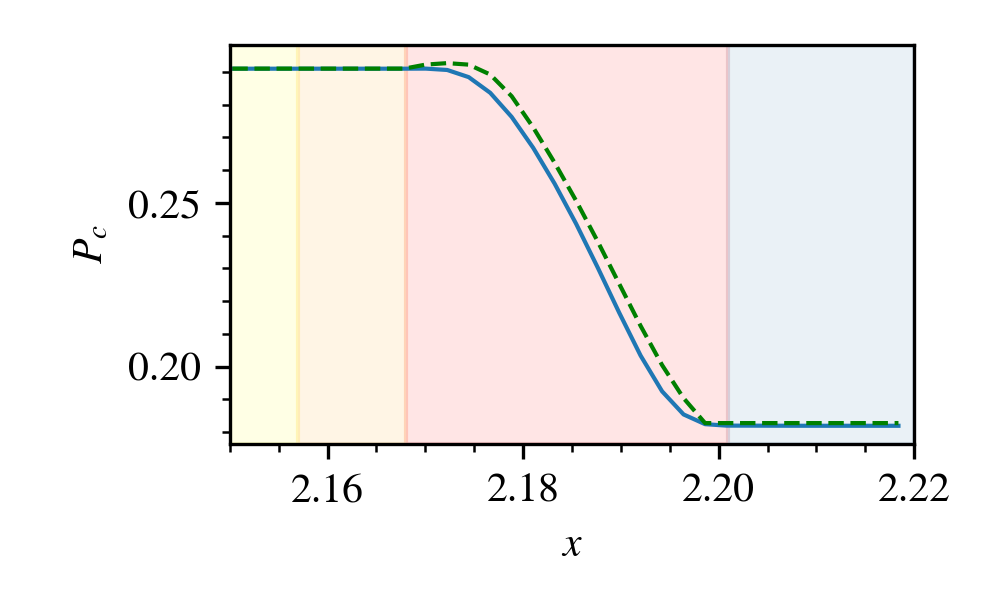} \\
    \includegraphics{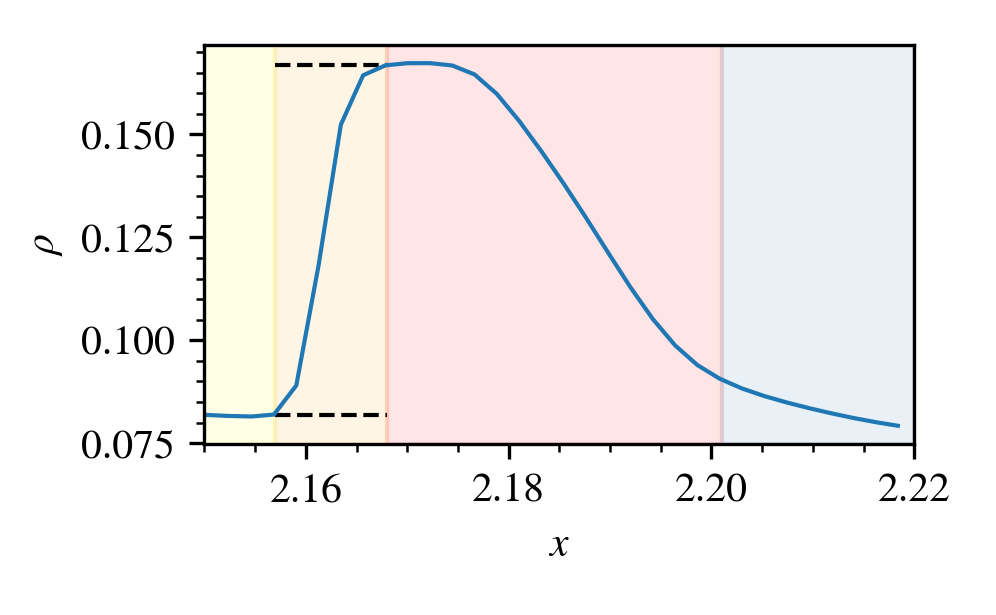} \\
    \includegraphics{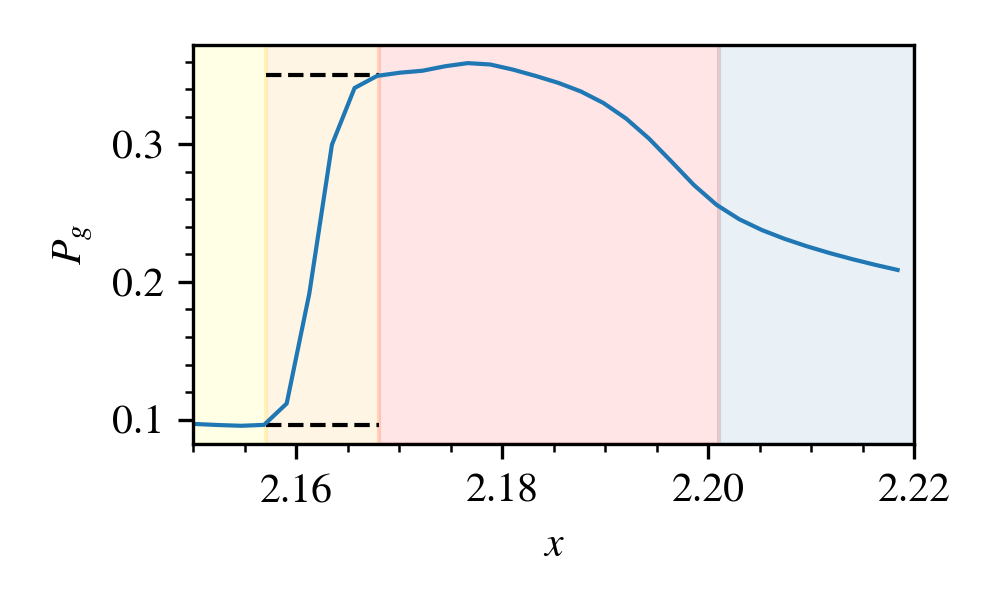}
    \caption{Zoom-in plot of the CR pressure (top), density (middle) and gas pressure (bottom) across a typical staircase jump that is propagating up the $P_c$ gradient (i.e. left in these plots). The blue solid curves are the simulation data. A stair jump in general consists of 4 sections, color coded by different background shades. The yellow section denotes the pre-jump plateau where CR and gas are uncoupled. The orange section denotes the hydrodynamic shock. The red section denotes the actual stair jump, where CR and gas are coupled. The blue section denotes the entailing plateau where CR and gas becomes uncoupled again. The green dashed curve in the $P_c$ plot (top panel) is the analytic $P_c$ profile calculated from equation \ref{eqn:bottleneck_with_flow} in the shock's rest frame for the simulated density profile. Given the upstream condition and the shock's Mach number, the Rankine-Hugoniot shock jump relations return the post-shock density and gas pressure, as displayed by the horizontal black dashed lines in the density plot, which closely match those in simulation.}
    \label{fig:panning}
\end{figure}

The $P_c$ jumps can provide intense local heating and momentum transfer as they propagate, potentially altering the overall dynamics of the gas-CR fluid. In this subsection we zoom-in onto a typical jump and explain the physics behind various features. 

Fig.\ref{fig:panning} shows the CR pressure, density and gas pressure profiles across one such jump. Since the instability is dominated by backward propagating waves (Fig \ref{fig:alpha1beta_5eta_01psi0_gauss}), like most others this jump is propagating to the left, up the CR gradient. We observe for other jumps the direction of propagation is always towards increasing $P_c$ in the \emph{rest} frame of the fluid, such that only in the supersonic part of the flow do the stairs propagate down the $P_c$ gradient in the \emph{lab} frame. Moving across the zoom-in profiles from left to right, the $P_c$ jump is preceded by sharp density and gas pressure increase. These are purely hydrodynamic shocks, across which $P_c$ remains constant and decoupled from the gas. The actual $P_c$ jump begins from the post-shock density peak, tracing the falling side of the acoustic disturbance. The jump is ensued by a CR plateau. 

Across a hydrodynamic shock, one can infer the shock speed $v_\mathrm{sh}$ by imposing mass continuity
\begin{equation}
    v_\mathrm{sh} = \frac{\rho_2 v_2 - \rho_1 v_1}{\rho_2 - \rho_1}, \label{eqn:shock_vel}
\end{equation}
where $v_1, v_2$ are the fluid velocities in the lab frame and the subscripts $1$ and $2$ denote the fluid quantities upstream and downstream of the shock respectively. The density and gas pressure increase follow the Rankine-Hugoniot shock jump relations, as shown by the black dashed lines. Proceeding down the jump, CR and gas are coupled. In the rest frame of the shock the bottleneck equation \ref{eqn:bottleneck_with_flow} is satisfied, as demonstrated by the green dashed line. The gas and CR profiles across other jumps also exhibit similar structure: a purely hydrodynamic shock at a CR plateau, followed by a jump in $P_c$ and an ensuing CR plateau.

The generation of gaseous shocks preceding each $P_c$ jump follows from wave steepening of acoustic waves, where differences in phase velocities between the wave crest and trough causes overtaking and a discontinuity to be formed. Waves generated in this manner are usually weak and propagate at approximately the sound speed in the fluid's rest frame (thus appearing to propagate down the CR gradient only for supersonic flows). However, with thermal cooling these initially weak shocks can evolve into strong shocks, as we describe below. 

The CR staircase is characterized by sudden drops in CR pressure (the jumps), connected by regions of constant CR pressure (the plateaus). CR and gas are decoupled at the plateaus and coupled at the jumps. Thus, there are no CR forces or CR heating at the plateaus, but very strong CR momentum and energy transfer to the gas at the jumps, where $\nabla P_c$ is much larger than in the background profile. This rearrangement of where CR momentum and heat is deposited causes the entire region to fall out of force and energy balance. Regions of excess cooling (the plateau) abut regions of intense CR heating (the jump). The cooling in plateaus causes gas pressure (and temperature) to fall, and pressure gradients between the plateau and jump drives a strong shock. This shock can be considerably stronger and different in character from simple steepening of an unstable acoustic wave. It is driven by the thermodynamics of the staircase structure when cooling is present. Cooling itself can create density peaks which create bottlenecks, and further alters the structure of the staircase. 

\subsubsection{Staircase Finder} \label{subsubsec:staircase_finder}

\begin{figure*}
    \centering
    \includegraphics{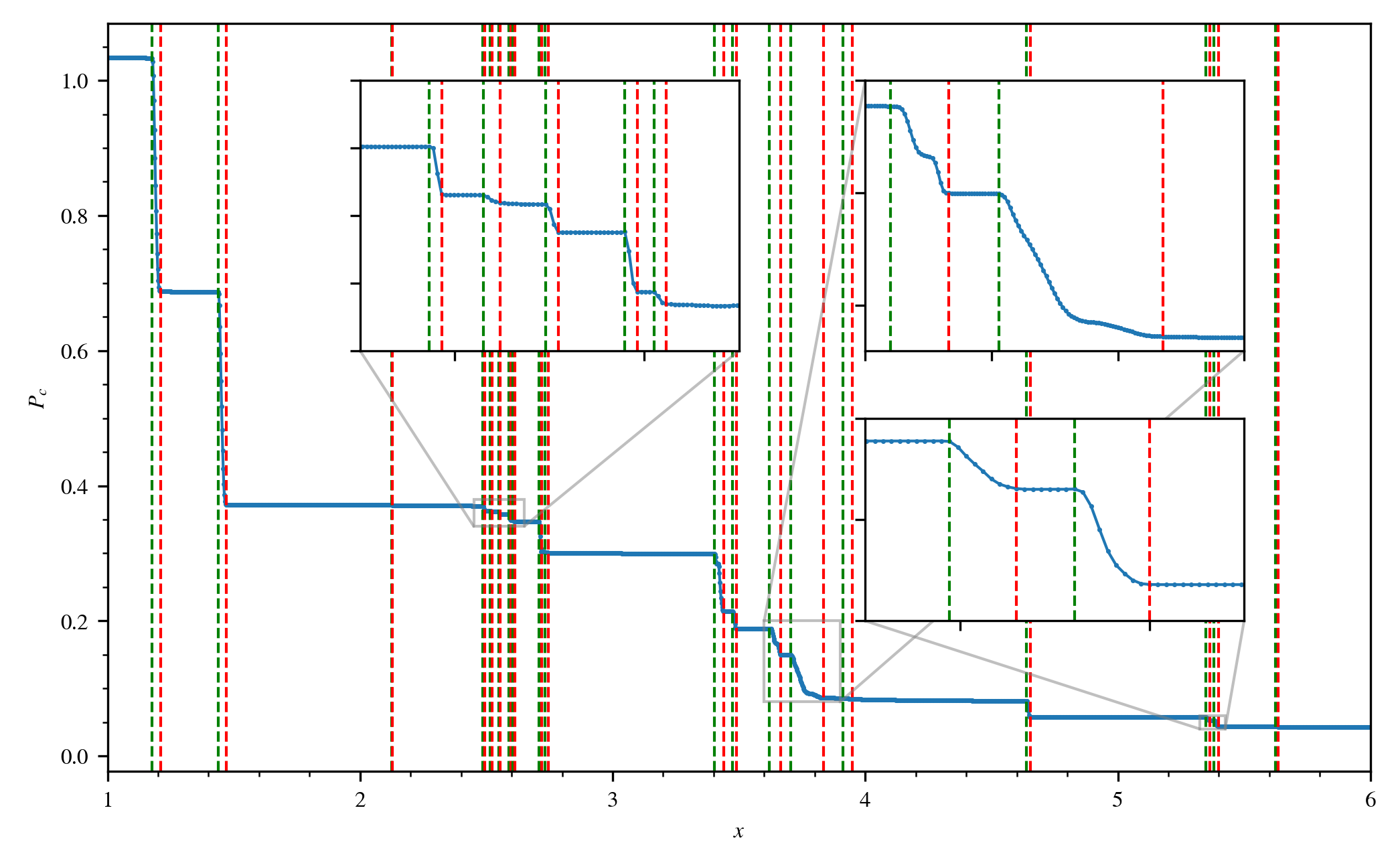}
    \caption{Staircases are identified using the algorithm described in \S\ref{subsec:non-linear}. The $P_c$ profile is plotted in blue solid line with green (red) dashed lines indicating the beginning (end) of a staircase jump. The zoom-in panels show with greater clarity parts of the $P_c$ profile with the identified jumps, showing the staircase finder to be robust. The case shown is a time slice taken from \texttt{NL4096alpha1beta.6eta.01ms.015psi0c200}.}
    \label{fig:staircase_pc}
\end{figure*}

Before we delve into the dynamical implications of the staircase, we shall determine the saturation of the non-linear staircase structure. To this end we have developed a simple staircase finder to identify staircase jumps in a $P_c$ profile. In light of equation \ref{eqn:coupling_cond}, we deem the gas to be coupled with CRs if the following condition holds: 
\begin{equation}
    \frac{\Delta x}{L_c} > \theta_\mathrm{thres}\frac{v_A}{c}, \label{eqn:finder_cond}
\end{equation}
where $\Delta x$ is the grid spacing (of order $c\Delta t$), $L_c$ is the local $P_c$ scale height and $\theta_\mathrm{thres}$ is some threshold parameter. Physically, this condition determines whether the time-dependent term in equation \ref{eqn:cr_flux} is negligible. If so, there is strong coupling, and the CR flux attains its steady state form (equation \ref{eqn:steady_state_flux}). We have found $\theta_\mathrm{thres}\approx 0.01$ to work well in identifying jumps in the staircase here, though note that this value is likely situation dependent. Every grid cell is categorized as `coupled' or `uncoupled' according to this criterion. If a `coupled' grid has an `uncoupled' grid on its left and a `coupled' grid on its right, it is deemed the start of a jump and vice versa for the end of a jump. Once the stair jumps have been identified we then record the number of jumps along the profile, as well as the jump widths, heights, etc. Fig.\ref{fig:staircase_pc} shows a snapshot of $P_c$ with vertical dashed green lines indicating the start of a jump and red dashed lines indicating the end of a jump. This method is quite robust in capturing staircase jumps.

\begin{figure*}
    \centering
    \includegraphics{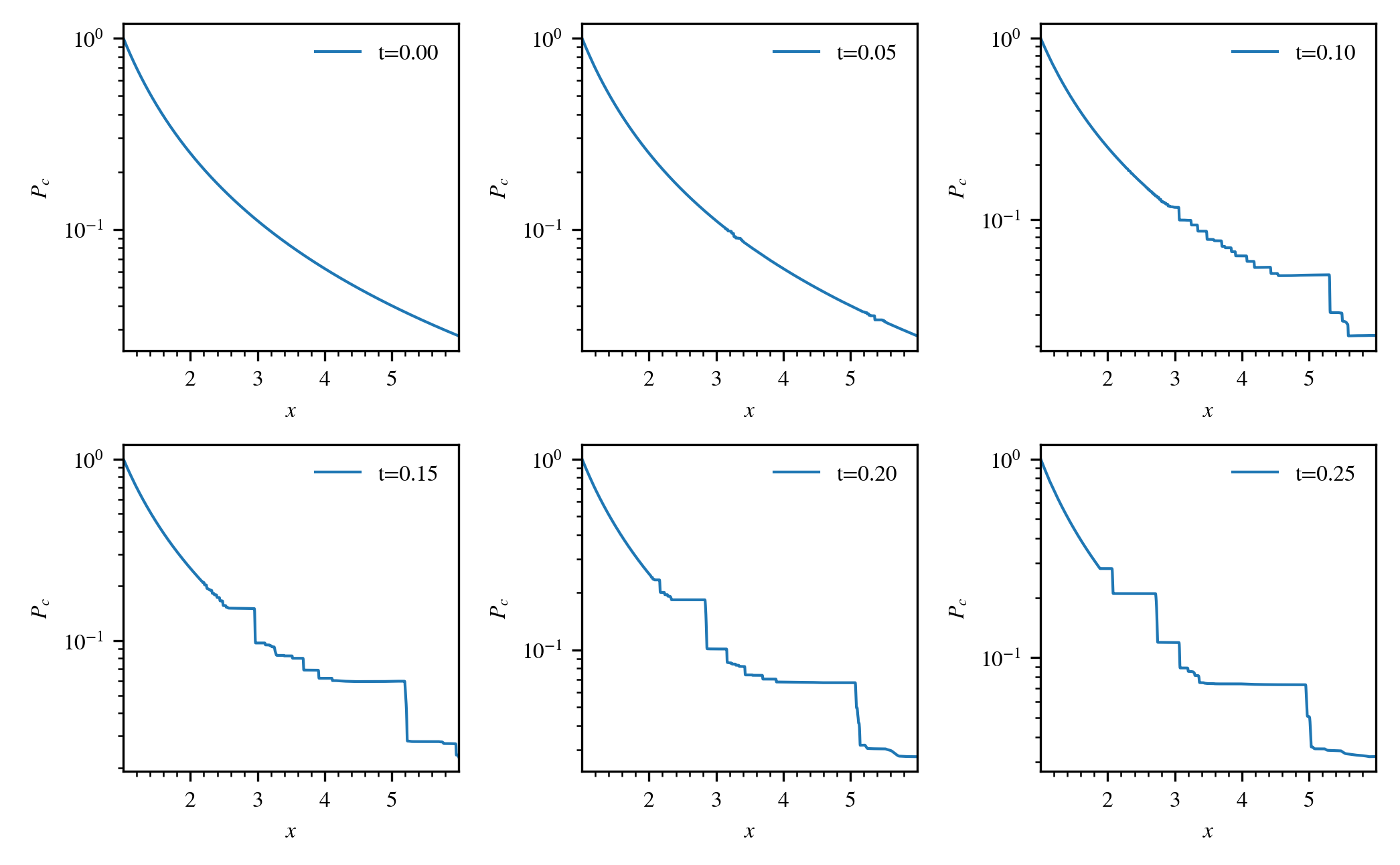}
    \caption{The evolution of the staircase at the first few time instances is displayed. Starting with smooth a background profile at $t=0$, $\nabla P_c$ zeros begin to appear due to the acoustic instability at $t=0.05$, followed by a surge of stairs at $t=0.1$. The stairs subsequently merge, propagate and fragment to new stairs. The case shown is \texttt{NL4096alpha1beta.6eta.01ms.015psi0c200}.}
    \label{fig:time_evolution}
\end{figure*}

\begin{figure}
    \centering
    \includegraphics{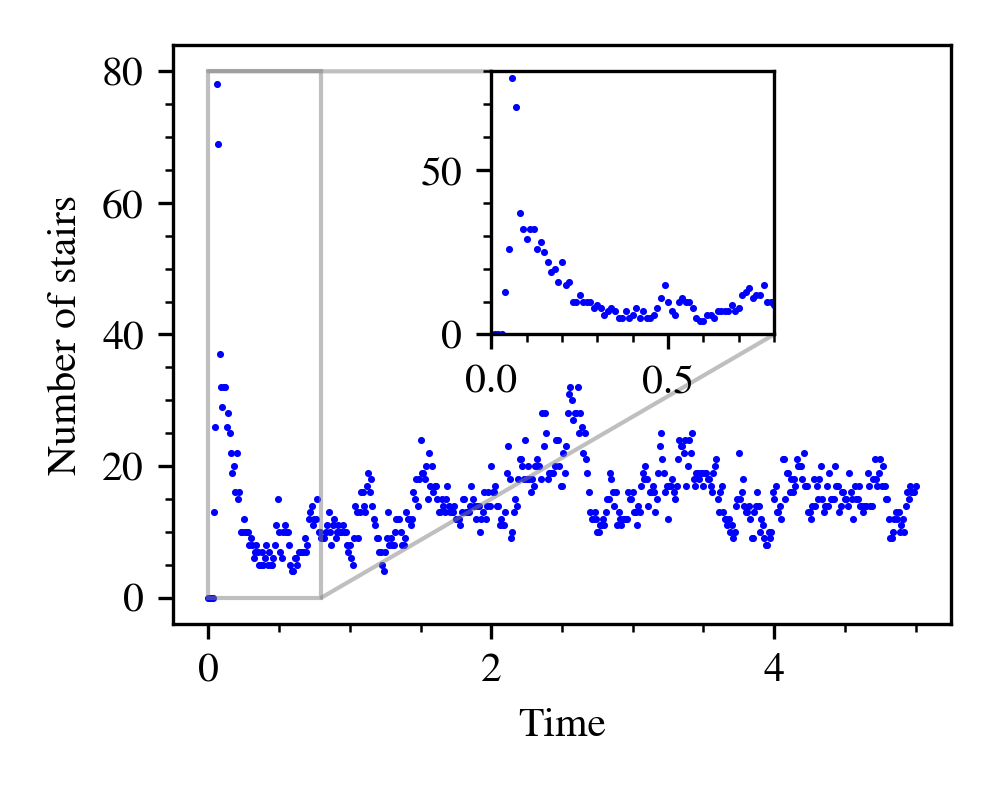}
    \caption{Number of staircases as a function of time. There is an initial surge of stairs from $t=0$ to $t=0.1$, followed by a merging phase from $t=0.1$ to $t=0.5$ and at last a quasi-steady state from $t=0.5$ onwards where the the number of staircases fluctuates about a constant value. The zoom-in panel is displays the $0<t<0.8$ section in greater detail, showing clearly an initial phase of staircase surge ($0<t<0.08$) followed by the merging phase ($0.08<t<0.5$). The case shown is \texttt{NL4096alpha1beta.6eta.01ms.015psi0c200}.}
    \label{fig:stair_number}
\end{figure}

\subsubsection{Quasi-Static State of the Staircase} \label{subsubsec:quasi_static}

The staircase finder was applied over time. Fig.\ref{fig:time_evolution} shows the evolution of the staircase at the first few time instances while Fig.\ref{fig:stair_number} shows the number of stairs (each pair of green and red dashed line is counted as one stair) captured as a function of time. From $t=0-0.1$ there is an initial surge of stair jumps seeded by numerical noise due to the acoustic instability. This time period is consistent with the growth time $t_\mathrm{grow}\sim\kappa/c_c^2\sim 0.01$ for the case displayed, where several e-folds are required to reach the non-linear stage. There is a large number of them because small scale perturbations from noise each grow until $\nabla P_c = 0$ is reached, forming plateaus. From $t=0.1-0.5$ the number of jumps drops drastically as the individual CR plateaus expand and merge. Since non-linearly steepened sound waves travel $\sim c_s$, we expect the difference in propagation speed between adjacent jumps to be $\sim c_s$, and the merging timescale $\sim H/c_s$, the sound crossing time across a plateau (the merging timescale in general scales as $H/v_\mathrm{bump}$, where $v_\mathrm{bump}$ is the jump propagation speed. In the presence of strong shocks due to cooling at the plateaus, $v_\mathrm{bump}$ does not scale as $c_s$. However, at the early stage of staircase formation, before cooling can take action, $v_\mathrm{bump}\sim c_s$ is generally true). Do all the CR plateaus simply merge into one big jump? The answer is no. From $t=0.5$ onwards the number of staircase steadied to around 15, fluctuating from 5 to 30. The number steadies due to two main reasons. First, merging of the CR plateaus have slowed down (the time for the stairs to merge lengthens with plateau width $H$). Second, newly seeded acoustic modes (seeded by numerical noise or propagating acoustic waves) at the CR jumps where CR and gas are still coupled lead to growth of a series of smaller CR stair jumps. This is similar to what happened at $t=0-0.1$, but occurring only at the jumps. This leads to a fragmentation of a stair jump into smaller sub steps. The relative independence of these two factors causes fluctuations in stair numbers for $t>0.5$. In this way the $P_c$ profile settles into a quasi-steady state marked by occasional merging, fragmentation and propagation of the staircase. In summary, the evolution of a staircase structure is characterized by: 1. an initial surge of jumps seeded by perturbations, scaled by the growth timescale $t_\mathrm{grow}$, followed by 2. merging of the jumps on some merger timescale $t_\mathrm{merge}$ and at last 3. a quasi-static state balancing fragmentation and merging of stairs. 

\begin{figure}
    \centering
    \includegraphics{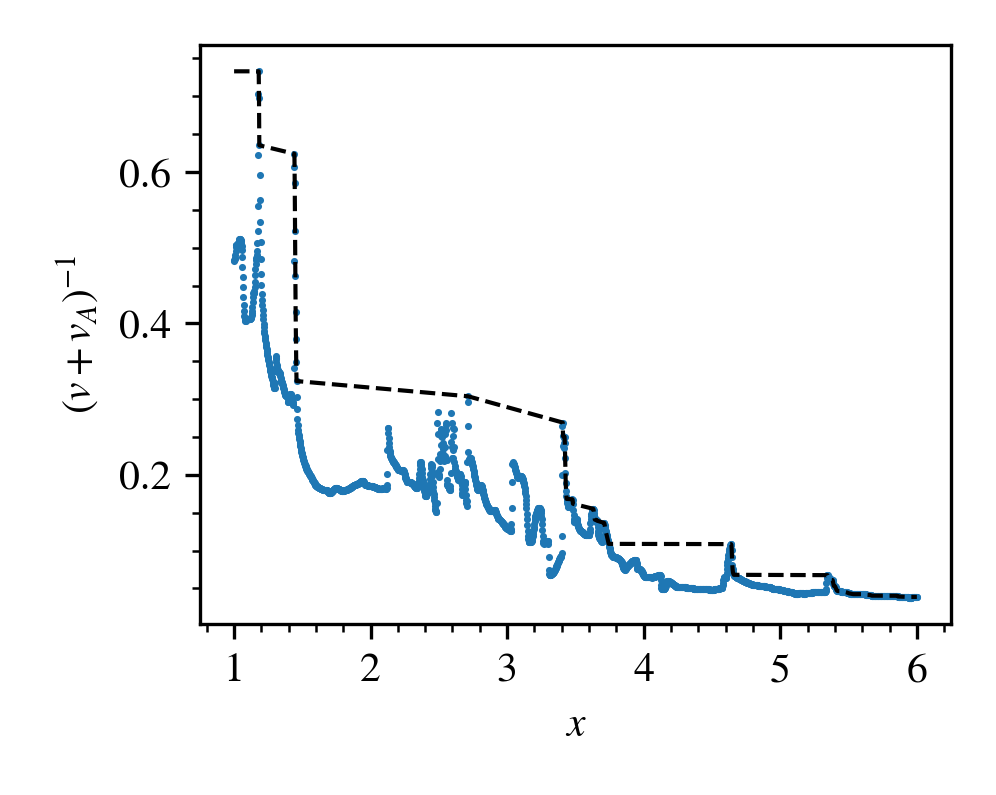} \\
    \includegraphics{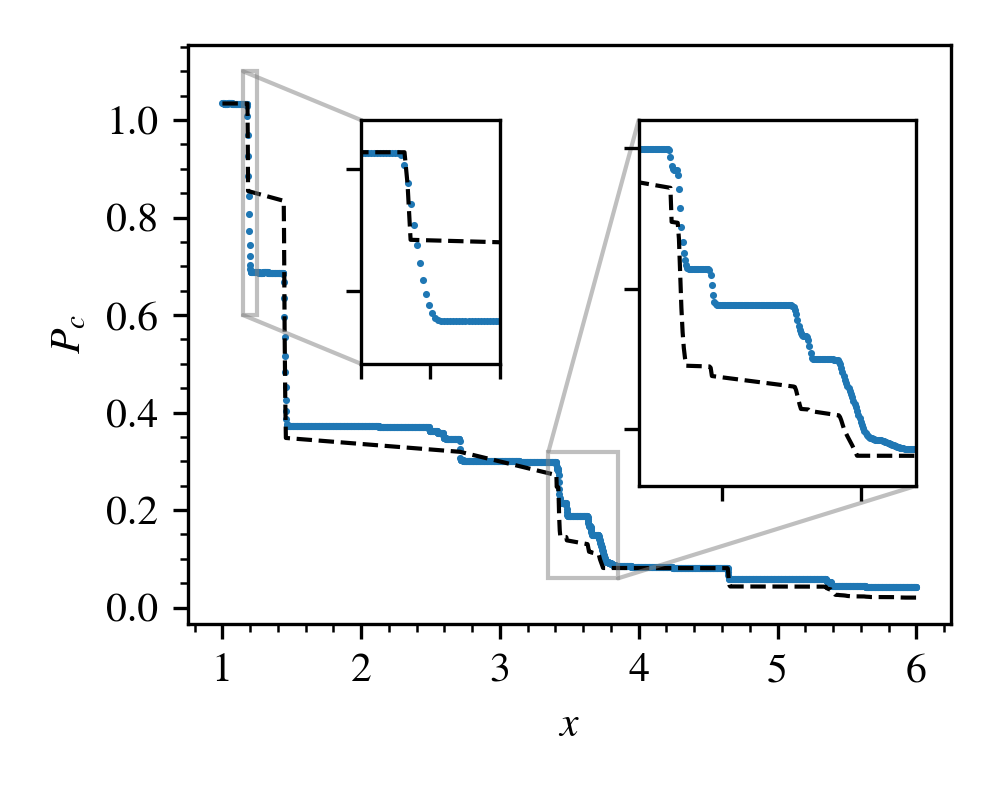} \\
    \caption{Top plot: $\qty(v + v_A)^{-1}$ (solid blue line) and its convex hull (black dashed line). Middle plot: The reconstructed $P_c$ profile from the convex hull (black dashed line) and the actual $P_c$ profile (blue) assuming profile stationarity. The zoom-in plots show with greater clarity how the convex hull procedure, assuming stationarity, fail in some instances to capture the correct jump heights. The case shown is a time slice taken from \texttt{NL4096alpha1beta.6eta.01ms.015psi0c200}.}
    \label{fig:simulation_convex_hull}
\end{figure}

\begin{figure}
    \centering
    \includegraphics{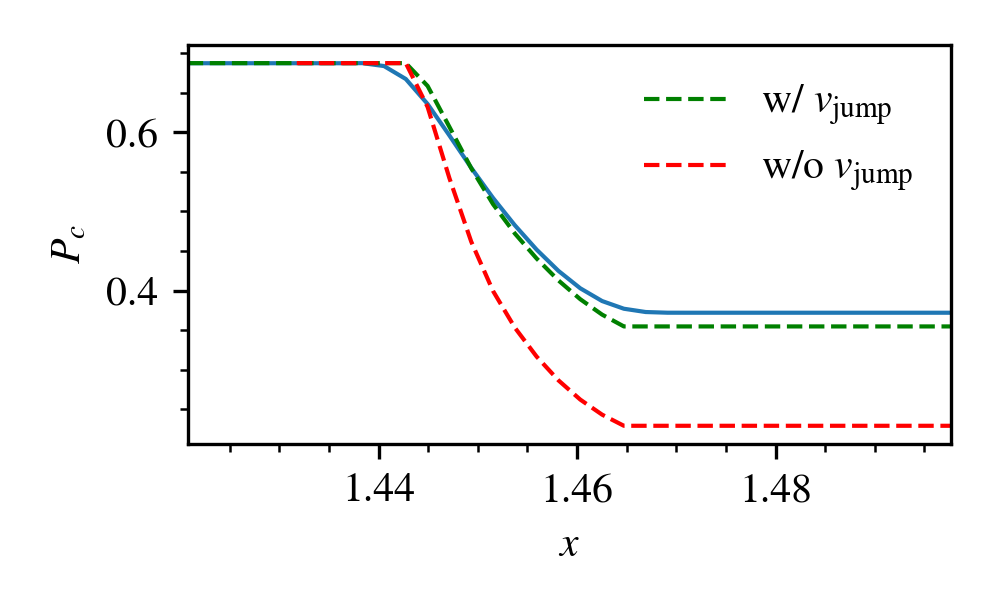}
    \caption{For a given density and velocity profile we evaluate the corresponding $P_c$ profiles from eqn.\ref{eqn:new_bottleneck_with_flow} and eqn.\ref{eqn:bottleneck_with_flow} with and without including $v_\mathrm{jump}$ respectively and compare them against $P_c$ from simulation. Blue solid line: Simulation data. Red dashed line: Estimated $P_c$ profile without $v_\mathrm{jump}$. Green dashed line: Estimated $P_c$ profile with $v_\mathrm{jump}$.}
    \label{fig:corrected_pc}
\end{figure}

\subsubsection{Bottleneck Effect with a Moving Staircase} \label{subsubsec:mergin_breaking}

In this section, we recall and extend our discussion of the bottleneck effect (\S\ref{subsubsec:bottleneck}) in the context of the non-linear profile arising from the acoustic instability (fig.\ref{fig:time_snapshot}). In the presence of non-linear acoustic disturbances, the bottleneck effect causes a CR plateau to be formed on the rising side of the disturbance (viewed from the standpoint of the streaming CRs). Meanwhile CR and gas are coupled on the falling side, forming CR jumps. The plateaus and jumps occur one after another, in conjunction with successively rising and falling acoustic disturbances, forming a staircase. If the density and velocity profiles were stationary, with all the peaks held fixed, $P_c$ would acquire a stationary profile as well, whose profile can be obtained through a `convex hull' procedure, as shown by the dashed curve in the top plot of Fig.\ref{fig:convex_hull}. The convex hull is the minimal surface that encompasses the entire $\qty(v + v_A)^{-1}$ profile\footnote{The steps to constructing a convex hull is described in greater detail here. 1. Identify the highest peak of the $\qty(v + v_A)^{-1}$ profile. Incoming CRs will bottleneck all the way up to here. 2. Trace the falling side of the $\qty(v + v_A)^{-1}$ peak while searching for the next highest peak. CRs will bottleneck up to here next. 3. By repeating this procedure over successively lower $\qty(v + v_A)^{-1}$ peaks a convex hull can be constructed for the $\qty(v + v_A)^{-1}$ profile. The convex hull is given by the dashed line in the top plot of fig.\ref{fig:simulation_convex_hull}. 4. Finally, the $P_c$ profile is obtained by applying equation \ref{eqn:bottleneck_with_flow} using the convex hull of $\qty(v + v_A)^{-1}$.}. $P_c$ can then be obtained via equation \ref{eqn:bottleneck_with_flow}. Fig.\ref{fig:simulation_convex_hull} shows one such example of reconstructed $P_c$ profile using the convex hull procedure. Comparing the reconstructed $P_c$ profile against actual simulations shows that even though the locations of the $P_c$ jumps can be identified reasonably, the magnitude of the individual jumps are incorrectly estimated. 

Clearly, the profiles are not stationary, since the jumps (and shocks) are propagating. Could this be the problem? Equation  (\ref{eqn:bottleneck_with_flow}) only holds in the rest frame of the jumps. In the lab frame, the conserved quantity is thus: 
\begin{equation}
    P_c\qty(v + v_A - v_\mathrm{bump})^{\gamma_c} = \text{const} 
\end{equation}
instead, where $v$ is the lab frame velocity profile and $v_\mathrm{bump}$ is the propagation velocity of the jump determined by imposing continuity across the preceding hydrodynamic shock (eqn.\ref{eqn:shock_vel}). This is the same as equation \ref{eqn:new_bottleneck_with_flow}, aforementioned in \S\ref{subsubsec:bottleneck}. In fig.\ref{fig:corrected_pc}, we show that once equation \ref{eqn:new_bottleneck_with_flow} is used, good agreement is restored. Since all the jumps propagate at different velocities, the frame transformation has to be applied separately to each jump to reconstruct an entire staircase, using the convex hull approach.  

\begin{figure}
    \centering
    \includegraphics{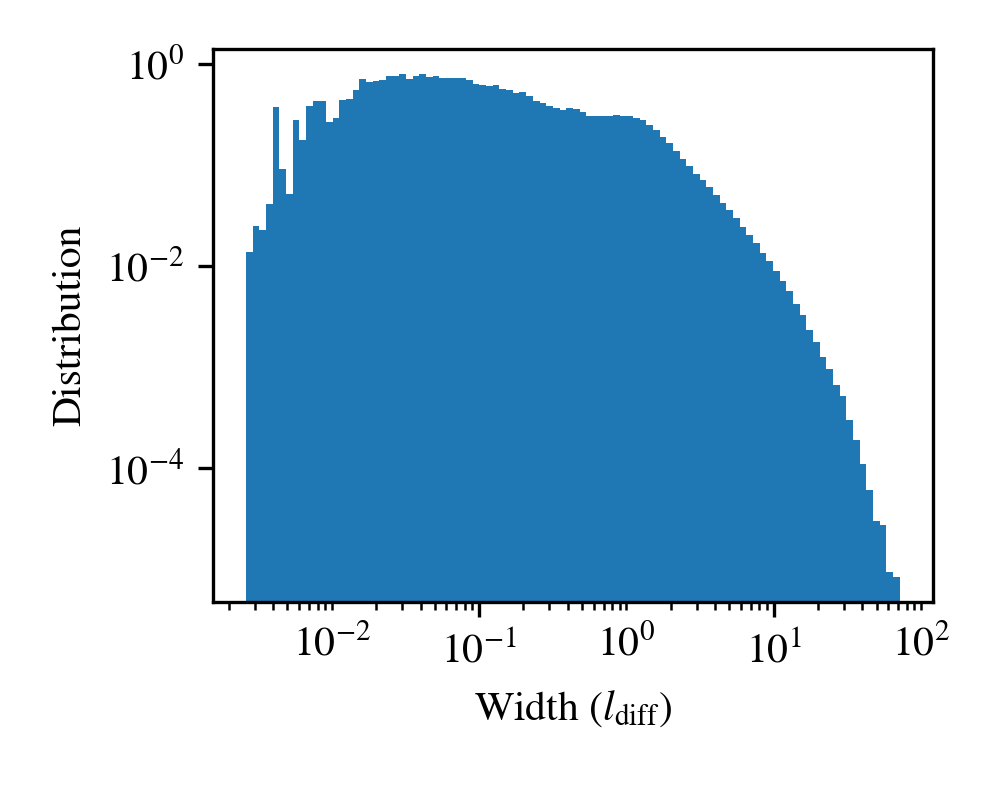}
    \caption{Distribution of jump widths $w$ (in units of $l_\mathrm{ldiff}$, i.e. $w\equiv\Delta x/l_\mathrm{diff}$), showing a broad peak about $l_\mathrm{diff}$ ($w\sim 1$) and a cutoff above. The case shown is \texttt{NL16384alpha1beta.6eta.01ms.015psi0c200}.}
    \label{fig:statwidth}
\end{figure}

\begin{figure}
    \centering
    \includegraphics{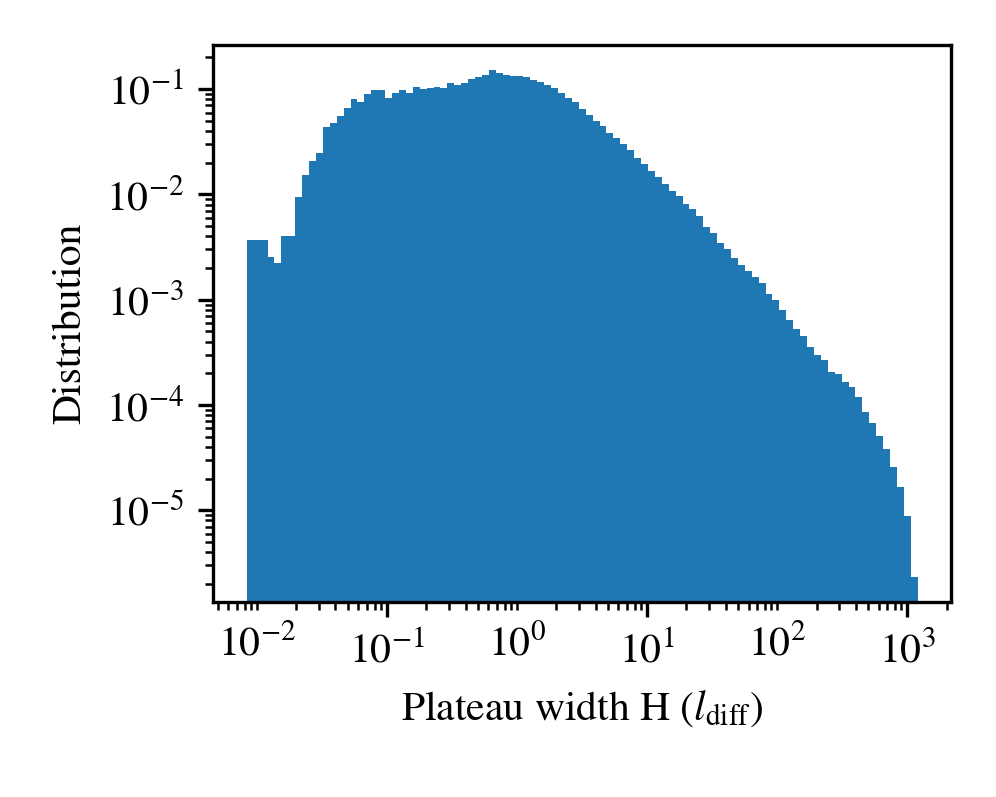}
    \caption{Distribution of plateau widths $H$ (in units of $l_\mathrm{ldiff}$), showing again a broad peak about $H\sim l_\mathrm{diff}$ (i.e. $\sim 1$ in the normalized scale shown). The plateau distribution is considerable dispersed in the part above $l_\mathrm{diff}$ compared to the jump width distribution (fig.\ref{fig:statwidth}). The case shown is \texttt{NL16384alpha1beta.6eta.01ms.015psi0c200}.}
    \label{fig:statplateau}
\end{figure}

\begin{figure}
    \centering
    \includegraphics{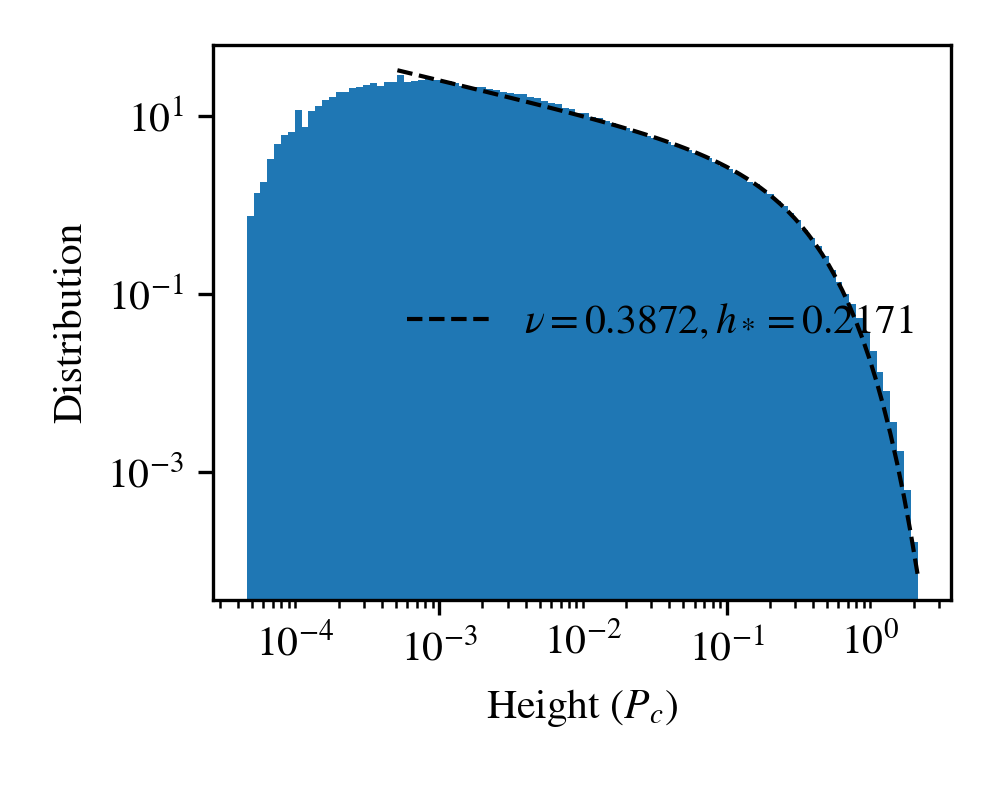}
    \caption{Distribution of jump heights (in units of the local $P_c$) with fitting parameters $\nu$ and $h_*$ (eqn.\ref{eqn:fitting_func}). The case shown is \texttt{NL16384alpha1beta.6eta.01ms.015psi0c200}.}
    \label{fig:statheight}
\end{figure}

\begin{figure}
    \centering
    \includegraphics{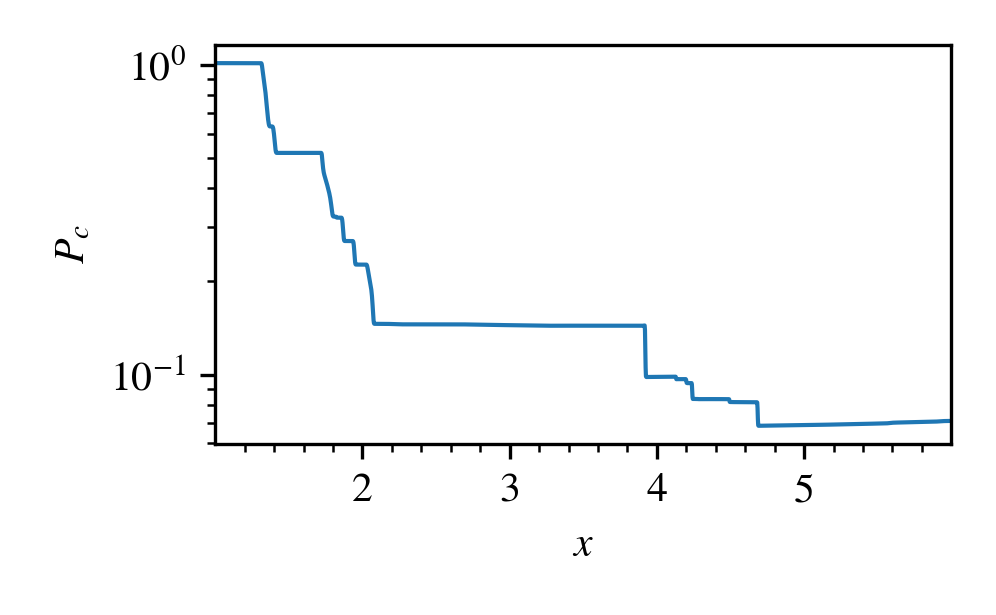} \\
    \includegraphics{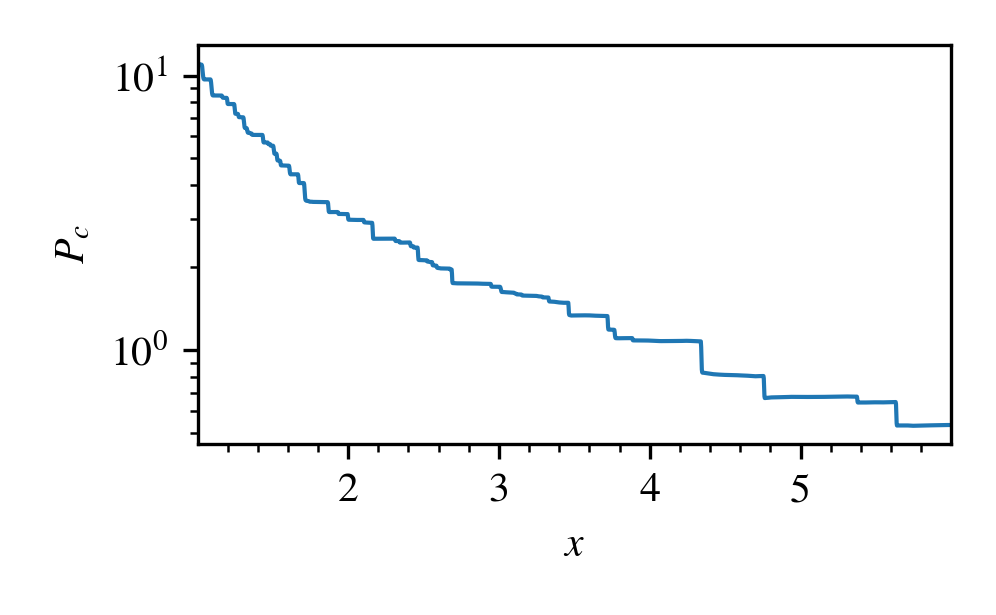}
    \caption{Two snapshots of $P_c$ taken at the same time in which the bottom test case has a CR pressure 10 times higher than the top case, all other parameters held constant. The bottom test case has considerably smaller plateau widths and jump height than the top case, consistent with the discussion in \S\ref{subsubsec:widths_heights}. The cases shown are \texttt{NL4096alpha1beta1eta.01ms.015psi0c200} and \texttt{NL4096alpha10beta1eta.01ms.015psi0c200}.}
    \label{fig:staircase_different_pc}
\end{figure}

\begin{figure}
    \centering
    \includegraphics{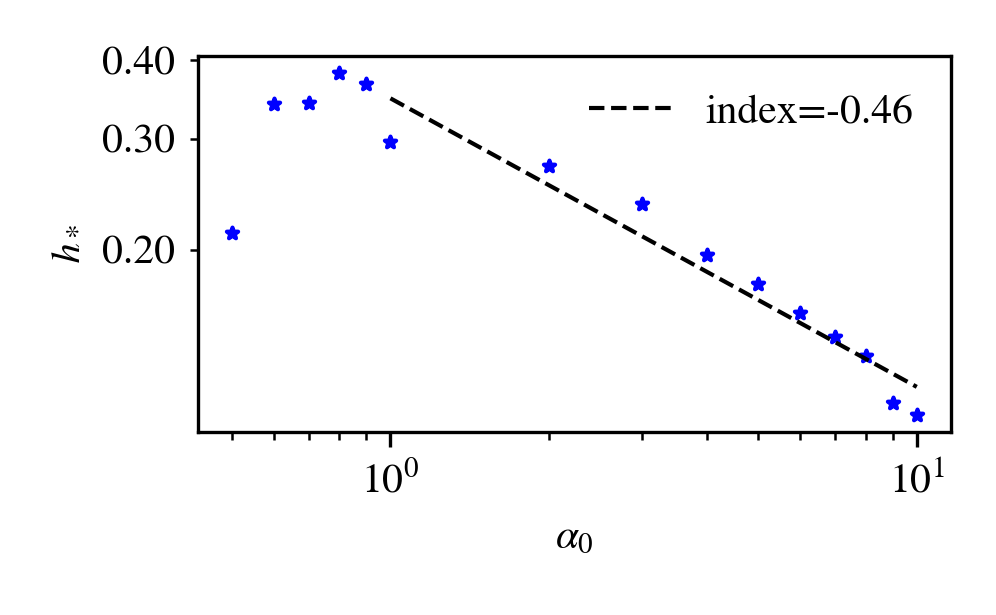} \\
    \includegraphics{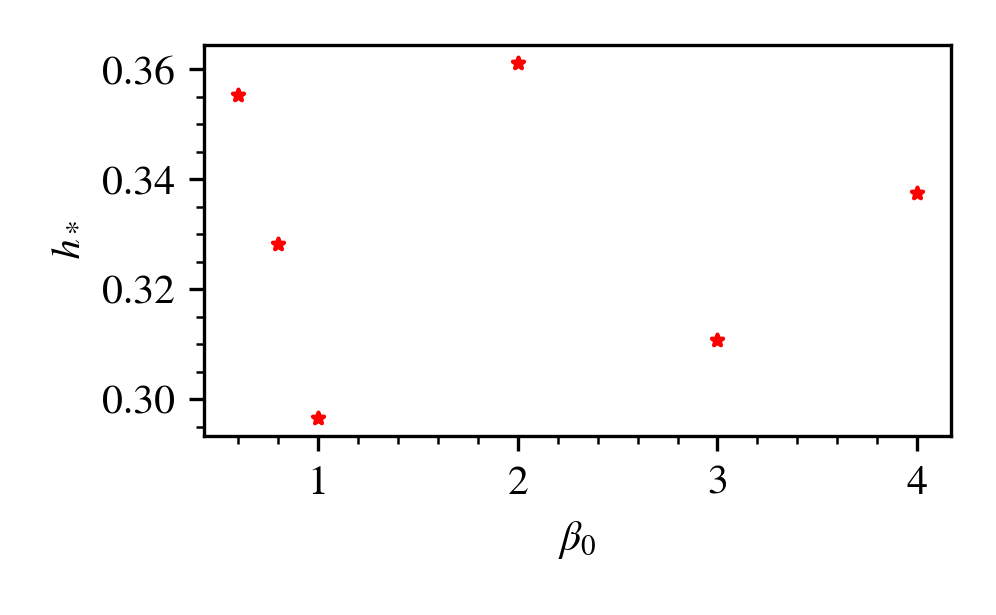} \\
    \includegraphics{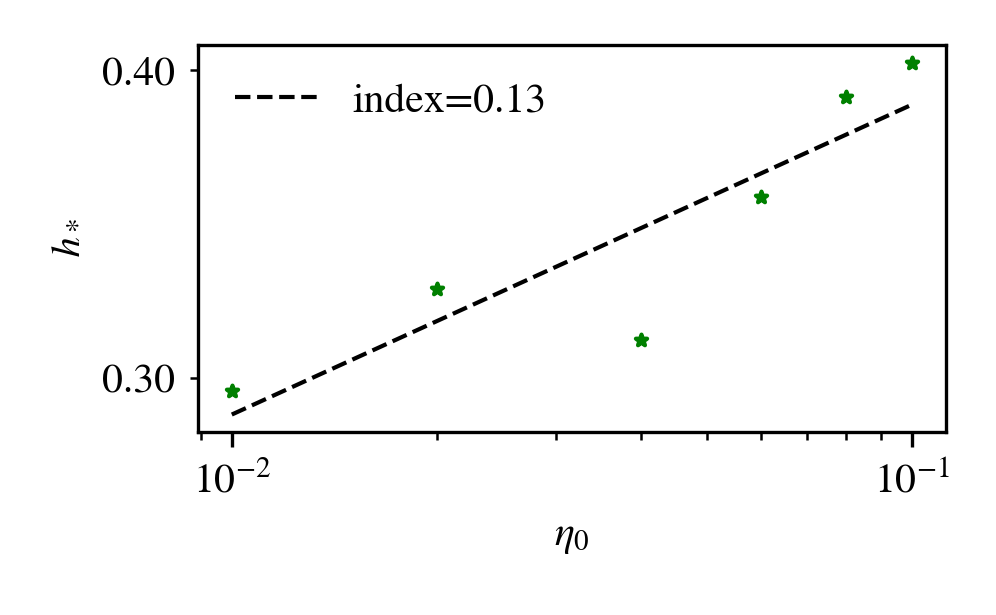}
    \caption{Variation of the characteristic jump height $h_*$ (defined in eqn.\ref{eqn:fitting_func}) for a range of $\alpha_0= P_{c0}/P_{g0}$ (top, fixing $\beta_0=1,\eta_0=0.01$), $\beta_0 = 8\pi P_{g0}/B^2$ (middle, fixing $\alpha_0=1,\eta_0=0.01$) and $\eta_0 = \kappa/\gamma_c L_{c0} c_{s0}$ (bottom, fixing $\alpha_0=1,\beta_0=1$). The legends indicate the power law index found from logarithmic fitting when there is a prevailing trend. Log-log plotting is used for the top and bottom panel.}
    \label{fig:stair_jump_scaling}
\end{figure}

\begin{figure}
    \centering
    \includegraphics{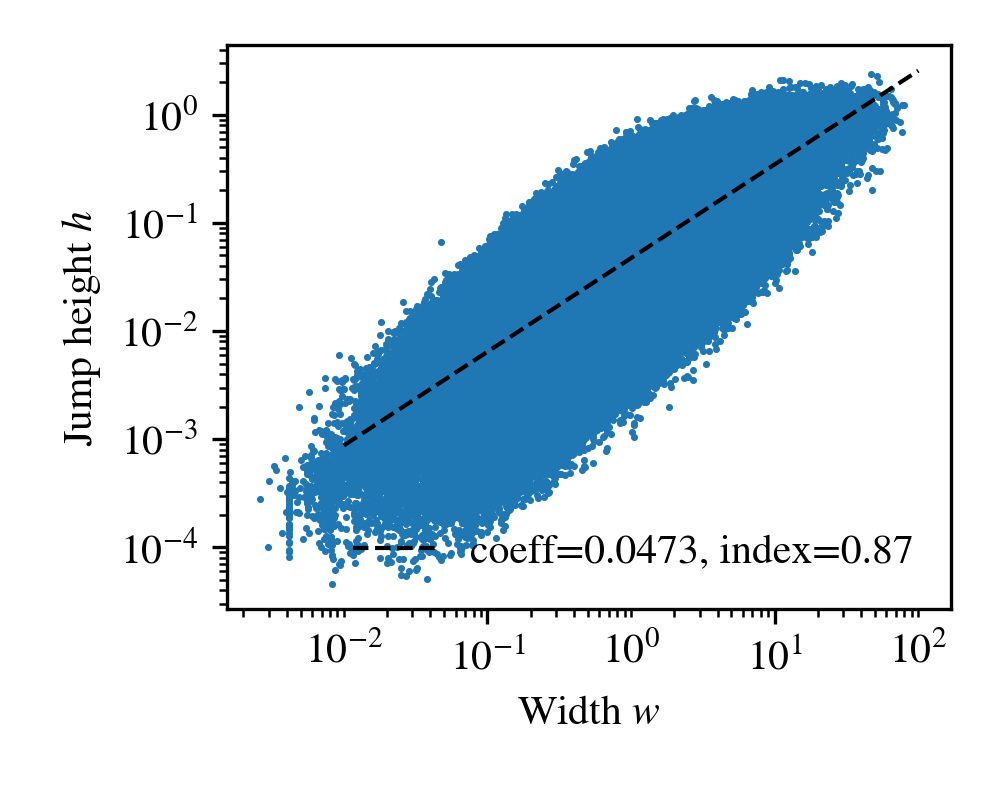}
    \caption{$h\equiv\Delta P_c/ P_c$ (logarithmic jump height) against $w\equiv\Delta x/l_\mathrm{diff}$ (normalized jump width) scatter plot (in code units). It is overplotted by a power law fit with the given power index. The cases shown is \texttt{NL16384alpha1beta.6eta.01ms.015psi0c200}.}
    \label{fig:scale_width_height}
\end{figure}

\subsubsection{Jump Widths, Heights and Plateau Widths} \label{subsubsec:widths_heights}

We now discuss some characeteristic scales in the staircase, such as the jump width, heights and plateau widths. We begin with the jump width $\Delta x$. As discussed in \S\ref{subsec:linear_theory}, the growth rate increases with wave number for $k\kappa/c_s\lesssim 1$, flattening to a constant value for $k\kappa/c_s\gtrsim 1$. With sufficient resolution, modes with wavelength less than $l_\mathrm{diff}\sim\kappa/c_s$ - the diffusion length, will grow the fastest and form non-linear stair jumps. Modes with wavelength close to the resolution grid size will be susceptible to numerical diffusion and damp. Thus we expect the distribution of stair width $\Delta x$ to have the following shape: a broad peak about the diffusion length $l_\mathrm{diff}\sim\kappa/c_s$, with a cutoff on larger scales due to long growth times, and another cutoff approaching the grid size, due to numerical diffusion. 

With the staircase finder one can also study the distribution of jump widths. We tally up the jump widths and display their distribution $\dv*{n}{w}$ in fig.\ref{fig:statwidth}, where $w\equiv\Delta x/l_\mathrm{ldiff}$ is the jump width normalized by the local diffusion length. The distribution has a broad peak at $w\sim 1$, truncating above $w\sim 1$ exponentially. The distribution below is relatively flat, but cuts off close to the grid scale. This shape is broadly consistent with expectations. In general, the jump width scales with the diffusion length, $\Delta x \sim l_\mathrm{diff}$.

Fig.\ref{fig:statplateau} shows a distribution of plateau widths H (in units of $l_\mathrm{diff}$). Again, a broad peak $\sim l_\mathrm{diff}$ can be observed, but the distribution above $l_\mathrm{diff}$ is considerably dispersed compared to that of the jump width distribution. Plateau widths of order $1000 l_\mathrm{diff}$ are detected. This is reasonable. Staircases are generated from acoustic waves growing non-linearly with the falling side of the wave becoming a jump and the rising side a plateau. The plateau width is therefore also an imprint of the wavelength of the growing wave, for which we have shown scales as the diffusion length $l_\mathrm{diff}$. However, plateaus can merge to become bigger, so plateaus with sizes much greater than the diffusion length could be present.

Finally, the distribution of jump heights $\Delta P_c/P_c$ is displayed in fig.\ref{fig:statheight}. It has a similar shape, cutting off sharply as $\Delta P_c/P_c$ approaches unity. This distribution can be roughly characterized as a power-law followed by an exponential cutoff at some characteristic scale, and be reasonably fitted with a Schechter function
\begin{equation}
    \dv{n}{h} = N_0\qty(\frac{h}{h_*})^{-\nu}e^{-h/h_*}, \label{eqn:fitting_func}
\end{equation}
where $h\equiv\Delta P_c/P_c = \Delta \ln{P_c}$ is the logarithmic jump height, with $\nu$ and $h_*$ denoting the power-law index and characteristic jump height respectively. 

How do these scales change as we change physical parameters? For instance, in Fig.\ref{fig:staircase_different_pc} we show the effects of a higher CR pressure. The stairs appear more clustered and there are many more of them, meaning that both the plateau widths and the jump heights are reduced. 
In Fig. \ref{fig:stair_jump_scaling}, we show how $h_*$ (the exponential cutoff as defined in equation \ref{eqn:fitting_func}) changes as we change parameters at the base ($\alpha_0,\beta_0,\eta_0$, defined in equation \ref{eqn:dimensionless}). Since our pressure profiles are power-law, this amounts to an overall rescaling; note in particular that $\alpha_0$ is independent of $x$. We find that $h_* \propto \alpha_0^{-1/2} = (P_{c0}/P_{g0})^{-1/2}$ for $\alpha_0 > 1$ (and saturates at $h_* = \Delta P_c/P_c \sim 0.4$ for $\alpha_0 < 1$). In addition, $h_*$ shows little dependence on $\beta_0,\eta_0$. 

These scaling relations are particular to our setup and likely sensitive to some key assumptions (e.g., about background profiles, as well as heating and radiative cooling). They should therefore be taken with a grain of salt; they are unlikely to be  universal for CR staircases. We can nonetheless understand some qualitative features. 
Suppose the number of staircases per scale height is $n_c = L_c/H_*$, so that $h_* = \Delta P_c/P_c \propto 1/n_c \propto H_*$, where both $h_*,H_*$ are representative values. The steady state number of staircases arises from a balance between staircase production (via the acoustic instability) and destruction (via merging). From equation \ref{eqn:growth_rate}, the linear growth rate of the acoustic instability is: 
\begin{equation}
    \Gamma_\mathrm{grow} \sim \frac{c_c^2}{\kappa}\qty(1 + \frac{1}{\beta^{1/2}})^2 + \frac{1}{\rho c_s}\qty(1 + \frac{1}{\beta^{1/2}})\dv{P_c}{x}. \label{eqn:rough_growth}
\end{equation}
$\dv*{P_c}{x}$ can be approximated as $\Delta P_c/\Delta x$. The jump width scales roughly as the diffusion length while $\Delta P_c$ is observed to be at most of order $P_c$ (e.g. in fig.\ref{fig:time_snapshot}). Therefore the term in equation \ref{eqn:rough_growth} involving $\dv*{P_c}{x}$ is at most of order $(c_c^2/\kappa)(1 + 1/\beta^{1/2})$. A close examination (not shown) of the jumps shows that the first term in \ref{eqn:rough_growth} usually dominates, and for simplicity we ignore the second term. On the other hand, the merger rate scales roughly as the shock crossing time across a plateau. We argued in \S\ref{subsubsec:zoom-in} that the shock is driven by pressure gradients. The free energy for the shocks comes from cosmic rays, such that $P_c \sim \rho v_{\rm sh}^{2}$. Thus, the characteristic shock propagation velocity is $v_{\rm sh} \sim c_c\sim \sqrt{P_c/\rho}$. Staircases `merge' when one shock (typically the stronger shock, which is propagating faster) overtakes another. If there is a distribution of shock speeds, and the characteristic spread is of order $\sim c_c$, then the merger rate is $\Gamma_{\rm merge} \sim H/c_c$. If we set $\Gamma_\mathrm{grow} \sim c_c^{2}/\kappa$ to $\Gamma_{\rm merge} \sim H/c_c$, we obtain $H \propto c_c^{-1} \propto P_c^{-1/2}$, which reproduces the scaling $h_* \propto \alpha^{-1/2}$ for $\alpha_0 > 1$. However, we caution that the growth and merger rates estimates we use are very crude, and this argument do not capture the relative independence with respect to $\beta_0,\eta_0$. Since it is unclear how universal these scalings are, we do not pursue this further. 

How are $h\equiv\Delta P_c/P_c$ (the logarithmic jump height) and $w\equiv\Delta x/l_\mathrm{diff}$ (the normalized jump width) related? Fig.\ref{fig:scale_width_height} shows a scatter plot of $h$ against $w$. A clear trend can be seen: $h$ generally increases with $w$, i.e. wider jumps are usually associated with larger jump heights.

\begin{figure*}
    \centering
    \includegraphics{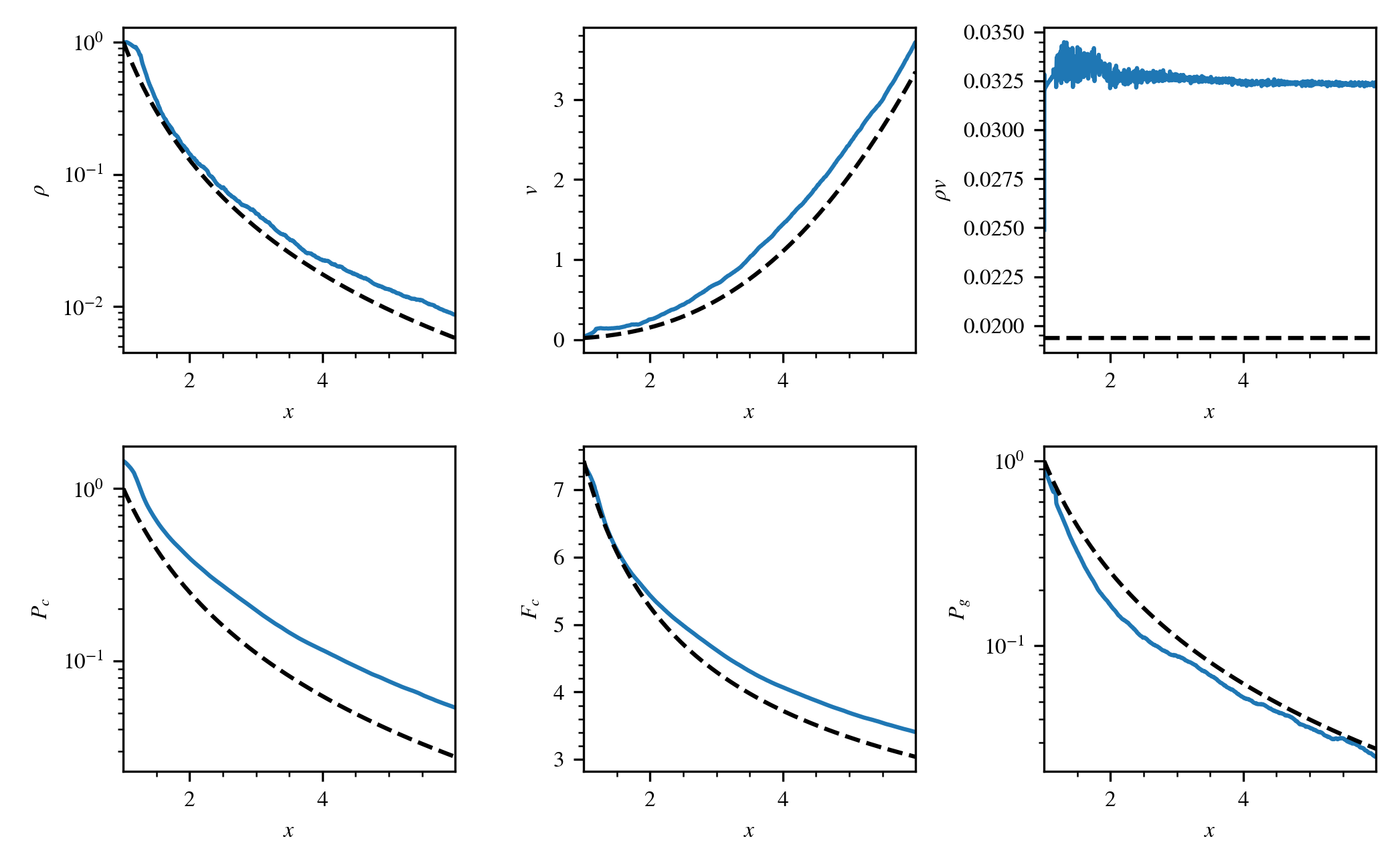}
    \caption{The blue solid lines denote the time averaged profile of density (top left), velocity (top middle), mass flux (top right), CR pressure (bottom left), CR flux (bottom middle) and gas pressure (bottom right). The black dashed lines show their respective initial profiles. The case shown is  \texttt{NL4096alpha1beta.6eta.01ms.015psi0c200}.}
    \label{fig:time_avg}
\end{figure*}

\begin{figure}
    \centering
    \includegraphics{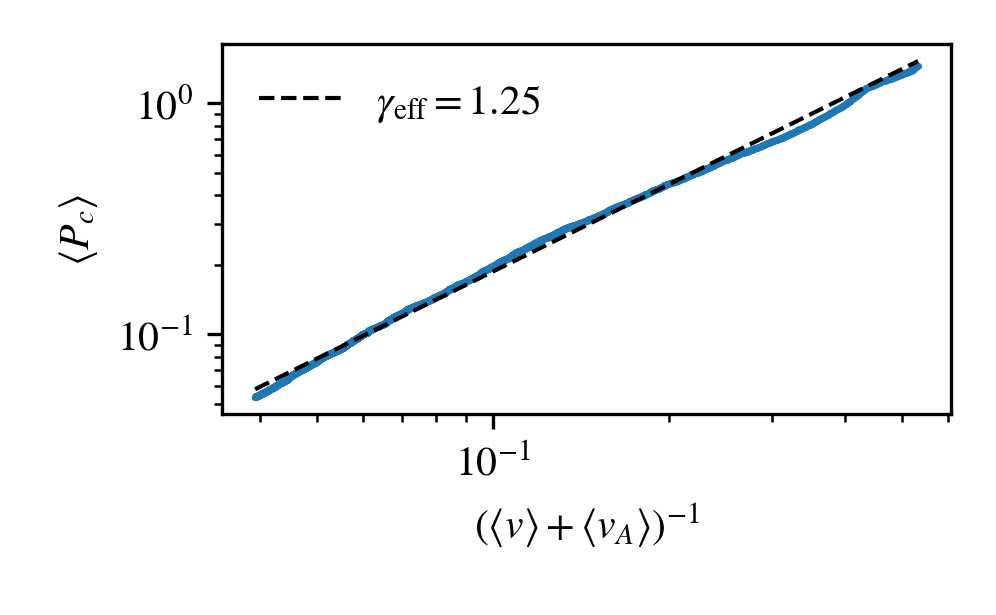} \\
    \includegraphics{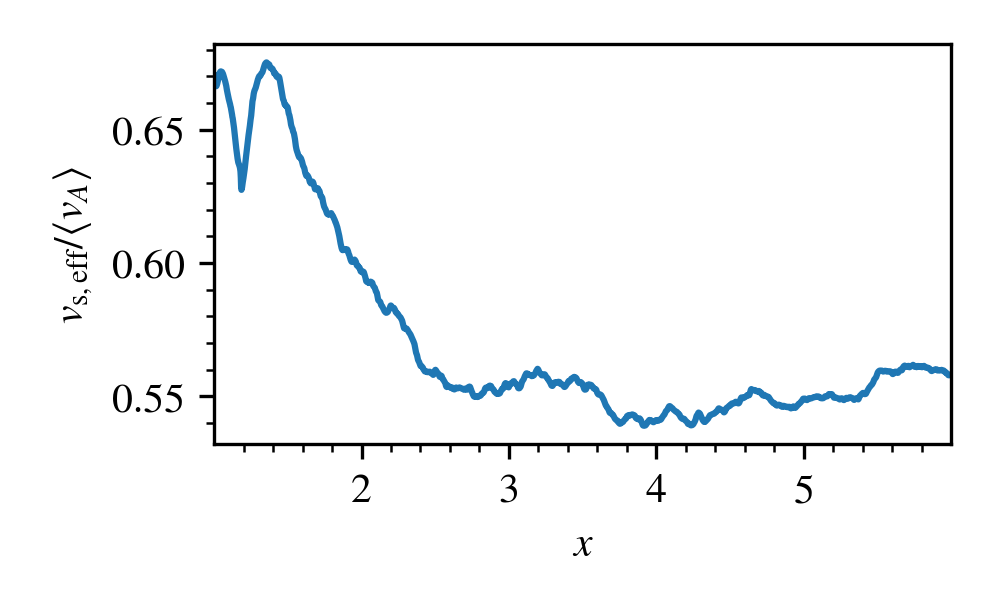}
    \caption{Top: Phase plot of $\langle P_c\rangle$ against $\qty(\langle v\rangle + \langle v_A\rangle)^{-1}$ with the effective adiabatic index $\gamma_\mathrm{eff}$ (eqn.\ref{eqn:eff_adiabatic_index}) found from fitting. Bottom: Plot of the effect streaming speed $v_\mathrm{s,eff}$ (in units of the local time averaged Alfven speed). The case shown is \texttt{NL4096alpha1beta.6eta.01ms.015psi0c200}.}
    \label{fig:two_suggestions}
\end{figure}

\begin{figure*}
    \centering
    \includegraphics{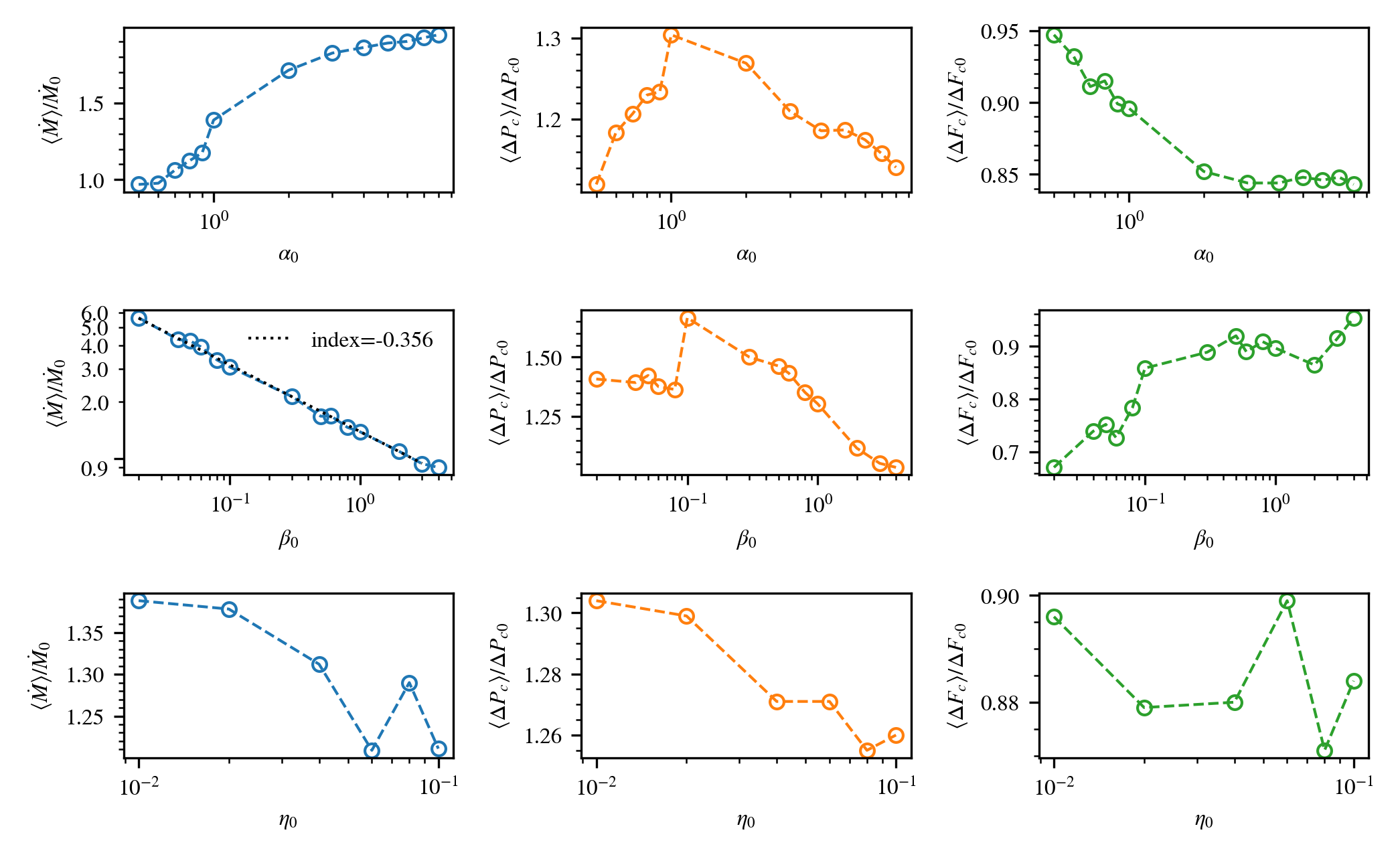}
    \caption{Time averaged quantities $\langle\dot{M}\rangle/\dot{M}_0$ (blue dashed line), $\langle\Delta P_c\rangle/\Delta P_{c0}$ (orange dashed line) and $\langle\Delta F_c\rangle/\Delta F_{c0}$ (green dashed line) for different $\alpha_0,\beta_0$ and $\eta_0$. All changes are with respect to the new background profile for a given set of parameters.}
    \label{fig:alpha_beta_eta_avg}
\end{figure*}

\subsubsection{Dynamical Effect and Averaged Properties} \label{subsubsec:dynamics}

The presence of staircases significantly changes outflow dynamics. The decoupling of gas from CRs at the plateaus deprives it of CR pressure support and Alfvenic heating. Great $P_c$ support and intense heating do occur, however, at the CR jumps, so a fluid parcel not co-propagating with the staircase experiences alternating pressure support and heating as it transverses plateaus and jumps. The question is: to what degree do the spasmodic pressure support and heating due to stair jumps balance the deficits at CR plateaus? And how does it affect the averaged profiles?

In \S\ref{subsubsec:mergin_breaking} we observed for a moving stair jump, it is the quantity given by equation \ref{eqn:new_bottleneck_with_flow} that is conserved. A moving jump, as shown in fig.\ref{fig:bottle_move} and \ref{fig:corrected_pc}, can cause the jump height to change as compared to when it is stationary\footnote{If one estimates the ratio of $P_c$ before and after the jump to be $P_{c,\mathrm{after}}/P_{c,\mathrm{before}}=A/B$, where $B>A$ then adding a positive constant $C$ to the numerator and denominator would lead to an increase in the ratio, i.e. $\qty(A+C)/\qty(B+C) > A/B$. For example, adding $2$ to the numerator and denominator of $1/4$ gives $3/5 > 1/4$. This means the jump height is lessened.}. In \S\ref{subsubsec:bottleneck} we discussed, for a steady state profile, the total momentum and energy transfer are given by $\Delta P_c$ and $\Delta F_c$. We also showed, in fig.\ref{fig:bottle_bumps} that provided none of the density bumps exceed the global maximum of the background and are stationary, there is no change in net momentum and energy transfer as compared to when there are no bumps. Now, the staircase is dynamically moving, merging and fragmenting, so a steady state profile in which all the time derivatives vanish is impossible. However, averaged over time, the time derivatives do vanish, and $\langle\Delta P_c\rangle$ and $\langle\Delta F_c\rangle$ do represent the time-averaged momentum and energy transfer (note that angle bracketed quantities are time averaged). Since $\Delta P_c$ is the sum of jump heights, in which each is affected by the jump velocity $v_\mathrm{jump}$, the time averaged momentum transfer therefore is deeply related to the jumps' motion, as is the time averaged energy transfer.

In addition to $\langle\Delta P_c\rangle$ and $\langle\Delta F_c\rangle$, the time averaged mass flux $\langle\dot{M}\rangle$ is also a quantity of interest as in winds it controls the mass loading and transport of materials out to the CGM. We report numerical results for these quantities from our simulations, and suggest physical motivations for our findings. We defer detailed modeling to future work.


In fig.\ref{fig:time_avg} we present an example of the time averaged profiles resulting from the staircase. The time averaged profiles (blue solid lines) are placed in juxtaposition to the initial profiles (black dashed lines). Overall the change is quite modest. Apart from the shifts in $\langle P_c\rangle$ and mass flux $\langle\dot{M}\rangle = \langle\rho v\rangle$, the other profiles remain relatively close to the initial profiles. In table \ref{tab:test_cases} and fig.\ref{fig:alpha_beta_eta_avg} we collect and display  $\langle\dot{M}\rangle$, $\langle\Delta P_c\rangle$ and $\langle\Delta F_c\rangle$ for the test cases we have performed. Overall, the changes to $\langle\Delta P_c\rangle$ and $\langle\Delta F_c\rangle$ are very modest, of order $\sim 10\%$ over 1-2 decades in the parameters probed. The main interesting change is to the mass outflow rate, which changes by a factor of $\sim 2$ over 1.5 decades in $\alpha_0$, and by a factor of $\sim 6$ over 2 decades in $\beta_0$. 

It is perhaps surprising that changes to global energy and momentum transfer are so modest. After all, the CR staircase produces a drastic rearrangement of CR forces and heating -- cutting it off through a majority of the profile, and leaving only a small fraction (the jumps) where the CRs are coupled, which receive intense forces and heating. If the staircase (and associated bottlenecks) were stationary, this state of affairs would indeed be deeply destabilizing. However, a flux tube threading {\it propagating bottlenecks} (in this case, shocks) still receives heat and momentum over its entire length, albeit in a very intermittent manner. Individual fluid elements experience brief periods of intense forcing and heating, followed by longer stretches without any CR interaction. But as we have seen, averaged over time, each fluid element receives heat and momentum comparable to the background profile. Thus, while there can be strong local fluctuations, the global flow is not destabilized. For instance, the timescale for a fluid element to fall out of force balance is the free fall time, which is of order the sound crossing time $t_{\rm sc} \sim L_{\rm P}/c_s$ in the quasi-hydrostatic part of the flow, where $L_{\rm P}$ is the pressure scale height. By contrast, the timescale to receive another `hit' of CR forces is $H/v_\mathrm{bump}$; thus, $t_{\rm stair}/t_{\rm sc} \sim H/L_{\rm P}(c_s/v_\mathrm{bump}) < 1$. If the bottlenecks were stationary (e.g., a cloud co-moving with a hot wind), their effects would be much more severe. 

Despite the modest changes in global momentum and energy transfer, it is interesting that the mass flux $\dot{M}$ can change so significantly. One way to understand this is as follows. We have a fixed flux of CRs at the base, which must be transported through the stratified atmosphere. Since CRs are trapped at bottlenecks, their effective streaming speed is reduced. In Fig. \ref{fig:two_suggestions}, we show: 
\begin{equation}
    v_\mathrm{s,eff} \equiv \frac{\langle F_c\rangle}{\langle P_c + E_c\rangle} - \langle v\rangle. \label{eqn:effective_streaming_velocity}
\end{equation}
which is reduced by a factor of $\sim 2$ for the simulation shown. Plugging the escape valve for CRs leads to a larger overall CR pressure, required to sustain the same flux $F_c \approx 4 P_c (v + v_\mathrm{s,eff})$. This increase in the normalization of $P_c \propto 1/v_\mathrm{s,eff}$ (already apparent at the base, where $v=0$) is seen in the lower left panel of Fig. \ref{fig:time_avg}; it drives a stronger outflow. The advective flux increases to compensate for the decrease in streaming flux. The situation is similar to increasing the opacity in a radiation pressure driven wind -- buildup in radiation pressure drives a stronger outflow. This increase in wind driving can be divorced from CR energy losses. For instance, consider purely diffusive models, where there are no CR heating losses. Nonetheless, for a fixed CR injection power, $\dot{M} \propto 1/\kappa$ increases as diffusivity $\kappa$ falls, since the base CR pressure scales as $P_c \propto 1/\kappa$ \citep{quataert21-diffusion}. Similar effects occur in streaming models as the effective streaming speed falls.  

In Fig \ref{fig:alpha_beta_eta_avg}, we see that $\dot{M} \propto \beta_{0}^{-0.36}$. Why is the impact of CR staircases sensitive to the background $\beta_{0}$? As B-fields (and hence $v_A$) increases, the streaming flux is increasingly dominant over the advective flux, and thus the impact of bottlenecks grows. Furthermore, as $v_A/v_{\rm jump}$ increases, the attenuation of the bottleneck due to bump motion is lessened (equation \ref{eqn:new_bottleneck_with_flow}); deeper bottlenecks imply greater build-up of CR pressure and stronger outflows. Accordingly, we find in our simulations that the suppression factor $f= v_\mathrm{s,eff}/\langle v_{\rm A} \rangle$ falls with decreasing $\beta$. 

\citet{quataert21} see a similar strong increase in $\dot{M}$ as CR bottlenecks develop in their isothermal wind simulations. This is consistent with an observed change in the apparent equation of state in the CRs, from the expected $P_c \propto \rho^{2/3}$ in their highly sub-Alfvenic flow to $P_c \propto \rho^{1/2}$. We also see this apparent change in the effective equation of state in our simulations. In Fig.\ref{fig:two_suggestions}, we show the effective CR adiabatic index $\gamma_\mathrm{eff}$, defined by 
\begin{equation}
    \gamma_\mathrm{eff} = \dv{\ln{\langle P_c\rangle}}{\ln{\qty(\langle v\rangle + \langle v_A\rangle)^{-1}}} \label{eqn:eff_adiabatic_index}. 
\end{equation}
We find that $\gamma_{\rm eff} \approx 1.2$ rather than $4/3$, which naively corresponds to $P_c \propto \rho^{\gamma_{\rm eff}/2} \propto \rho^{0.6}$ in the sub-Alfvenic limit. \citet{quataert21} note that over a large radial range, $F_c \approx 4 P_c v_A \approx$const, which is consistent with $P_c \propto v_A^{-1} \propto \rho^{0.5}$. They also note that heating losses were $\sim 1/3$ of what one might expect from the time-averaged profile; if heating losses were negligible compared with the cosmic ray energy flux over a majority of the volume, this would explain $F_c \approx$ const. 


In our simulations, the change in energy losses is mild, even when $\dot{M}$ changes significantly. Here, we offer a slightly different interpretation, which relies on the role of moving bottlenecks in the CR flux. By themselves, bottlenecks do not change the equation of state (e.g., consider the stationary flow in Fig \ref{fig:bottle_bumps}, where $P_c \propto v_{\rm A}^{-4/3}$). However, the motion of the bottlenecks can change the apparent CR flux divergence and equation of state if not taken into account. For instance, as noted in \S\ref{sec:analytic}, bump motion reduces $\nabla \cdot F$, with $\nabla \cdot F_c \rightarrow 0$, $F_c \rightarrow$const for $v_{\rm bump} \gg v,v_A$. Consider highly sub-Alfvenic motion (e.g., in very low $\beta$ flows) , where one might expect $P_c \propto v_A^{-4/3} \propto \rho^{2/3} $. Instead, $v_{\rm A}$ falls at density jumps in shocks and can become comparable to $v-v_{\rm bump}$. Indeed, since the CRs are only coupled in dense regions, $v_{\rm A}$ should be evaluated here. From equation \ref{eqn:new_bottleneck_with_flow}, we have: 
\begin{equation}
    \tilde{\gamma}_\mathrm{eff} \equiv \dv{\ln{P_c}}{\ln(v + v_A)^{-1}} = \gamma_c\frac{v + v_A}{v + v_A - v_\mathrm{bump}}, 
    \label{eq:gamma_tilde}
\end{equation}
where we have defined $\tilde{\gamma}_\mathrm{eff}$ separately from $\gamma_\mathrm{eff}$ as it is not derived from time averaged quantities. Only for stationary bumps $v_\mathrm{bump} = 0$ do we recover $\tilde{\gamma}_\mathrm{eff} = 4/3$. If the bumps propagate {\it up} the gradient (i.e. $v_\mathrm{bump} < 0$), the bottlenecks reduce the CR flux compared to the pure streaming case and $\tilde{\gamma}_\mathrm{eff} < \gamma_c$. This is the canonical case for the acoustic instability. Conversely, if the bumps propagate {\it down} the gradient (i.e. $v_\mathrm{bump} > 0$), the bottlenecks enhance outward CR transport relative to the pure streaming case and $\tilde{\gamma}_\mathrm{eff} > \gamma_c$.  
However, if $(v+v_{\rm A})\gg v_{\rm bump}$, then $\tilde{\gamma}_{\rm eff} \rightarrow \gamma_{c}$. 
This is potentially at play in Fig.8 of \citet{quataert21}, which shows that while $P_c\propto\rho^{0.5}$ at the mid-range densities, at low densities (the outskirts, where flow becomes highly supersonic, with $v \gg v_{\rm bump}$), the effective adiabatic index steepens. While these effects are definitely present, whether they fully determine the change in apparent equation of state requires further quantitative study.

In summary, our simulation results are as follows: except in low $\beta$ environments, the changes in net heating and mass flux are generally modest, reaching at most 85-90\% and a factor of 2 respectively compared to no staircases. However, at low $\beta$, $\langle\dot{M}\rangle \propto \beta^{-0.36}$ changes more significantly, and can increase by an order of magnitude. This arises from the build-up of CR pressure due to stronger bottlenecks in low $\beta$ flows. Our simulation results are consistent with the higher $\beta$ ($\sim 1$) study by \citet{huang21} and low $\beta$ ($\ll 1$) study by \citet{quataert21}, the former reporting heating rates 95\% of the background profile, and the latter finding a change of a factor of $\sim 10$ for $\langle\dot{M}\rangle$. Note that these three studies all make different assumptions about cooling/thermodynamics, as well as geometry, so the overall broad agreement is reassuring. 

In our simulations, the time-averaged rate of global momentum and energy transfer is constrained if equilibrium is to hold. For instance, our cooling rates are time-steady, i.e. the total cooling luminosity of the simulation box is fixed. Hence, in global equilibrium, the time-averaged heat input from CRs -- either in the form of direct $v_A \cdot \nabla P_c$ heating, or from shocks (which are ultimately powered by CRs) must balance this constant rate, and cannot deviate too much. In simulations with realistic radiative cooling, the global cooling luminosity and the density profile could change significantly. This could strongly affect momentum/energy transfer from the CRs. This will be the subject of future work. 


\section{Discussion and Conclusions} \label{sec:discussion}

\subsection{Brief Summary} 
\label{subsec:summary}

In this paper, we carried out simulations of a CR driven acoustic instability \citep{begelman94}, focussing on the streaming-dominated limit. The condition for this instability is strong B-fields ($\beta < 0.5$), so that CR heating $v_{\rm A} \cdot \nabla P_c$, which drives the instability, is sufficiently important. In addition, a diffusion length $l_{\rm diffuse} \sim \kappa/c_s$ shorter than the background scale height $L_c$ is required. If this is not satisfied, sound waves will still be unstable, but the staircase structure we focus on is washed out by diffusion. The instability becomes stronger at smaller lengthscales, with the growth time $t_{\rm grow} \sim \kappa/c_c^2 \sim \kappa \rho/P_c$ becoming independent of wavelength at scales below the diffusion length  $l_{\rm diffuse}$. 

As sound waves steepen and become non-linear, they turn into a quasi-periodic sequence of shocks. The density jumps at the shocks in turn create bottlenecks for CR streaming,  resulting in a CR staircase structure. The jump widths are of order the diffusion length, while the jump heights depend on an equilibrium between staircase creation and mergers, and decrease with $P_c$. The CRs are uncoupled at staircase plateaus, but exert intense forces and heating at the staircase jumps. This rearrangement of CR pressure profiles has important consequences, which we now discuss. 

\subsection{Physical Significance}

Some key physical consequences the CR acoustic instability and ensuing CR staircases are: 
\begin{itemize} 
\item{{\it Shocks; density and velocity fluctuations.} The non-linear CR acoustic instability creates a propagating shock train. In our simulations, the shocks are initially fairly weak $\mathcal{M} \sim 1, \delta \rho/\rho \sim 1$, but they become stronger with the onset of cooling. The free energy for these shocks come from CRs, which thus result in significant density and velocity fluctuations. We anticipate this will drive turbulence in 2D and 3D simulations. These shocks are an important potential observational signature of the CR acoustic instability.}

\item{{\it Spatial and temporal fluctuations in CR forces and heating.} CRs provide a steady body force $\nabla P_c$ and heating $v_A \cdot \nabla P_c$ when there is a global background gradient. The CR staircase breaks this up into patchy, highly intermittent momentum and energy transfer where (at any given instant) the CRs are uncoupled with the gas throughout most of the volume, but exert intense forces and heating over narrow regions with widths of order the diffusion length. Since these stair steps and associated shocks are rapidly propagating, averaged over time the entire gas volume does gain momentum and energy from the cosmic rays, but in an intermittent and stochastic fashion. We expect the intermittency---similar to the highly intermittent and fluctuating nature of turbulent dissipation -- to become more apparent in 2D and 3D simulations. The departure from local momentum and energy balance can drive dynamical and thermal instability, which deserve in depth investigation. In our simulations, the sudden loss of CR heating in plateaus drives rapid cooling and large gas pressure fluctuations.} 

\item{{\it Changes in net momentum and energy transfer.} CR staircases also affect the net momentum and energy transfer averaged over space and time once the system has reached a steady state, $\Delta P_c, \Delta F_c$. 
In our simulations, these changes are relatively modest, although they could potentially be more significant in simulations with realistic radiative cooling where the energy source terms evolve. More importantly, the CR staircase can significantly change mass outflow rates $\dot{M}$, as also seen by \cite{quataert21}. We interpret this as due to the build up in CR pressure due to reduced streaming speeds at bottlenecks, which ultimately drives a stronger outflow as advective flux outcompetes CR streaming flux; this becomes progressively more important at lower $\beta$ where the bottlenecks are deeper and changes to CR streaming are stronger.}  
\end{itemize} 

\subsection{Applications}
\label{sect:applications} 

Can the acoustic instability and CR staircases arise in the CGM\footnote{It is likely to also be relevant in the ISM, but our focus here is on the CGM.}? Depending on gas pressure profiles, this requires $B \sim 0.5-{\rm few} \, \mu$G in the CGM. Observations of the galaxy halo magnetic fields are challenging and sparse. Recent observations using an FRB burst to observe Faraday rotation measured a parallel magnetic field $B_{\parallel} \sim 1 \mu$G of order the estimated equipartition magnetic field, such that $\beta \sim 1$ \citep{prochaska19}, modulo uncertainties such as field geometry. For instance, field reversals reduce the rotation measure and lead to an underestimate of $B_{\parallel}$. 
\citet{voort20} show from a suite of zoom-in cosmological simulations of galaxy formation that the plasma beta can reach as low as $0.01$ in regions that coincide with the biconical outflow. The magnetic field can acquire such dominance from turbulent dynamo action and metal enriched cooling. It is quite likely that $\beta$ fluctuates spatially in the CGM. Some regions may be unstable to the acoustic instability, while others are not.

If the acoustic instability is present, it has a very short growth time: 
\begin{equation}
    t_\mathrm{grow} = 15 \ \mathrm{Myr}\qty(\frac{\kappa}{10^{29}\ \mathrm{cm}^2\,\mathrm{s}^{-1}}) \qty(\frac{c_s}{150 \ \mathrm{km}\,\mathrm{s}^{-1}})^{-2}  \qty(\frac{P_c/P_g}{1}),
\end{equation}
where we have normalized to the (large) diffusion efficient $\kappa \sim 10^{29} {\rm cm^{2} \, s^{-1}}$ that appears necessary to avoid overproducing $\gamma$-rays at a level inconsistent with observations \citep{chan19}. This growth time is far shorter than the $0.1-1$ Gyr dynamical times typical of CGM processes (e.g., $L_c/c_s \sim 0.1$ Gyr for our fiducial parameters). The ratio of the diffusion length to the background scale height in galaxy halos is: 
\begin{equation}
    \eta = \frac{\kappa}{c_s L_c} \sim 0.1  \left(\frac{\kappa}{10^{29} {\rm cm^2 s^{-1}}} \right)
    \left(\frac{c_s}{150 \, {\rm km \, s^{-1}}} \right)^{-1} 
    \left(\frac{L_c}{20 \, {\rm kpc}} \right)^{-1} 
    \label{eq:eta_CGM} 
\end{equation}
which means that one can expect sharp staircase steps. 

Of course, the CGM is multi-phase, and the cooler $T \sim 10^{4}$K component is a critical component. Indeed, it is generally the only component we directly observe. At face value, it might appear from equation \ref{eq:eta_CGM} that we will not see the CR staircase in cooler $T \sim 10^{4}$K clouds, where both the sound speed
$c_s$ and CR scale height $L_c$ are much smaller. This is not correct, because the ambient diffusion coefficient adjusts to local conditions. In the self-confinement picture, diffusion expresses transport relative to the Alfven wave frame, and can be written as: 
\begin{equation}
\frac{\kappa}{v_A L_c} =  \frac{v_D}{v_A} -1 = \frac{l_{\rm mfp} c}{3 v_A L_c} \ll 1 \ \ \ {\rm (strong \ coupling)} 
\label{eq:kappa_linear} 
\end{equation} 
where $v_D$ is the drift speed relative to the Alfven wave frame, and $l_{\rm mfp}$ is the CR mean free path $l_{\rm mfp} \sim r_{g}/(\delta B/B)^{2}$, where $r_g$ is the CR gyroradius and the CR-excited Alfven wave amplitude $(\delta B/B)^{2}$ can be calculated in quasi-linear theory by balancing wave growth and damping rates \citep{farmer04,wiener13}. At $\sim$GeV energies (where most of the CR energy resides and the gyro-resonant streaming instability is strong), we expect $(v_D/v_A -1) \sim 0.01-0.1$; i.e., the CRs are tightly locked to the Alfven wave frame. See \citet{wiener17} for expressions relevant to coronal gas, and \citet{wiener17_cloud} for expressions relevant to $T\sim 10^{4}$K clouds and their interfaces with coronal gas. Our parameter $\eta$ is directly related to this measure of CR coupling: 
\begin{equation}
    \eta = \frac{\kappa}{c_s L_c} \sim 0.1 \left( \frac{v_D/v_A-1}{0.1} \right) \beta^{-1/2} 
\end{equation}
As a sanity check, note that for our fudicial assumptions of $c_s \sim 150 \, {\rm km \, s^{-1}}$, $L_c \sim 20$ kpc, $\beta \sim 1$ in the coronal gas, equation \ref{eq:kappa_linear} gives $\kappa \sim 10^{29} {\rm cm^{2} s^{-1}}$ for $(v_D/v_A -1) \sim 0.1$

It is also important to remember that CR staircases are not unique to the acoustic instability. They are seeded by density fluctuations, since overdense regions serve as streaming bottlenecks. They are agnostic as to the origin of these density fluctuations. Thus, overdensities created by thermal instability, or a network of overdense clouds in a multi-phase medium, can have similar effects. For this reason, CR staircases can show up in a wide range of scenarios. 

Some potential applications include: 
\begin{itemize} 
\item{{\it Galactic Winds.} Galactic winds driven by CRs have often been simulated in two limits: diffusive `extrinsic confinement', where CRs are scattered by extrinsic turbulence, and streaming-dominated `self confinement', where CRs are confined by Alfven waves they produce via the gyroresonant streaming instability. In the diffusive `extrinsic confinement' case, CRs do not heat the gas\footnote{The only energy exchange is slow Fermi II acceleration of the CRs.}. In the streaming dominated `self confinement' case, CR transport heats gas at a rate $v_A \cdot \nabla P_c$. The diffusive case fits $\gamma$ ray observations better, because CRs can propagate out of the galaxy faster \citep{chan19}. It is also much better at driving winds, because the CRs do not suffer strong energy losses via Alfven wave heating \citep{wiener17,hopkins20}. However, we expect self-confinement to be very strong at the $\sim$GeV energies where CR energy peaks \citep{kulsrud69,farmer04,wiener13}, while extrinsic compressible turbulence is strongly damped at small scales, and unlikely to efficiently scatter $\sim$GeV CRs \citep{yan02}. Thus, CR winds should be streaming dominated and relatively inefficient. The CR staircase changes these dichotomies by changing the structure of the wind. We have seen how CR pressure can build up in streaming dominated simulations, due to trapping at bottlenecks. This  increases mass outflow rates, similar to the effect of increased opacity in radiative outflows. In CR streaming simulations of isothermal winds where the CR acoustic instability arose, \citet{quataert21} found an increase in wind mass loss rates by an order of magnitude, compared to analytic models without a CR staircase, illustrating the potential impact of CR staircases.  High resolution cosmological zoom simulations of CR staircases are actually well within reach. As seen in Appendix \S\ref{app:resol_rspl}, all that is required is that the diffusion length $l_{\rm diff} \sim \kappa/c_s \sim 2 \, {\rm kpc} \, \left(\frac{\kappa}{10^{29} {\rm cm^2 s^{-1}}} \right)
    \left(\frac{c_s}{150 \, {\rm km \, s^{-1}}} \right)^{-1}$ is resolved. However, to date only the FIRE collaboration has implemented the two moment method (capable of dealing with CR streaming) in such simulations, and-- in contrast to, for instance, \citet{voort20} -- the plasma $\beta$ in their winds is too high for the acoustic instability to develop \citep{hopkins20}. But alternate setups where CR staircases appear are certainly numerically feasible.}

\item{{\it Thermal Instability.} As already seen in this paper, the patchy nature of heating due to a staircase structure can play an important role in thermal instability, particularly if CR heating is significant in the background equilibrium profile. The sudden loss of CR heating at plateaus triggers rapid cooling. The large gas pressure gradients and density fluctuations provide unusually non-linear, large-scale perturbations. It would be particularly interesting to see in 2D and 3D simulations if the high gas pressure gradients trigger `shattering' of condensing large scale patches of cold gas, creating a `fog' of cloudlets \citep{mccourt18,gronke20-mist}. The train of shocks which propagating over condensing cold gas can also play a role in subsequent dynamics, breaking up the cold gas further and driving baroclinic vorticity.}

\item{{\it Thermal Interfaces.} CRs provide pressure support and heating to the interfaces between warm ($T \sim 10^4$K) photoionized gas and hot ($T \sim 10^6$K) coronal gas, thickening them and setting a characteristic temperature scale height. Similar to the case with thermal conduction, it is possible to solve for the steady state structure of CR mediated fronts \citep{wiener17_cloud}. These fronts are currently unresolved in simulations of cloud acceleration \citep{bustard20,bruggen20} and their structure influences the strength of the `bottleneck' and hence the momentum that is deposited towards cloud acceleration. It is therefore important to understand them in detail. The interfaces can be magnetically dominated due to flux freezing as hot gas condenses onto the interface \citep{gronke20,butsky20}. Therefore they are a likely breeding ground for the CR acoustic instability. If a CR staircase appears, the spatially fluctuating pressure and thermal balance triggers mixing, shocks and turbulence, which in term create dissipation and diffuse heat transport. The long term stability and structure of such fronts could change significantly, affecting the mass flux between the phases as well as observational diagnostics such as the ratio of low to high ionic species (e.g. N(CIV)/N(OVI)). }

\item{{\it Observational Signatures.} Although the study of CR driven winds have become an intense area of activity, observational constraints are unfortunately few and far between. If seen, the quasi-periodic network of shocks due to the CR acoustic instability could provide a sorely needed observational diagnostic of the presence of cosmic rays in galaxy halos. For instance, they could potentially create wide-spread radio synchrotron emission from CR acceleration at shocks, at a level and with spectral indices inconsistent with transport of CR electrons out of galaxies, due to rapid synchrotron and inverse Compton cooling. The resultant density fluctuations could also potentially be probed by frequency-dependent temporal broadening of radio waves from Fast Radio Bursts \citep{macquart13,prochaska19} passing through intervening galaxy halos. The challenge is in disentangling these effects from other sources of shocks and turbulence. Presumably the closely spaced, wide-spread nature of the shock train, as well as accompanying signatures of CRs (gamma-rays, synchrotron emission) help, but this must be studied in more detail. For instance, the passage of multiple weak shocks leaves a distinct spectral signature, with the spectrum flattening and the shock acceleration efficiency increasing at each shock \citep{kang21}.}
\end{itemize} 

\subsection{Looking Forward}

This paper is a first detailed study of CR staircases, which we expect to have broad applicability. Indeed, CR staircases due to the acoustic instability have just appeared in two recent preprints \citep{huang21,quataert21}. More work is needed to clarify the impact of CR staircases on the interaction between gas and CRs. Some of the most pressing improvements include: (i) 2D and 3D MHD simulations, to assess the role of B-field geometry (particularly tangled magnetic fields, spatially varying B-fields, MHD forces and MHD acoustic modes), as well as the role of turbulence. For instance, in winds, one might expect the flow to develop significant anisotropy, depending on where bottlenecks develop and how field lines warp in response. (ii) Better treatment of the thermodynamics, and more realistic cooling functions. This is particularly important in assessing cooling at CR plateaus and the development of thermal instability. (iii) Exploring parameter space with a wider range of background profiles which are less highly idealized.

\section*{Acknowledgements}

We thank Chad Bustard, Shane Davis, Eliot Quataert, Huang Xiaoshan for helpful discussions. We acknowledge NSF grant AST-1911198 and XSEDE grant TG- AST180036 for support. This research was supported in part by the National Science Foundation under Grant No. NSF PHY- 1748958 to KITP. The Center for Computational Astrophysics at the Flatiron Institute is supported by the Simons Foundation.

\section*{Data Availability}

The data underlying this article will be shared on reasonable request to the corresponding author. 



\bibliographystyle{mnras}
\bibliography{main} 




\appendix

\section{Linear Growth Rates in 1D including Background Gradient} \label{app:linear_growth_rates}

Here, we provide a concise derivation of linear growth rates for the acoustic instability. More details can be found in \citet{begelman94}. 

\subsection{Adiabatic EOS for Finite Diffusion Coefficient} \label{subsec:adiabatic_finite_kappa}

In the well coupled limit, the time-dependent flux term in equation \ref{eqn:cr_flux} can be ignored, reducing equations \ref{eqn:continuity}-\ref{eqn:cr_flux} to the one-moment equations. Expressing the equations in 1D and in primitive form,
\begin{gather}
    \pdv{\rho}{t} + \pdv{x}\qty(\rho v) = 0 \label{eqn:continuity_1d} \\
    \pdv{v}{t} + v\pdv{v}{x} = -\frac{1}{\rho}\pdv{x}\qty(P_g + P_c) + \rho g \label{eqn:momentum_1d} \\
    \pdv{P_g}{t} + v\pdv{P_g}{x} + \gamma_g P_g \pdv{v}{x} = -\qty(\gamma_g - 1) v_A\pdv{P_c}{x} + \qty(\gamma_g - 1)\mathcal{L} \label{eqn:energy_1d} \\
    \pdv{P_c}{t} + \qty(v + v_A)\pdv{P_c}{x} = -\gamma_c P_c\pdv{x}\qty(v + v_A) + \pdv{x}\kappa\pdv{P_c}{x} \label{eqn:cr_flux_1d} \\
\end{gather}
For simplicity we assume the diffusion coefficient $\kappa$ is constant. We perform a WKB analysis similar to \citet{drury86}. Assume all quantities $\hat{Y}$ can be expanded as a background plus fluctuating part
\begin{equation}
    \hat{Y}\qty(x, t) = Y\qty(x) + \tilde{Y}\qty(x, t), \label{eqn:sum_background_fluc}
\end{equation}
where $\tilde{Y}\ll Y$. Keeping terms up to the first order in the fluctuating quantities gives
\begin{gather}
    \pdv{\tilde{\rho}}{t} + \pdv{x}\qty(\rho\tilde{v} + \tilde{\rho} v) = 0, \label{eqn:perturbed_continuity} \\
    \pdv{\tilde{v}}{t} + v\pdv{\tilde{v}}{x} + \tilde{v}\pdv{v}{x} = - \frac{1}{\rho}\pdv{\tilde{P}_g}{x} - \frac{1}{\rho}\pdv{\tilde{P}_c}{x} + \frac{\tilde{\rho}}{\rho^2}\pdv{P_g}{x} + \frac{\tilde{\rho}}{\rho^2}\pdv{P_c}{x}, \label{eqn:perturbed_momentum} \\
    \pdv{\tilde{P}_g}{t} + v\pdv{\tilde{P}_g}{x} + \tilde{v}\pdv{P_g}{x} + \gamma_g P_g\pdv{\tilde{v}}{x} + \gamma_g\tilde{P}_g\pdv{v}{x} = \nonumber \\
    \qquad -\qty(\gamma_g - 1) v_A\pdv{\tilde{P}_c}{x} + \qty(\gamma_g - 1)\frac{v_A}{2\rho}\tilde{\rho}\pdv{P_c}{x} + \qty(\gamma_g - 1)\qty(\tilde{\rho}\pdv{\mathcal{L}}{\rho} + \tilde{T}\pdv{\mathcal{L}}{T}), \label{eqn:perturbed_energy} \\
    \pdv{\tilde{P}_c}{t} + \qty(v + v_A)\pdv{\tilde{P}_c}{x} + \qty(\tilde{v} + \tilde{v}_A)\pdv{P_c}{x} = \nonumber \\
    \qquad -\gamma_c P_c\pdv{x}\qty(\tilde{v} + \tilde{v}_A) - \gamma_c \tilde{P_c}\pdv{x}\qty(v + v_A) + \kappa\pdv[2]{\tilde{P_c}}{x}. \label{eqn:perturbed_cr_energy} 
\end{gather}
In WKB analysis we assume the fluctuating length and timescales are much smaller than the scales on which the background varies. We express the fluctuating quantities as
\begin{equation}
    \tilde{Y}\qty(x, t) = \sum_{n=0}^{\infty}\epsilon^n Y_n\qty(x, t) e^{i\theta/\epsilon}, \label{eqn:WKB_expansion}
\end{equation}
where $\epsilon$ is a small parameter and $\pdv*{\theta}{t} = \omega, \pdv*{\theta}{x} = -k$. Note that $\pdv*{\omega}{x} + \pdv*{k}{t} = 0$. Substituting into equation \ref{eqn:perturbed_continuity}-\ref{eqn:perturbed_cr_energy}, we find to the lowest order $\epsilon^{-2}$,
\begin{equation}
    k^2\kappa P_{c0} = 0, \implies P_{c0} = 0. \label{eqn:order_epsilon_2}
\end{equation}
To order $\epsilon^{-1}$,
\begin{gather}
    \bar{\omega}\rho_0 = k\rho v_0, \label{eqn:order_epsilon_1_continuity} \\
    \bar{\omega}\rho v_0 = k P_{g0}, \label{eqn:order_epsilon_1_momentum} \\
    \bar{\omega} P_{g0} = k\gamma_g P_g v_0, \label{eqn:order_epsilon_1_energy} \\
    k^2\kappa P_{c1} = i k\gamma_c P_c\qty(v_0 - \frac{v_A}{2\rho}\rho_0), \label{eqn:order_epsilon_1_cr_energy}
\end{gather}
where $\bar{\omega} = \omega - k v$. Solving for $\bar{\omega}$ from equation \ref{eqn:order_epsilon_1_continuity}-\ref{eqn:order_epsilon_1_energy} we obtain the dispersion relation of a sound wave
\begin{equation}
    \bar{\omega} = \pm k c_s, \label{eqn:dispersion} 
\end{equation}
where $c_s = \sqrt{\gamma_g P_g/\rho}$. To order $\epsilon^0$, using relations \ref{eqn:order_epsilon_2} through \ref{eqn:dispersion}, an equation for the action density $\mathcal{A}$, defined by 
\begin{equation}
    \mathcal{A} = \frac{\rho v_0^2}{\bar{\omega}}, \label{eqn:action_density}
\end{equation}
can be derived
\begin{gather}
    \pdv{\mathcal{A}}{t} + \pdv{x}\qty[\qty(v\pm c_s)\mathcal{A}] = \frac{\mathcal{A}}{\rho c_s^2}\gamma_g\qty(\gamma_g - 1)\qty(v_A\pdv{P_c}{x} - \mathcal{L}) \nonumber \\
    \qquad -\frac{c_c^2\mathcal{A}}{\kappa}\qty[1\pm\qty(\gamma_g - 1)\frac{v_A}{c_s}]\qty(1\mp\frac{v_A}{2 c_s}) \nonumber \\
    \qquad \pm\frac{\mathcal{A}}{\rho c_s}\qty(1\pm\qty(\gamma_g - 1)\frac{v_A}{2 c_s})\pdv{P_c}{x} \nonumber \\
    \qquad + \mathcal{A}\frac{\qty(\gamma_g - 1)}{c_s^2}\qty(\pdv{\mathcal{L}}{\rho} + \qty(\gamma_g - 1)\frac{T}{\rho}\pdv{\mathcal{L}}{T}), \label{eqn:wave_action_adiabatic}
\end{gather}
where $c_c = \sqrt{\gamma_c P_c/\rho}$. This equation governs the evolution of the wave action density as it propagates through a background. The LHS describes the adiabatic change due to a varying background whereas the RHS describes genuine growth/damping. Without loss of generality, we group the prefactors of $\mathcal{A}$ on the RHS into a term $\mathcal{G}\qty(x)$ such that
\begin{equation}
    \pdv{\mathcal{A}}{t} + \pdv{x}\qty[\qty(v\pm c_s)\mathcal{A}] = \mathcal{G}\qty(x)\mathcal{A}. \label{eqn:simplified_action}
\end{equation}
Growth occurs when $\mathcal{G} > 0$ while damping occurs otherwise. For purpose of linear analysis assume the velocity perturbation has a form
\begin{equation}
    v_0\qty(x, t) = \hat{v}\qty(x)\exp{i\omega t - i k x} \label{eqn:velocity_perturbation}
\end{equation}
and the background gradients can be neglected over some region $x_\mathrm{inj}$ to $x$ such that $\omega, k$ can be considered constants, it can be easily shown that
\begin{equation}
    \pdv{x}\ln{\rho\hat{v}^2} = \pm\frac{\mathcal{G}}{c_s}.  \label{eqn:amplitude_equation}
\end{equation}
Solving gives
\begin{equation}
    \hat{v}\qty(x) = \hat{v}\qty(x_\mathrm{inj})\exp{\frac{1}{2}\ln{\frac{\rho_\mathrm{inj}}{\rho}} + \frac{1}{2} \mathcal{I}\qty(x, x_\mathrm{inj})}, \label{eqn:amplitude_track}
\end{equation}
where $\mathcal{I}\qty(x, x_\mathrm{inj})$, given by
\begin{equation}
    \mathcal{I}\qty(x,x_\mathrm{inj}) = \int_{x_\mathrm{inj}}^{x}\pm\frac{\mathcal{G}}{c_s}\dd{x}, \label{eqn:integral_tracking} 
\end{equation}
is the integral of the RHS of \ref{eqn:amplitude_equation} from the location where the wave is injected $x_\mathrm{x,inj}$ to some location $x$ later in its path. The first term within the brace bracket of \ref{eqn:amplitude_track} denotes the adiabatic change in wave amplitude due to background profile change while the second term represent that due to genuine growth. The phase velocity of a sound wave is $\dv*{x}{t} = \pm c_s$, so $\mathcal{I}$ in \ref{eqn:integral_tracking} is equivalent to integrating the function $\mathcal{G}$ over time from the moment of injection to some later time $t$
\begin{equation}
    \mathcal{I}\qty(x,x_\mathrm{inj}) = \int^t_{t_\mathrm{inj}}\mathcal{G}\dd{t'}. \label{eqn:integral_equivalent}
\end{equation}
Differentiating the expression within the brace bracket by time $t$ we obtain an expression for the growth rate $\Gamma_\mathrm{grow}$
\begin{align}
    \Gamma_\mathrm{grow} = \frac{\mathcal{G}}{2}. \label{eqn:growth_rate_full}
\end{align}

\subsection{Adiabatic EOS with a Small Diffusion Coefficient} \label{subsec:adiabatic_small_kappa}

If the diffusion coefficient $\kappa$ were small such that the term $k^2 \kappa P_{c0}$ is of the same order as the other perturbed terms in the CR energy equation, equation \ref{eqn:order_epsilon_2} may not be valid. This implies $P_{c0}\neq 0$. Including this term at order $\epsilon^{-1}$ yields
\begin{gather}
    \bar{\omega}\rho_0 = k\rho v_0, \label{eqn:order_epsilon_1_continuity_small_diff} \\
    \bar{\omega}\rho v_0 = k P_{g0} + k P_{c0}, \label{eqn:order_epsilon_1_momentum_small_diff} \\
    \bar{\omega} P_{g0} = k\gamma_g P_g v_0 + \qty(\gamma_g - 1) k v_A P_{c0}, \label{eqn:order_epsilon_1_energy_small_diff} \\
    \qty(\bar{\omega} - k v_A - i k^2\kappa) P_{c0} = k\gamma_c P_c\qty(v_0 - \frac{v_A}{2\rho}\rho_0). \label{eqn:order_epsilon_1_cr_energy_small_diff}
\end{gather}
Rearranging, we obtain
\begin{gather}
    \bar{\omega}\qty(\bar{\omega}^2 - k^2 c_s^2)\qty(\omega - k v_A - i k^2\kappa) = \nonumber \\
    \qquad k^2 c_c^2\qty[\bar{\omega} + \qty(\gamma_g - 1) k v_A]\qty(\bar{\omega} - \frac{k v_A}{2}) \label{eqn:small_diff_dispersion}
\end{gather}
as the dispersion equation. In the limit where $k\kappa/c_s\to\infty$ we recover the gas acoustic mode $\omega \approx \pm  k c_s$, though at moderate values of $k\kappa/c_s$ the gas acoustic mode is clearly not a solution. This equation has been solved in various limits in \citet{begelman94}. In particular, in the limit $v_A\gg c_c\gg c_s$, an unstable hybrid mode with phase speed intermediate between the gas sound speed and the Alfven speed appears
\begin{equation}
    \bar{\omega}^3 = \frac{\qty(\gamma_g - 1) k^3 v_A^2 c_c^2}{2}\frac{v_A - i k\kappa}{v_A^2 + k^2\kappa^2}. \label{eqn:hybrid}
\end{equation}
For $k\kappa\ll v_A$
\begin{equation}
    \bar{\omega} = \qty[\frac{\qty(\gamma_g -1) k^3 v_A c_c^2}{2}]^{1/3}\qty(-\frac{1}{2} - \frac{\sqrt{3}}{2} i), \label{eqn:hybrid_non_diff}
\end{equation}
while for $k\kappa\gg v_A$
\begin{equation}
    \bar{\omega} = \qty[\frac{\qty(\gamma_g -1) k^2 v_A^2 c_c^2 }{2\kappa}]^{1/3}\qty(\pm\frac{\sqrt{3}}{2} - \frac{1}{2} i). \label{eqn:hybrid_diff}
\end{equation}
These modes are mediated by gas pressure perturbations, but are driven unstable by CR heating. The growth rate scales as the wavenumber so higher resolution simulations can potentially seed faster growth. The transition from the acoustic mode to these hybrid modes occurs at $k\kappa/c_s \sim 1$. 

Solving equation \ref{eqn:small_diff_dispersion} numerically, one finds that the growth rate for $k\kappa/c_s \lesssim 1$ increases with wavenumber (equation \ref{eqn:hybrid_non_diff}) and then flattens off with respect to wavenumber for $k\kappa/c_s\gtrsim 1$ (as one would expect from looking at the RHS of equation \ref{eqn:wave_action_adiabatic}, which is independent of $k$). As discussed in \S\ref{sec:simulation}, for converged simulations, the diffusion length must be resolved. This implies that in the simulations, our fastest growing modes are always in the limit $k\kappa/c_s\gtrsim 1$, and hence we are dominated by acoustic modes. 

\subsection{Isothermal EOS with Finite Diffusion Coefficient} \label{subsec:isothermal_finite_kappa}

For isothermal EOS, equation \ref{eqn:perturbed_energy} is ignored. The gas pressure relates to the density by
\begin{equation*}
    P_g = c_s^2\rho, \label{eqn:isothermal_eos_pg}
\end{equation*}
with the sound speed $c_s$ a constant. Repeating the calculation above gives
\begin{equation}
    \bar{\omega} = \pm k c_s \label{eqn:dispersion_isothermal}
\end{equation}
as the dispersion relation and 
\begin{equation}
    \pdv{\mathcal{A}}{t} + \pdv{x}\qty[\qty(v\pm c_s)\mathcal{A}] = \pm\frac{\mathcal{A}}{\rho c_s}\pdv{P_c}{x} - \frac{c_c^2}{\kappa}\mathcal{A}\qty(1\mp\frac{v_A}{2 c_s}) \label{eqn:wave_action_isothermal}
\end{equation}
as the wave action equation, which is simply equation \ref{eqn:wave_action_adiabatic} with $\gamma_g = 1$ and without the heating/cooling terms. Condition for genuine growth is again
\begin{equation}
    \Gamma\qty(x) = \pm\frac{1}{\rho c_s}\pdv{P_c}{x} - \frac{c_c^2}{\kappa}\qty(1\mp\frac{v_A}{2 c_s}) > 0. \label{eqn:iso_growth_condition}
\end{equation}

\section{Resolution and Reduced Speed of Light Study} \label{app:resol_rspl}

Acoustic waves with wavelengths much shorter than the diffusion length $l_\mathrm{diff} = \kappa/c_s$ grow in the linear phase at a rate independent of the wavelength, as discussed in \S\ref{subsec:linear_theory} and \S\ref{app:linear_growth_rates}. If the diffusion length is well resolved, the characteristic staircase scales should $\sim l_\mathrm{diff}$ (see \S\ref{subsubsec:widths_heights}). As the resolution decreases, so that the diffusion length is no longer resolved, the wavelength of the growing modes will also increase. In particular, for $k l_\mathrm{diff} \lesssim 1$, the acoustic mode will bifurcate into hybrid modes which propagate at some modified sound speed, with growth rate that decreases linearly with the wavenumber $k$ (see \S\ref{subsec:adiabatic_small_kappa}). Thus, decreasing resolution will 1. cause slower growth of the staircase and 2. smooth out small scale stairs and render stair sizes larger.

In this section we rerun the test case \texttt{NLalpha1beta1eta.01phi2} (table \ref{tab:test_cases}) with several resolutions and reduced speed of light $c$, comparing their time averaged mass flux $\dot{M}$, $\Delta P_c$ and $\Delta F_c$. We shall also discuss the effect of resolution on the distributions of stair width, plateau width and jump height. A summary of the resolution, reduced speed of light and time averaged quantities is drawn up in table \ref{tab:resol_c_list}.

\begin{table*}
    \centering
    \begin{tabular}{c|c|c|c|c}
        \multicolumn{5}{c}{Test case: \texttt{NL4096alpha1beta1eta.01ms.015phi2c200}} \\
        \hline
        \hline
        Resolution $\Delta x$ ($\langle l_\mathrm{diff}\rangle/\Delta x$) & $c$ & $\langle\dot{M}\rangle/\dot{M}_0$ & $\langle\Delta P_c\rangle/\Delta P_{c0}$ & $\langle\Delta F_c\rangle/\Delta F_{c0}$ \\
        \hline
        $7.03\times 10^{-2}$ ($0.0588$) & $200$ & 1.155 & 1.204 & 0.951 \\
        $3.52\times 10^{-2}$ ($0.1168$) & $200$ & 1.282 & 1.270 & 0.963 \\
        $1.76\times 10^{-2}$ ($0.233$) & $200$ & 1.257 & 1.319 & 0.982 \\
        $8.79\times 10^{-3}$ ($0.465$) & $200$ & 1.355 & 1.339 & 0.955 \\
        $4.39\times 10^{-3}$ ($0.926$) & $200$ & 1.365 & 1.353 & 0.933 \\
        $2.20\times 10^{-3}$ ($1.85$) & $200$ & 1.388 & 1.304 & 0.896 \\
        $1.10\times 10^{-3}$ ($3.70$) & $200$ & 1.339 & 1.379 & 0.914 \\
        $5.49\times 10^{-4}$ ($7.41$) & $200$ & 1.449 & 1.407 & 0.924 \\
        $5.49\times 10^{-4}$ ($7.41$) & $400$ & 1.408 & 1.395 & 0.918 \\
        $1.37\times 10^{-4}$ ($25.9$) & $400$ & 1.465 & 1.339 & 0.900 \\
        \hline
    \end{tabular}
    \caption{Re-running with different resolutions and reduced speed of light. Column 1: Resolution given in grid spacing with (the bracketed quantities show the number of grids the mean diffusion length is resolved with, i.e. $\langle l_\mathrm{diff}\rangle/\Delta x$). Column 2: Reduced speed of light. Column 3-5: Time averaged mass flux $\dot{M}$, $\Delta P_c$ and $\Delta F_c$ (in units of the initial, unperturbed $\dot{M}_0$, $\Delta P_{c0}$ and $\Delta F_{c,0}$).}
    \label{tab:resol_c_list}
\end{table*}

\begin{figure}
    \centering
    \includegraphics{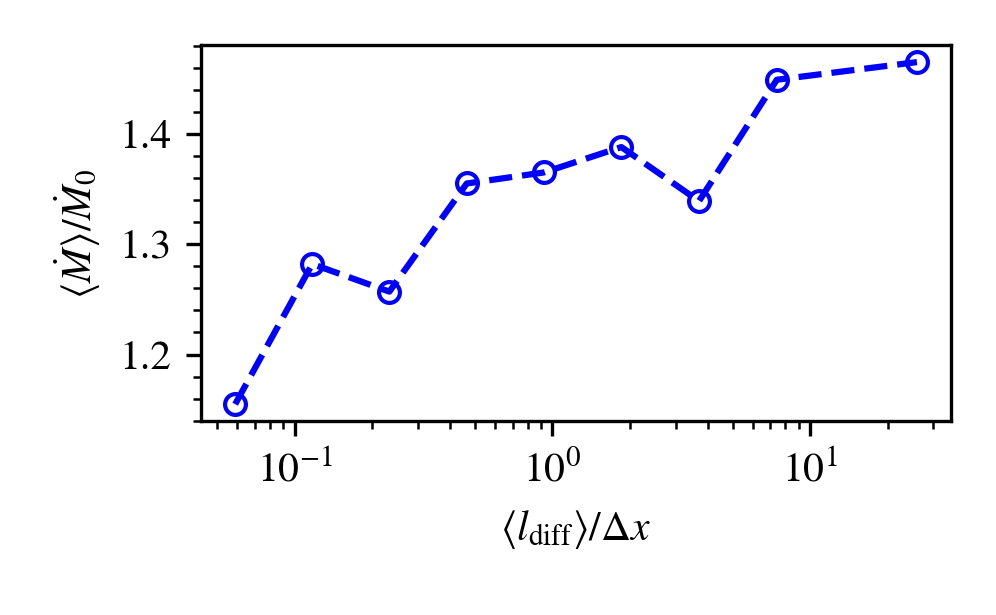} \\
    \includegraphics{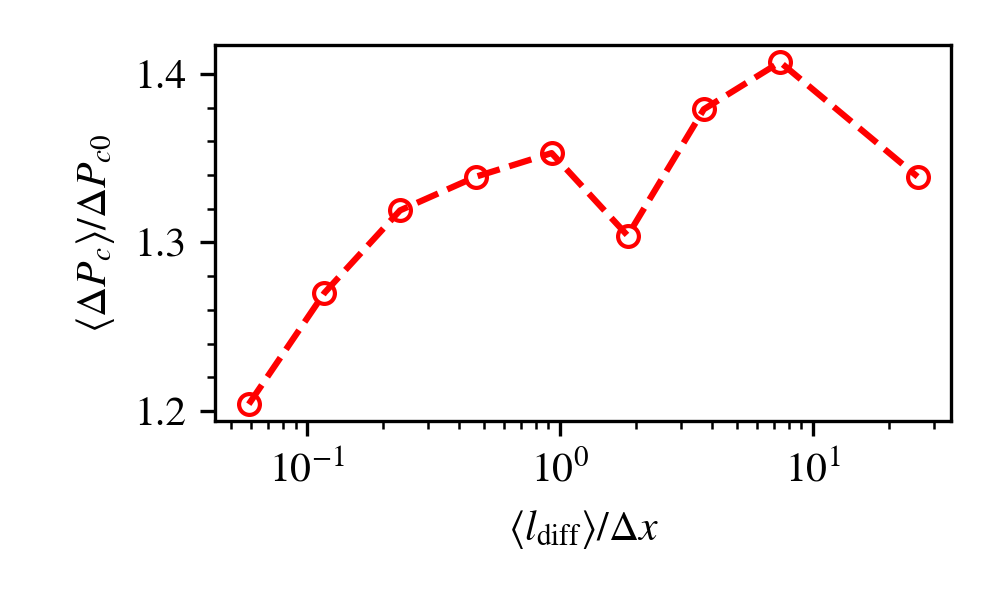} \\
    \includegraphics{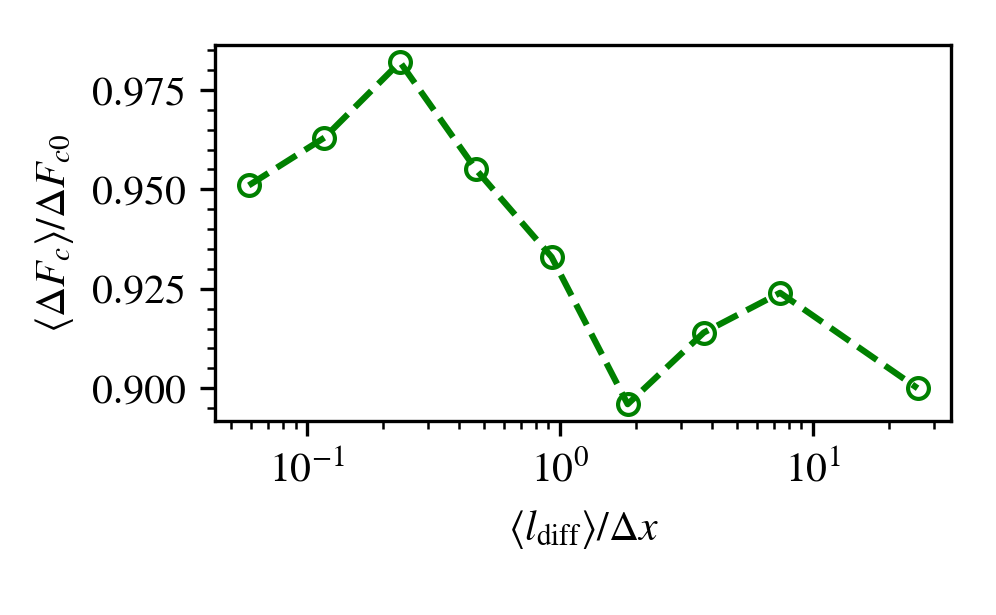}
    \caption{Time average quantities ($\dot{M}/\dot{M}_0$, $\Delta P_c/\Delta P_{c0}$, $\Delta F_c/\Delta F_{c0}$) as function of resolution. Resolution given in the x-axis denotes the number of grids the mean diffusion length is resolved with ($\langle l_\mathrm{diff}\rangle/\Delta x$), the larger this is the higher the resolution.}
    \label{fig:resol_avg}
\end{figure}

\begin{figure}
    \centering
    \includegraphics{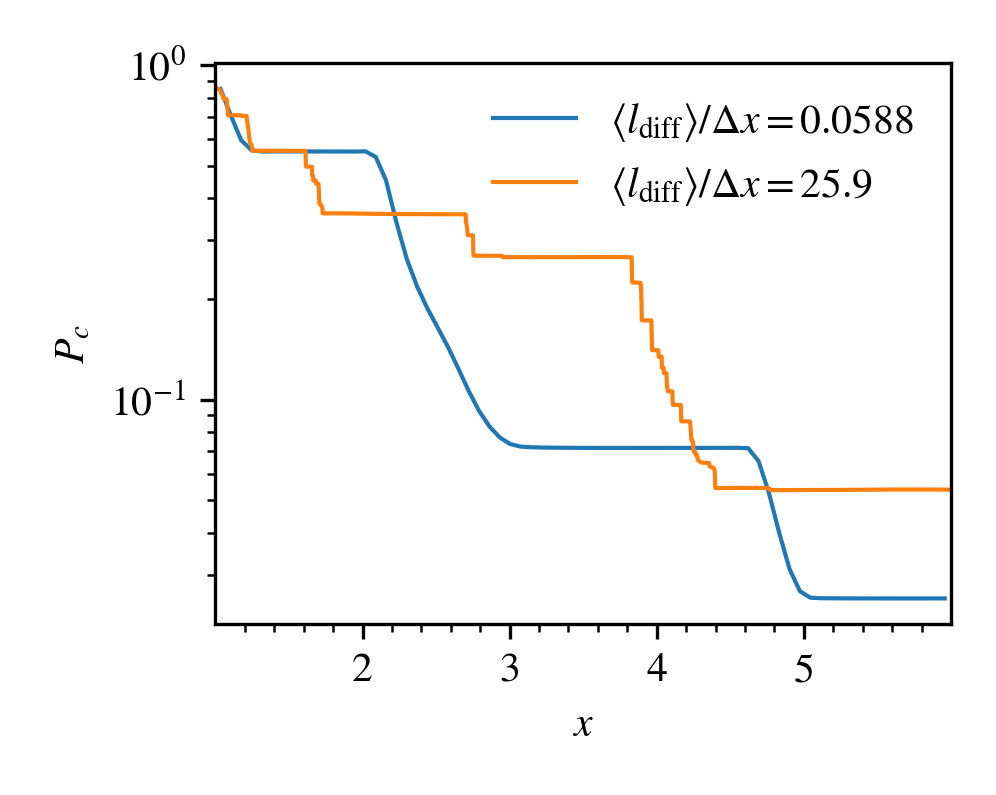}
    \caption{$P_c$ profile taken at the same time instance for a low ($\langle l_\mathrm{diff}\rangle/\Delta x = 0.0588$) and high resolution run ($\langle l_\mathrm{diff}\rangle/\Delta x = 25.9$).}
    \label{fig:pc_resol}
\end{figure}

In fig.\ref{fig:resol_avg} we plot $\dot{M}/\dot{M}_0$, $\Delta P_c/\Delta P_{c0}$ and $\Delta F_c/\Delta F_{c0}$ as function of $\langle l_\mathrm{diff}\rangle/\Delta x$, the number of grids the mean diffusion length is resolved with. Overall, despite small fluctuations at large $\langle l_\mathrm{diff}\rangle/\Delta x$, the time averaged quantities appear reasonably converged. Deviations appear when the mean diffusion length is under-resolved, i.e. $\langle l_\mathrm{diff}\rangle/\Delta x < 1$, yet even in the lowest resolution explored (i.e. $\langle l_\mathrm{diff}\rangle/\Delta x = 0.0588$), a staircase structure can be clearly seen (fig.\ref{fig:pc_resol}). Generally, effects of the staircase on $\dot{M}/\dot{M}_0,\Delta P_c, \Delta F_c$ dwindle with resolution in the under-resolved regime, yet even in the lowest resolution explored the time-averaged quantities deviate from the resolved runs by less than 20\%. This suggests effects on the time averaged quantities is due mainly to the bigger stairs, with minor modifications from the small stairs.

Visually inspecting fig.\ref{fig:pc_resol}, which shows the $P_c$ profile taken at the same time for the lowest and highest resolutions explored, it is observed that more small scale structures arise when the resolution is high. Only the largest jumps are resolvable at low resolution, details of the small scale jumps smoothed out. 

\begin{figure}
    \centering
    \includegraphics{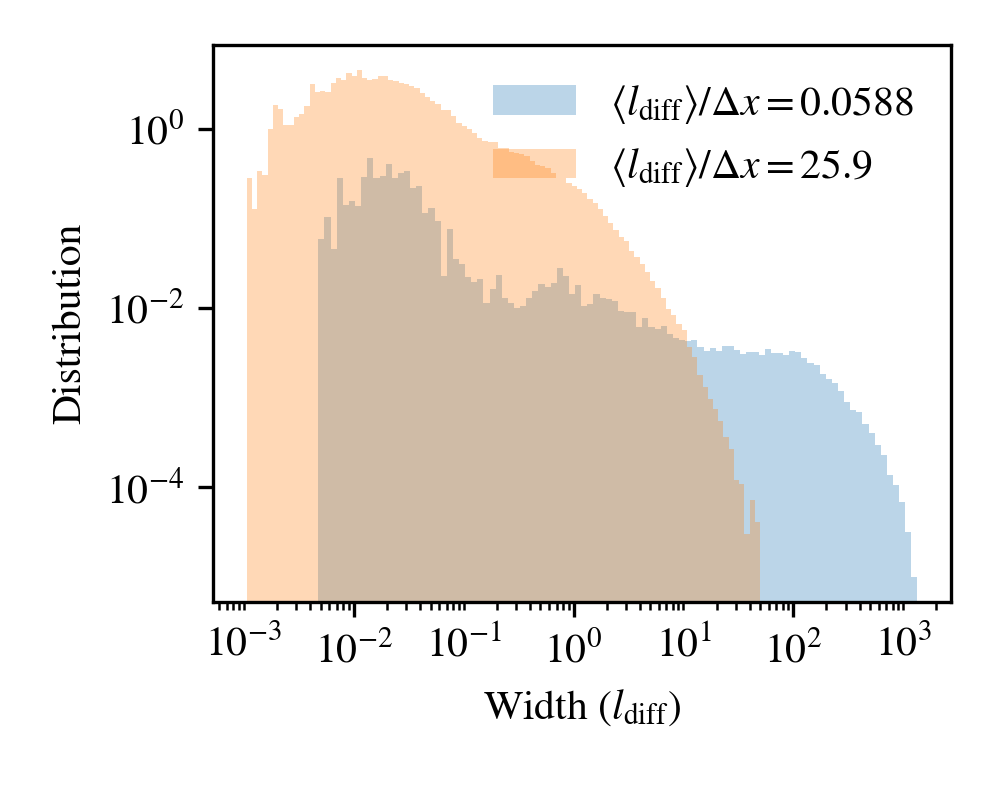} \\
    \includegraphics{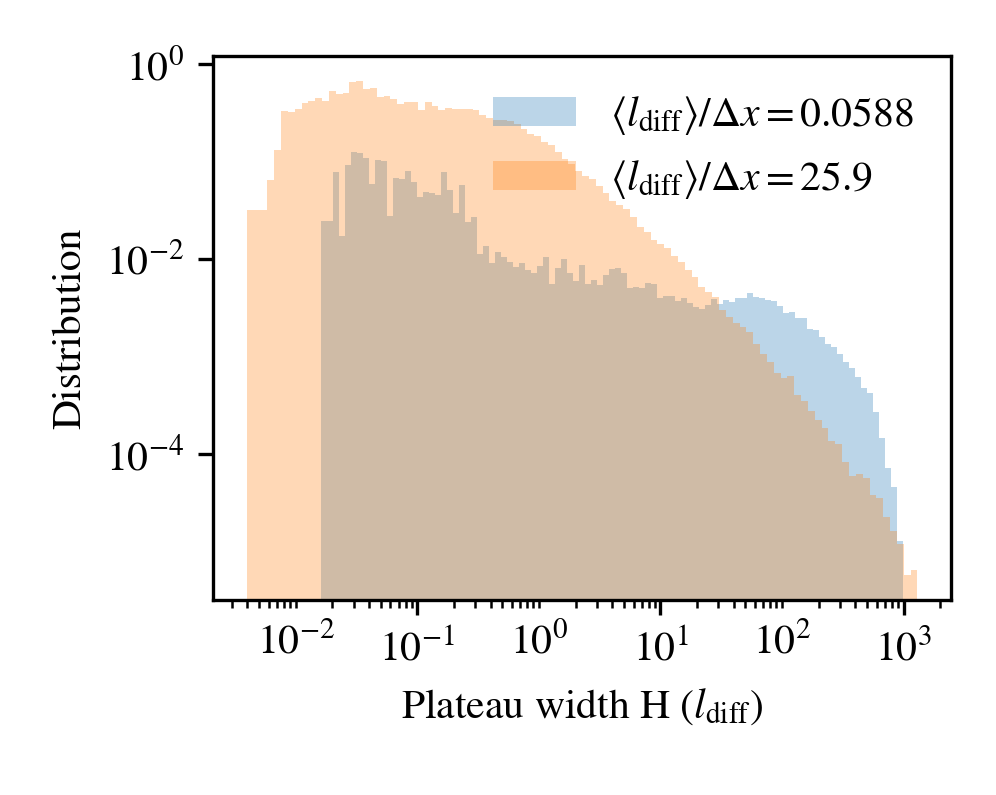} \\
    \includegraphics{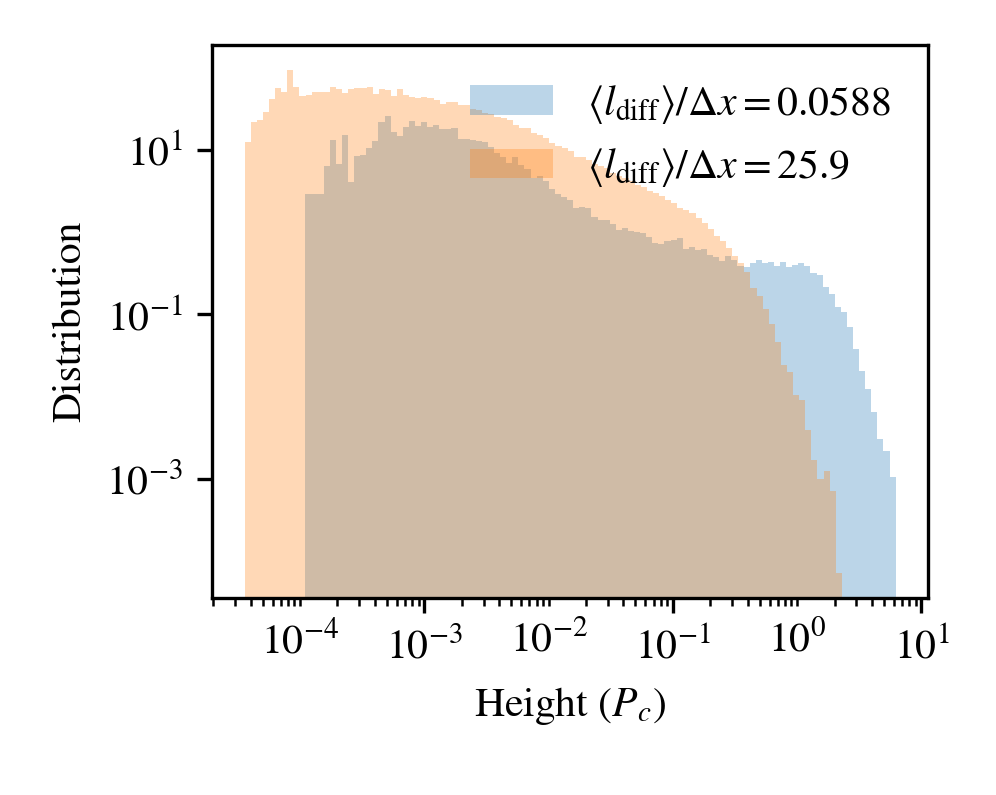}
    \caption{Distributions of jump width (top), plateau width (middle) and jump height (bottom) for low ($\langle l_\mathrm{diff}\rangle/\Delta x = 0.0588$) and high resolutions ($\langle l_\mathrm{diff}\rangle/\Delta x = 25.9$).}
    \label{fig:stair_reso}
\end{figure}

In fig.\ref{fig:stair_reso} we plot the distributions of stair width, plateau width and jump height for the highest and lowest resolutions explored, finding there to be more small scale structures (smaller widths and heights) for the more resolved run while the low resolution run have more large scale structures (larger widths and heights). This lies within expectation as under-resolving the diffusion length would cause small scale jumps (typically having size of the diffusion length) to be smoothed out into a bigger jump.

All in all, in practice (e.g. in galaxy scale simulations), for the purpose of eliciting the staircase and its time averaged effects, it appears acceptable to resolve the diffusion length by a few cells. However, should effects of individual stairs be important (e.g. cloud survival under bombardment of a few of these stairs), higher resolution is probably necessary.

On a shorter note, changing the reduced speed of light $c$ appears to have little effect on our results as long as it is much greater than any other velocity scales present (e.g. $c, c_s, c_c, v_A$). This is consistent with \citet{jiang18}, and we shall not pursue this further.


\bsp	
\label{lastpage}
\end{document}